\documentclass[aps,prd,preprint,superscriptaddress,tightenlines,nofootinbib]{revtex4}

\usepackage{epsf}% Include figure files
\usepackage{dcolumn}% Align table columns on decimal point
\usepackage{bm}% bold math
\usepackage{ifthen}

\def\epjC#1,#2(#3){{\rm Eur.\ Phys.\ J.\ }{\bf C#1}, {\rm#2} {\rm(#3)}}

\begin{document}
\newboolean{cbxNote}
\setboolean{cbxNote}{false}
%
%
%%%%%%%%%%%%%%%%%%%%%%%%%%%%%%%%%%%%%%%%%%%%%%%%%%
%                                                %
%    Title page info                             %
%                                                %
%%%%%%%%%%%%%%%%%%%%%%%%%%%%%%%%%%%%%%%%%%%%%%%%%%
\ifthenelse{\boolean{cbxNote}}{
\preprint{CBX 03-03}
}{
\preprint{CLNS 03/1819}
\preprint{CLEO 03-05}
}

\title{Study of the $q^2$--dependence of
$B\to\pi\ell\nu$ and $B\to\rho(\omega)\ell\nu$ Decay and Extraction of $|V_{ub}|$}

\ifthenelse{\boolean{cbxNote}}{
\author{V.~Boisvert}
\author{N.~E.~Adam}
\author{L.~Gibbons}
\author{T.~O.~Meyer}
\author{A.~Warburton}
\author{E.~H.~Thorndike}
}{
\author{S.~B.~Athar}
\author{P.~Avery}
\author{L.~Breva-Newell}
\author{V.~Potlia}
\author{H.~Stoeck}
\author{J.~Yelton}
\affiliation{University of Florida, Gainesville, Florida 32611}
\author{K.~Benslama}
\author{B.~I.~Eisenstein}
\author{G.~D.~Gollin}
\author{I.~Karliner}
\author{N.~Lowrey}
\author{C.~Plager}
\author{C.~Sedlack}
\author{M.~Selen}
\author{J.~J.~Thaler}
\author{J.~Williams}
\affiliation{University of Illinois, Urbana-Champaign, Illinois 61801}
% ...
%\author{(CLEO Collaboration)} %FOR PRD_SPECIAL_CHANGEME
\collaboration{CLEO Collaboration} %FOR PRL,CLNS
\noaffiliation
}

\date{April 10, 2003}

%:abstract
\begin{abstract}
We report on determinations of $|V_{ub}|$ resulting from studies of the 
branching fraction and $q^2$ distributions in exclusive semileptonic $B$ decays 
that proceed via the $b\to u$ transition. Our data set consists of the $9.7\times 10^6$ $B\overline{B}$ 
meson pairs collected at the $\Upsilon (4S)$ resonance with the CLEO II detector.  
We measure ${\cal B}(B^0\to\pi^-\ell^+\nu)=(1.33\pm 0.18 \pm0.11 \pm 0.01\pm 0.07)\times 10^{-4}$
and ${\cal B}(B^0\to\rho^-\ell^+\nu)=(2.17\pm 0.34\;^{+0.47}_{-0.54} \pm 0.41\pm 0.01)\times 10^{-4}$, 
where the errors are statistical, experimental systematic, systematic due to residual 
form--factor uncertainties in the signal,  and systematic due to residual form--factor uncertainties
 in the cross--feed modes, respectively. We also find ${\cal B}(B^+\to\eta\ell^+\nu)=(0.84 \pm 0.31 \pm 0.16 \pm0.09)\times 10^{-4}$, 
 consistent with what is expected from the $B\to\pi\ell\nu$ mode and quark model symmetries. We extract 
 $|V_{ub}|$ using Light-Cone Sum Rules (LCSR) for $0\le q^2<16$ GeV$^2$ and 
 Lattice QCD (LQCD) for 16 GeV$^2$ $\le q^2<q^2_{\rm max}$. Combining both intervals 
 yields $|V_{ub}|=(3.24\pm 0.22 \pm 0.13 \;^{+0.55}_{-0.39}\pm 0.09)\times 10^{-3}$ 
 for $\pi \ell\nu$, and $|V_{ub}|=(3.00\pm 0.21 \;^{+0.29}_{-0.35} \;^{+0.49}_{-0.38}\pm0.28)\times 10^{-3}$ 
 for $\rho\ell\nu$, where the errors are statistical, experimental systematic, 
 theoretical, and signal form-factor shape, respectively. Our combined value from both 
 decay modes is $|V_{ub}|=(3.17\pm 0.17 \;^{+0.16}_{-0.17} \;^{+0.53}_{-0.39}\pm0.03)\times 10^{-3}$.
\end{abstract}

\ifthenelse{\boolean{cbxNote}}{}{
%\vspace{1cm}
}
\maketitle

\clearpage

%
%
%%%%%%%%%%%%%%%%%%%%%%%%%%%%%%%%%%%%%%%%%%%%%%%%%%
%                                                %
%    BEGINNING OF TEXT                           %
%                                                %
%%%%%%%%%%%%%%%%%%%%%%%%%%%%%%%%%%%%%%%%%%%%%%%%%%

\ifthenelse{\boolean{cbxNote}}{
\section{Prologue}
This CBX note is identical to the PRD draft to be submitted for this analysis, though results
in some tables are reported here to an extra significant figure.  Almost every
experimental requirement is reported in the PRD, and the systematic evaluation is laid
out in some detail. All details that are not reported in this CBX are identical to those
laid out in V. Boisvert's thesis (http://www.lns.cornell.edu/public/THESIS/2002/THESIS02-2/boisvert.ps).
Note that the final results presented here differ from the preliminary results presented
in the thesis because of minor changes in the final analysis.  This analysis furthers the
work described in detail in CBX 95-6, CBX 96-26, CBX 96-27 and CBX 96-103.
}{}

\section{Introduction}
The element $V_{ub}$ remains one of the most poorly constrained parameters of
the Cabbibo-Kobayashi-Maskawa (CKM) matrix \cite{bb:CKM}.  Its
magnitude, $|V_{ub}|$, plays a central role in constraints based on the
unitarity of the CKM matrix and inputs from both $CP$--conserving processes in the
$B$ meson decay and $CP$--violating processes in the neutral kaon and $B$
systems.  The value of $|V_{ub}|$ and, in particular, the accuracy to which we have
measured this important parameter, have been the subjects of considerable debate
over the past decade \cite{bb:PDG_vub_minireview}.    An accurate determination of 
$|V_{ub}|$ with well-understood uncertainties remains one of the fundamental priorities
for heavy flavor physics.

A number of $|V_{ub}|$ measurement approaches have been attempted, and are
reviewed in reference~\cite{bb:PDG_vub_minireview}.  Inclusive techniques
are hampered by a mismatch in kinematic regions where the large experimental backgrounds
from $b\to c\ell\nu$ can be suppressed versus regions in which the theoretical
uncertainties can be reliably determined.  For exclusive reconstruction of particular
final states, the primary challenge is calculation of the form factors for those channels.
The first measurements of exclusive charmless semileptonic branching fractions
\cite{bb:lkg_cleo_exclusive}, including
evaluation of $|V_{ub}|$, were performed by the CLEO experiment at the Cornell Electron 
Storage Ring (CESR) using the modes $B^0\to\pi^-\ell^+\nu$, $B^+\to\pi^0\ell^+\nu$,
$B^0\to\rho^-\ell^+\nu$, $B^+\to\rho^0\ell^+\nu$, $B^+\to\omega\ell^+\nu$,
and charge-conjugate decays, where $\ell=e\text{ or }\mu$.  A second measurement of 
the $\rho\ell\nu$ modes by CLEO \cite{bb:lange_cleo_exclusive}, using
similar techniques but a much different signal to background optimization, provided
consistent, essentially independent, results with a similar total uncertainty.  The
combined analyses yielded $|V_{ub}|=(3.25 \pm 0.14^{+0.21}_{-0.29} \pm 0.55)\times 10^{-3}$,
where the errors are statistical, experimental systematic and estimated theoretical uncertainties,
respectively.  The $\pi$ and $\rho$ modes contribute about equally to this result.   

This paper presents an update of the original exclusive $B\to X_u\ell\nu$ analysis 
\cite{bb:lkg_cleo_exclusive}, and is based on a total data sample of $9.7\times10^6$ $B\bar{B}$
pairs collected on the $\Upsilon(4S)$ resonance.  The results presented here supersede
those of reference \cite{bb:lkg_cleo_exclusive}.
In addition to using a larger data set, the analysis has been modified to minimize uncertainties
arising from the momentum-transfer ($q^2$) dependence of the form factors.  Most notably, the lower bounds
on the charged-lepton momentum for both the pseudoscalar and the vector modes have
 been lowered, and the branching fractions are determined independently in
three $q^2$ regions.  For the $\rho$ modes, the branching fractions as a function of $q^2$
were first determined by the second CLEO $\rho\ell\nu$ analysis \cite{bb:lange_cleo_exclusive}.
The present analysis has a significantly broader accepted range for the charged lepton momentum,
which  allows for better discrimination among models.  A detailed description of this analysis
can be found in reference~\cite{bb:vb_thesis}.

\section{Exclusive charmless semileptonic decays}

The semileptonic transition of a $B$ meson (a pseudoscalar) to a 
final state with a single pseudoscalar meson $P$ can,
in the limit of a massless charged lepton, be described
by a single form factor $f_1(q^2)$:
\begin{equation}
\frac{d\Gamma(B^0\to P^-\ell^+\nu)}{dy\,d\cos\theta_{W\ell}} =
|V_{ub}|^2\frac{G_F^2 k_P^3 M_B^2}{32\pi^3}\sin^2\theta_{W\ell}|f_1(q^2)|^2,
\end{equation}
where $y = q^2/M_B^2$, $M_B$ is the mass of the $B$ meson, $G_F$ is the Fermi constant, $k_P$ is the meson momentum,
 and $\theta_{W\ell}$ is the angle between the charged lepton
direction in the virtual $W$ ($\ell+\nu$) rest frame and the direction of the
virtual $W$ in the $B$ rest frame.
For a transition to a final state with a single vector meson $V$,
three form factors ($A_1$, $A_2$, and $V$) are necessary:
\begin{eqnarray}
\frac{d\Gamma(B^0\to V^-\ell^+\nu)}{dy\,d\cos\theta_{W\ell}} & = & |V_{ub}|^2
   \frac{G_F^2 k_V M_B^2 y}{128\pi^3} \times \\ \nonumber
 & & \left[
      (1-\cos\theta_{W\ell})^2\frac{|H_+|^2}{2} +
      (1+\cos\theta_{W\ell})^2\frac{|H_-|^2}{2} +
             \sin^2\theta_{W\ell} |H_0|^2 \right],
\end{eqnarray}
where $k_V$ is the meson momentum
and the three helicity amplitudes are given by
\begin{eqnarray}
H_\pm &= &\frac{1}{M_B+m_V}\left[ A_1(q^2) \mp 2M_B k_V V(q^2)\right]
\ ,\ \ {\rm and}  \\
H_0   &=  &\frac{1}{\sqrt{y}}\frac{M_B}{2m_V(M_B+m_V)} \left[
\left(1-\frac{m_V^2}{M_B^2}-y\right)A_1(q^2) - 4k_V^2A_2(q^2) \right]. 
\end{eqnarray}

The structure of these differential decay rates immediately allows us to
draw some general conclusions regarding the properties of the semileptonic decays
that we reconstruct in this analysis.   For the $\rho(\omega)\ell\nu$ transitions,
the left-handed, $V-A$, nature of the charged current at the quark level
manifests itself at the hadronic level as $|H_-|>|H_+|$.
 The $H_-$ contribution is also expected to dominate the $H_0$ contribution,
leading to a forward-peaked distribution for $\cos\theta_{W\ell}$.  For $\pi(\eta)\ell\nu$,
there is a $\sin^2\theta_{W\ell}$ dependence, independent of the
form factor.  The pseudoscalar modes also contain an extra
factor of the meson momentum squared, which suppresses the rate near $q^2_{\text{max}}$ ($k_P=0$).
Taken together, these two effects give the pseudoscalar modes a softer
charged lepton momentum spectrum than the vector modes.

\begin{figure}[t]
\centering
\leavevmode
\epsfxsize=6.5in
\epsfbox{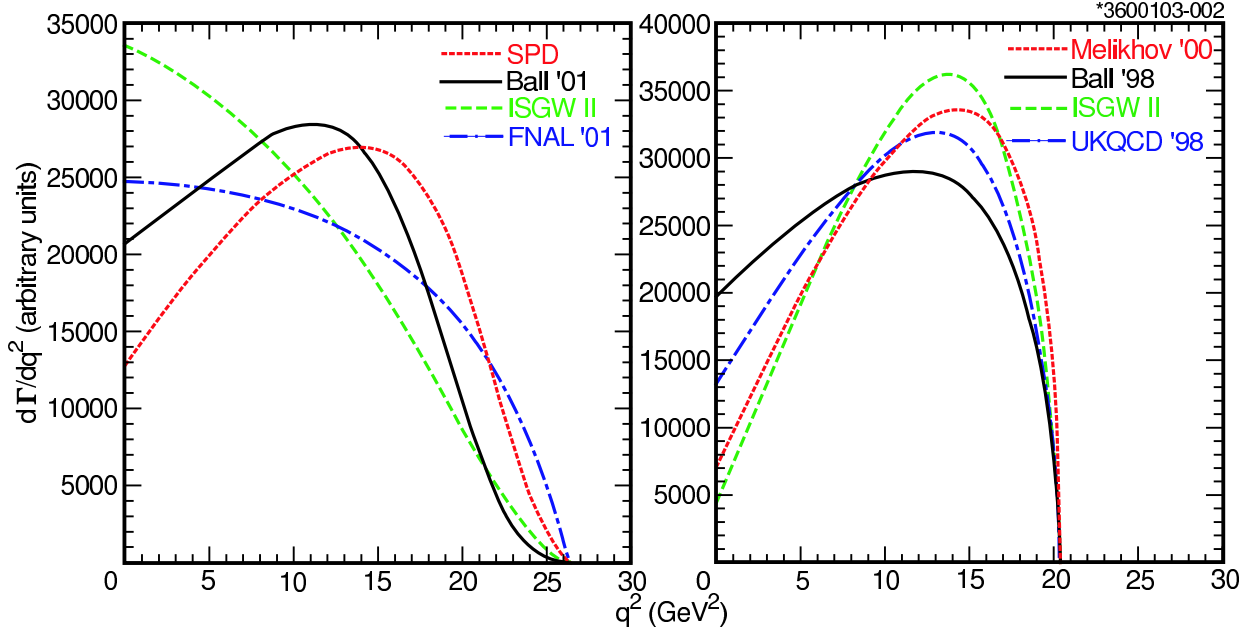}
\caption{Predictions for $d\Gamma(B\to\pi\ell\nu)/dq^2$ (left) and
for $d\Gamma(B\to\rho\ell\nu)/dq^2$ (right) for a variety of calculations, illustrating
the range of variation of the predicted $q^2$--dependence.  See Section~\protect\ref{sec:formfacts}
for further discussion of the calculations.}
\label{fig:formfactors}
\end{figure}

Calculation of the form factors has become a considerable
theoretical industry, with a variety of techniques now being employed.
Form factors based on lattice QCD (LQCD) calculations 
\cite{Abada:1993dh,Allton:1994ui,DelDebbio:1997kr,Hashimoto:1997sr,Ryan:1998tj,Ryan:1999kx,Lellouch:1999dz,Bowler:1999xn,Becirevic:1999kt,Aoki:2000by,El-Khadra:2001rv,Aoki:2001rd,Abada:2000ty}
and on light-cone sum rules (LCSR)
\cite{Ball:1997rj,Ball:1998kk,Khodjamirian:1997ub,Khodjamirian:2000ds,Bakulev:2000fb,Huang:2000hs,Wang:2001mi,Wang:2001bh,Ball:2001fp} 
currently have uncertainties in the $15\%$ to $20\%$ range.  A variety
of quark-model calculations exist 
\cite{Wirbel:1985ji,Korner:1987kd,Isgur:gb,Scora:1995ty,Melikhov:1995xz,Beyer:1998ka,Faustov:1995bf,Demchuk:1997uz,Grach:1996nz,:2000ae,Melikhov:2000yu,Feldmann:1999sm,Flynn:2000gd,Beneke:2000wa,Choi:1999nu}. Finally,
a number of other approaches 
\cite{Kurimoto:2001zj,Ligeti:1995yz,Aitala:1997cm,Burdman:1996kr,Lellouch:1995yv,Mannel:1998kp}, such as 
dispersive bounds and experimentally--constrained models based on heavy quark symmetry, all
seek to improve the range of $q^2$ over which the form factors can be estimated without
introduction of significant model dependence. Figure~\ref{fig:formfactors} illustrates
the broad variation in shape that arises in the literature. Unfortunately, all the
form-factor calculations currently have contributions to the uncertainty that are uncontrolled.
The light-cone sum rules calculations assume quark--hadron duality, offering
a ``canonical'' contribution to the uncertainty of $10\%$, but with no known
means of rigorously estimating that uncertainty.  The LQCD calculations to date remain
in the ``quenched'' approximation (no light quark loops in the propagators),
which limits the ultimate precision to the 15\% to 20\% range.  With the quark-model
calculations it is difficult to quantify the uncertainty of a particular calculation
by their very nature.  These uncertainties in the form factors translate
directly into the same fractional uncertainty on $|V_{ub}|$.

In the $\pi\ell\nu$ modes, with only a single form factor in the massless lepton
approximation,  we expect that the rates extracted in the $q^2$ intervals that
we have chosen will be largely independent of the form-factor shapes.  In the
vector modes, however, the three form factors interfere and differences in this
interference among models, particularly at lower $q^2$ values, can
lead to a residual model dependence.  To investigate this effect, we will analyze
the vector modes with three separate charged lepton momentum requirements.

\section{Event reconstruction and selection}

The CLEO detector \cite{bb:CLEO-nim,bb:silicon-nim} contains three concentric tracking
devices within a 1.5 T superconducting solenoid that detect charged particles over
95\% (93\%) of the solid angle for the first third (last two thirds) of the data.
For the last two thirds of the
data, a silicon vertex detector replaced a straw-tube wire chamber. The momentum 
resolution at 2 GeV/$c$ is 0.6\%.  A
CsI(Tl) electromagnetic calorimeter, also inside the solenoid, covers 98\% of
$4\pi$.  A typical $\pi^0$ mass resolution is 6 MeV.  Charged tracks are
assigned the most probable mass based on specific ionization, time of flight,
and the relative rates as a function of momentum for proton, $K^+$, and $\pi^+$ 
production in $B$ decay.

The undetected neutrino complicates analysis of semileptonic decays.  Because
of the good hermeticity of the CLEO detector, we can reconstruct the neutrino via the missing energy
($E_{\text{miss}}\equiv 2E_{\text{beam}}-\sum E_i$) and missing momentum
($\vec{P}_{\text{miss}}\equiv -\sum\vec{p}_i$) in each event.  In the process
$e^+e^-\to\Upsilon(4S)\to B\bar{B}$, the total energy of the beams is
imparted to the $B\bar{B}$ system; at CESR, that system is at, or nearly at, rest.
(A small crossing angle has been in use at CESR for most of the running.) 
The missing mass, $M^2_{\text{miss}} \equiv E^2_{\text{miss}} - |\vec{P}_{\text{miss}}|^2$, must
be consistent, within resolution, with a massless neutrino. Specifically, we
require $-0.5 < M^2_{\text{miss}} / 2E_{\text{miss}} < 0.3\ \text{GeV}$ for
events with a total charge $\Delta Q=0$, and 
$|M^2_{\text{miss}}| / 2E_{\text{miss}} < 0.3\ \text{GeV}$ for events with 
$|\Delta Q|=1$.

Signal Monte Carlo (MC) events 
show a $|\vec{P}_{\text{miss}}|$ resolution of 85 MeV/$c$.  
The resolution on $E_{\text{miss}}$ is about three times larger  than
the momentum resolution \cite{bb:emissres}.
Significant effort has been
devoted to minimizing multiple counting of charged particles in the track
reconstruction ({\em e.g.}, particles that curl multiple times within the
tracking volume), and to suppressing clusters in the calorimeter from charged
hadrons that have interacted.

With an estimate of the neutrino four--momentum in hand, we can employ full
reconstruction of our signal modes.
Because the resolution on $E_{\text{miss}}$ is so much larger than that for 
$|\vec{P}_{\text{miss}}|$, we use
$(E_\nu,\vec{p}_\nu)=(|\vec{P}_{\text{miss}}|,\vec{P}_{\text{miss}})$ for
full reconstruction.  
The neutrino combined with the signal charged lepton ($\ell$) and meson ($m$) should
satisfy, within resolution, the constraints on energy, $\Delta E \equiv
(E_\nu+E_\ell+E_m)-E_{\text{beam}} \approx 0$, and on momentum, $M_{m\ell\nu}
\equiv [E_{\text{beam}}^2 - |\alpha\vec{p}_\nu+\vec{p}_\ell+\vec{p}_m|^2]^\frac12 \approx
M_B$, where $\alpha$ is chosen to force  $\Delta E=0$.  The neutrino
momentum resolution dominates the $\Delta E$ resolution, so the momentum
scaling corrects for the mismeasurement of the magnitude of the neutrino momentum
in the $M_{m\ell\nu}$ calculation. Uncertainty in the neutrino direction then remains as the
dominant source of smearing in this mass calculation.

We reconstruct $q^2=M_{W^*}^2=(p_\nu+p_\ell)^2$ for each decay from the
reconstructed charged lepton four--momentum and the missing momentum.  In addition
to using the scaled reconstructed momentum $\alpha\vec{p}$ described
above, the direction of the missing momentum is changed through the
smallest angle consistent with forcing $M_{m\ell\nu}= M_B$.  This procedure
results in a $q^2$  resolution of 0.3 GeV${^2}$, independent of $q^2$.  The $\pi\ell\nu$ and the
$\rho\ell\nu$ modes are analyzed separately in the intervals $q^2 < 8$ GeV$^2$,
$8\le q^2 < 16$ GeV$^2$, and $q^2\ge 16$ GeV$^2$.  For the $\omega\ell\nu$
and $\eta\ell\nu$ modes, for which we have low statistics, we sum over all $q^2$.

Information from specific ionization is combined with calorimetric and
tracking measurements to identify electrons with $p_\ell>600\text{ MeV}/c$ over
90\% of the solid angle.  Particles registering hits in counters deeper than
5 interaction lengths over the polar angle range $|\cos\theta|<0.85$ are
considered muons. Those with hits beyond 3 interaction lengths over
$|\cos\theta|<0.71$ are used in a multiple-lepton veto, described below.
Candidate leptons must have $p_\ell>1.0 \text{ GeV}/c$ for the $\pi$ and 
$\eta$ (pseudoscalar) modes,
and $p_\ell>1.5 \text{ GeV}/c$ for the $\rho$ and $\omega$ (vector) modes,
which can couple to the $W$ helicities $\pm1$ and hence have a harder
spectrum.   This momentum requirement for the vector modes defines
the nominal analysis. We also analyze the vector modes with the
lepton momentum requirements $p_\ell>1.75 \text{ GeV}/c$
and $p_\ell>2.0 \text{ GeV}/c$.
The identification efficiency above $1.5 \text{ GeV}/c$ averages
over 90\%; the probability that a hadron is misidentified as an
electron (muon), a fake lepton, is about 0.1\% (1\%).  

The 5-interaction-length requirement for muons causes the muon
acceptance to fall rapidly below 1.4 GeV/$c$.  As a result, only
electrons contribute at the low end of the  momentum range we accept
for $\pi\ell\nu$, and electrons dominate the measurement in the lowest
$q^2$ interval.

A $\pi^0$ candidate must have a $\gamma\gamma$ mass within 2 standard
deviations of the $\pi^0$ mass.  We reconstruct the $\omega$ via its
$\pi^+\pi^-\pi^0$ decay, reducing combinatoric background by rejecting
combinations away from the center of the $\omega$ Dalitz plot.   We reconstruct
$\eta$ in both the $\gamma\gamma$ and the $\pi^+\pi^-\pi^0$ decay modes.
For the $\gamma\gamma$, we require the reconstructed mass to be within
 2 standard deviations of the $\eta$ mass (within about 26 MeV).
For the $\pi^+\pi^-\pi^0$,  we require $|m_{\pi^+\pi^-\pi^0} - m_\eta| < $ 10 MeV
(about 1.7 times the resolution).  We impose a kinematic mass constraint
on the momentum  of all  $\pi^0$ or $\eta$  candidates in the $\gamma\gamma$ final state.

Backgrounds arise from the $e^+e^-\to q\bar{q}$ and
$e^+e^-\to \tau^+\tau^-$ continuum, fake
leptons, $b\to c\ell\nu$, and $B\to X_u\ell\nu$ modes other than the signal
modes.
Backgrounds from continuum processes are suppressed by use of two
event-shape variables. The selection criteria were optimized using
background and signal Monte Carlo samples, rather than data, to avoid potential
bias.  The first variable is the angle ($\cos\theta_{\mathrm{thrust}}$)
between the thrust axis evaluated
for the candidate signal--mode particles (not including the neutrino) and that for the rest
of the event.  (The thrust axes are signed by picking the hemisphere containing
the most energy.)
For $B\bar{B}$ events at CESR, the distribution in this variable is flat
because the $B$'s are nearly at rest and thus their decay orientations
are independent.  For continuum events the distribution is strongly forward and
backward peaked. The ratio $R_2$ of the second to the zeroth
Fox-Wolfram moment \cite{bb:fox_wolfram},
which distinguishes spherical from jetty topologies, is also utilized.  
The continuum background tends to have a small reconstructed $q^2$.  We therefore
tune the continuum cut employed in the $R_2$--$\cos\theta_{\mathrm{thrust}}$ plane
separately in each $q^2$ interval, and separately for the $\pi$ and $\rho$ modes.
Signal events with low $q^2$  appear rather jetty, so a cut
using $R_2$, when data is binned over a broad $q^2$ range, would introduce
an efficiency bias.  So for the $\omega$
and $\eta$ modes, for which all $q^2$ regions are combined, only a
$\cos\theta_{\mathrm{thrust}}$ cut is applied, reducing uncertainties from the
$q^2$--dependence of
the form factors.  
%The selection requirements are shown graphically
%in Figure~\ref{fig:cont_suppress} for the $B^0\to\pi^-\ell^+\nu$ reconstruction.
Our criteria suppress the continuum background by
over a factor of 10 and are about 80\% efficient.

%\begin{figure}[tb]
%\begin{center}
%%\includegraphics[width=6in]{file_path}
%\vspace{5 cm}
%\caption{The distribution of $R2$ versus $|\cos\theta_{\rm thrust}|$ in the three $q^2$ intervals
%$q^2 < 8$ GeV$^2$, $8\le q^2 < 16$ GeV$^2$ and $q^2\ge 16$ GeV$^2$.  The dots are
%$B^0\to\pi^-\ell^+\nu$ MC (ISGW2), and the open squares are combined continuum 
%and $e^+e^-\to \tau^+\tau^-$ MC (???).  The
%line indicates the optimum requirement in each interval for the $\pi^-\ell^+\nu$ reconstruction.}
%\label{fig:cont_suppress}
%\end{center}
%\end{figure}

The $|p_\ell|$ cuts greatly reduce background from
$b\to c\to s\ell\nu$ and bias mildly against $b\to c\ell\nu$.   For the vector modes, we
further require $\cos\theta_{W\ell}>0$, since the signal rate is largely suppressed
by $V-A$ outside this region, while the background is roughly flat in the region
excluded, and falls off in the region accepted.

Backgrounds, particularly $b\to c\ell\nu$, can smear into the signal region in $\Delta
E$ and $M_{m\ell\nu}$ when $\vec{P}_{\text{miss}}$ misrepresents
$\vec{p}_\nu$.  Such backgrounds are highly suppressed by rejecting events
with multiple charged leptons or a total event charge $|\Delta Q|>1$, both of which indicate  missing
particles. Requiring $M^2_{\text{miss}}$ to be consistent with zero also provides
powerful background suppression.  Still, Monte Carlo
studies show that the dominant remaining $b\to c\ell\nu$ events contain
either a $K_L$ meson or a second neutrino (from $c\to s\ell\nu$, with the lepton not
identified) that is roughly collinear with the primary neutrino.

Our selection criteria studies, based on statistical considerations, indicated that
keeping the $|\Delta Q|=1$ sample as well as
the $\Delta Q=0$ was favorable in spite
of the poorer signal-to-background ratio.  Further systematic considerations
indicated that the use of the $|\Delta Q|=1$ sample remained advantageous
for the pseudoscalar modes.  For the vector, in particular the $\rho$ modes,
however, the overall poorer signal-to-background ratio made the  $|\Delta Q|=1$
sample overly sensitive to systematic effects in both the modelling of the $B\to X_u\ell\nu$
backgrounds and the simulation of the detector.  Therefore for the vector modes
we require $\Delta Q=0$.

\section{Extraction of branching fractions}

\subsection{Method and binning}

To extract the branching fraction information, we performed
a binned maximum likelihood fit that was extended to include the 
finite statistics of the Monte Carlo, off-resonance, and fake-lepton samples following
the method of Barlow and Beeston \cite{bb:BarlowBeeston}.
The data in each mode were coarsely binned  over the two dimensional region 
$5.175\le M_{m\ell\nu}<5.2875 \text{ GeV}, |\Delta E|<0.75 \text{ GeV}$. 
We further binned the data in the reconstructed $2\pi$ and $3\pi$ 
masses in the $\rho$ and $\omega$ modes.  The  $|\Delta Q|=1$
samples were binned separately from $\Delta Q=0$ samples.  Separation
of the net charge samples allowed us to take advantage
of the better signal-to-noise ratio of the $|\Delta Q|=0$ sample while reducing our
dependence on our knowledge of the absolute tracking efficiency. Finally,
we binned the data in $q^2$ for the two $\pi\ell\nu$ and the
two $\rho\ell\nu$ modes.  For the $\omega\ell\nu$ and the $\eta\ell\nu$
modes, we combined all $q^2$ information into a single bin.

Our fitting strategy was designed to minimize dependence of the results on the
details of the simulation -- both from detector and physics standpoints.
The choice of binning balanced separation of signal and background
against reliance on detailed MC shape predictions.   
To help minimize the model dependence of the branching fraction
determinations, we did not use information from the lepton momentum spectrum or
from $\cos\theta_{W\ell}$ within the fit.  Extraction of rates in the
separate $q^2$ intervals further reduces reliance on the form factors.

The $\Delta E$ bin intervals used in the nominal fit were $-0.75\le \Delta E<-0.45$ GeV,
$-0.45\le \Delta E<-0.15$ GeV, and $-0.15\le \Delta E<0.25$ GeV (the $\Delta E$ signal band).
The $M_{m\ell\nu}$ bin intervals were 
$5.175\le M_{m\ell\nu}< 5.2425 \text{ GeV}$ and 
$5.2425\le M_{m\ell\nu}< 5.2875 \text{ GeV}$. 
In the $\Delta E$ signal band, this second mass interval is divided into two equal bands.
Hence we used a total of seven bins in these two variables.  In the $\rho\ell\nu$
($\omega\ell\nu$) modes, we used three equal
bins over the 2$\pi$ ($3\pi$) mass range within $\pm285$ MeV ($\pm 30$ MeV)
of the nominal $\rho$ ($\omega$) mass.  The three $q^2$ intervals in the
$\pi\ell\nu$ and the $\rho\ell\nu$ modes were
$q^2<8\ \mathrm{GeV}^2$, $8 \le q^2<16\ \mathrm{GeV}^2$,
 and $q^2\ge 16\ \mathrm{GeV}^2$.   The
 number of bins for each mode in the nominal fit is summarized in
 Table~\ref{tab:nbins}.  The nominal fit had a total of 259 bins.    
 For studies in which the $|\Delta Q|=1$
sample is included in the $\rho$ and $\omega$ modes, the fit had
an additional 147 bins for a total of 406 bins.

\begin{table}
 \centering 
 \caption{Summary of the number of bins used in each mode
 for the nominal fit.}\label{tab:nbins}
 \begin{tabular}{rccccc} \hline\hline
&  $\Delta E$, $M_{m\ell\nu}$ & $\Delta Q$ & $M_{2\pi,3\pi}$ & $q^2$ & total \\
\hline
$\pi^-\ell^+\nu$                                   & 7 &  2 & 1 & 3 & 42 \\
$\pi^0\ell^+\nu$                                  & 7 &  2 & 1 & 3 &  42 \\
$\rho^-\ell^+\nu$                                 & 7 &  1 & 3 & 3 & 63 \\
$\rho^0\ell^+\nu$                                & 7 &  1 & 3 & 3 & 63 \\
$\omega\ell^+\nu$                              & 7 &  1 & 3 & 1 & 21 \\
$\eta_{\gamma\gamma}\ell^+\nu$   & 7 &  2 & 1 & 1 & 14 \\
$\eta_{3\pi}\ell^+\nu$                          & 7 &  2 & 1 & 1 & 14 \\
\hline \hline
\end{tabular}
\end{table}

To examine yields, efficiency, and kinematics in this paper, we use the
most sensitive bin (the ``signal bin'') $5.265\le M_{m\ell\nu}< 5.2875
\text{ GeV}$ and $-0.15\le\Delta E<0.25\text{ GeV}$, though neighboring bins also
contribute information to the fit.  For comparison, the $M_{m\ell\nu}$ and
$\Delta E$ resolutions are about 7 MeV and 100 MeV, respectively,
dominated by the resolution on $|\vec{p}_\nu|$. 
The 2$\pi$ (or 3$\pi$) mass
intervals $\pm 95$ MeV and $\pm 10$ MeV, centered on the nominal masses,  are used for 
figures for $\rho$ and $\omega$ candidates, respectively.  

To simplify the statistical interpretation of the results, we limited the number of
multiple entries per event.  For each individual mode, the candidate with the
smallest $|\Delta E|$ among those satisfying $M_{m\ell\nu}> 5.175$ GeV
was chosen, independent of $q^2$.  A given event could contribute to multiple
modes, although contribution near the signal region in more than one
mode was rare.  In the $\rho$ and $\omega$ modes, each of the mass bins
described above was considered a separate mode.

\subsection{Fit components and parameters}

MC simulation provided the distributions in each mode for signal, the $b\to
c$ background, the cross feed among the modes, and the feed down from higher
mass $B\to X_u\ell\nu$ decays. It included a full description of the $b\to c$
and charm decay modes and a GEANT-based \cite{bb:GEANT} detector model.  The
$X_u\ell\nu$ feed down was evaluated with a simulation of the $B\to X_u\ell\nu$ process
based on an inclusive operator product expansion (OPE) calculation \cite{bb:inclusive_theory} of
$d\Gamma(B\to X_u\ell\nu)/dM_{X_u}$,
using parameters determined from the CLEO analysis of the $B\to X_s\gamma$ photon spectrum 
\cite{bb:bsgamm_theor,bb:bsgamm_exp}
(also used in the recent CLEO lepton-momentum end-point analysis \cite{bb:new_endpoint}). The
nominal analysis combined this inclusive spectrum with the ISGW~II model \cite{Scora:1995ty}
for all mesons  through the $\rho(1450)$. For each exclusive mode, we ``subtracted rate'' from the inclusive
calculation with a weight of the form $\exp{[-\alpha(M_{X_u}-M_R)^2/\Lambda_{QCD}^2]}$, where $M_R$ is
the central mass of the resonance $R$.  At any given $M_{X_u}$, the rate remaining after
this subtraction of the exclusive modes is hadronized nonresonantly.  Variations of the inclusive
parameters based on the uncertainties in the $B\to X_s\gamma$ analysis and variations of
the hadronization model ({\em e.g.}, fully nonresonant but with $\pi\pi\ell\nu$ removed from the 
$\rho$ mass region) are included in the systematic uncertainties.  The signal modes are excluded
from these $B\to X_u\ell\nu$ samples.

The contributions from events in which hadrons have faked the signal leptons and 
from continuum are evaluated using data.  The electron and muon identification
fake rates from pions, kaons, and protons are measured in data using a variety of tagged
samples.  The analysis is performed on a sample of hadronic events with no identified leptons,
treating each track in turn as a signal electron and then a signal muon. The contribution
in each mode is weighted according to the fake rate. 

    We determined the residual continuum background using data collected 60 MeV
below the $\Upsilon$(4S) energy.  The center-of-mass energy and cross--section
differences were taken into account as necessary.  For each combination of mode,
reconstructed $q^2$ bin, and for each $\Delta Q $ value,
we determined the {\em rate} over the full $\Delta E - M_{m \ell \nu}$ plane by
applying all cuts, including continuum-suppression cuts, and then scaling according to
the relative on--resonance and off--resonance luminosities.  To smooth the
statistical fluctuations within each combination, we determined the {\it shape}
over the $\Delta E - M_{m \ell \nu}$ plane by the following procedure.  First,
we dropped the continuum-suppression cuts, and obtained the shape over the
$\Delta E - M_{m \ell \nu}$ plane, for each combination from data.  Then, 
from continuum $q \bar q$ MC,
$\tau^+ \tau^-$ MC, and our fake lepton samples, we determined the change in
shape over the $\Delta E - M_{m \ell \nu}$ plane caused by application of the
continuum-suppression cuts, {\it i.e.}, we obtained the ratio of yields, with to
without cuts,  for each 
$\Delta E - M_{m \ell \nu}$ bin, for each combination.
Applying the ratios so obtained to the off-resonance data without
continuum-suppression cuts, we obtained the {\it shape} of the background over
the $\Delta E - M_{m \ell \nu}$ plane, for each combination.

For each signal mode, we generated a sample of signal Monte Carlo that is flat
in phase space and processed these samples with our GEANT-based detector simulation.  As
we analyze each reconstructed event, we reweight the event to correspond to a particular calculation for
the form factors involved in the decay.  This procedure allowed us to sample a variety of
form factor calculations.  For each mode, we determine the efficiency matrix for reconstructed versus
true $q^2$.  Given our resolution and binning, the matrix is essentially diagonal, as
Table~\ref{tab:q2_eff_matrix} shows for the $\pi\ell\nu$ form-factor calculation of Ball and Zwicky (Ball'01)
\cite{Ball:2001fp}.

\begin{table}[t]
\caption{The efficiency matrix in percent describing the probability that an event from a  given 
generated $q^2$ interval reconstructs in a given $q^2$ interval  for $B^0\to\pi^-\ell^+\nu$ events
that pass all cuts and reconstruct within the ``signal region'' of $\Delta E$ versus $M_{m\ell\nu}$.
The efficiencies are based on Ball'01.}
\label{tab:q2_eff_matrix}
\begin{tabular}{l|rrr} \hline\hline
true $q^2$  & \multicolumn{3}{c}{reconstructed $q^2$} \\ %\cline{2-4}
(GeV$^2$)  & 0 -- 8  & 8 -- 16 & $\ge16$ \\ \hline
0 -- 8            & 2.5 & 0.07  &  0.001 \\
8 -- 16          & 0.07 & 4.6  & 0.06 \\
$\ge16$       & 0.000 & 0.15  & 4.4  \\ \hline\hline
\end{tabular}
\end{table}

For these results, we have examined the following form
factors for the signal modes and cross--feed rates. For $\pi\ell\nu$:
Ball and Zwicky  (light-cone sum rules) \cite{Ball:2001fp},
ISGW~II (a nonrelativistic quark model) \cite{Scora:1995ty},  and
the skewed parton distributions (SPD) of Feldmann and Kroll \cite{Feldmann:1999sm}.
Other LQCD and LCSR calculations are also considered in extracting $|V_{ub}|$.
For $\rho\ell\nu$: 
Ball and Braun (light-cone sum rules -- Ball'98) \cite{Ball:1998kk},
ISGW~II,
Melikhov and Stech (a relativistic quark model -- Melikhov'00) \cite{Melikhov:2000yu},
and UKQCD (a  LQCD calculation -- UKQCD 98) \cite{DelDebbio:1997kr}.
For $\eta\ell\nu$, we have only
considered the  ISGW~II form factor. 
The above choices for $\pi\ell\nu$ and $\rho\ell\nu$ bracket the extremes in the variation of the
shape of $d\Gamma/dq^2$ and hence provide a conservative estimate
of the theoretical uncertainty on the branching fractions.  In general, 
the theory references provide minimal guidance  on the theoretical uncertainty in the
form-factor shapes, and the variation among the chosen calculations appears larger than the
variation expected within a given calculation.  For nominal yields and
figures, we use Ball'01 for the $\pi$ modes and Ball'98 for the vector modes.

We fit all the signal modes simultaneously.  The parameters for the
three $\pi^-\ell^+\nu$ $q^2$ intervals, the three $\rho^-\ell^+\nu$
$q^2$ intervals, and the total $\eta\ell\nu$ branching fraction
floated as free parameters in the fit, for a total of 7 signal parameters.
The isospin and quark symmetry relations
$\Gamma(B^0\to\pi^-\ell^+\nu) = 2\Gamma(B^+\to\pi^0\ell^+\nu)$ and
$\Gamma(B^0\to\rho^-\ell^+\nu) = 2\Gamma(B^+\to\rho^0\ell^+\nu) =
2\Gamma(B^+\to\omega\ell^+\nu)$ constrain the rates for $B^+$ relative to
$B^0$, and are assumed to hold for each $q^2$ region.  We combined the 
three $\omega\ell\nu$ rate predictions that result from the quark symmetry
assumption and the three $\rho\ell\nu$ rates to obtain the fit prediction for 
the total observed reconstructed  $\omega\ell\nu$ yield.  As mentioned above, 
only this integrated yield for 
$\omega\ell\nu$ contributes to the likelihood.
The two $\eta$ submodes are tied to the total $\eta\ell^+\nu$ branching
fraction by the measured $\eta$ branching fractions and the submode reconstruction
efficiencies. To implement the isospin constraints, we assume equal charged 
and neutral $B$ production,  $f_{+-}=f_{00}$, and input a lifetime ratio of  
$1.083\pm0.017$ \cite{bb:pdg2002}. 
For  self-consistency,  the cross--feed rates are constrained to the observed yields.  

The $b\to c$ normalization in the fit varies independently
for each mode, and within each mode for $\Delta Q=0$ and $|\Delta Q|=1$.
The normalizations obtained are generally within 10\% of  those derived
from luminosity and cross sections.   The nominal fit therefore has an additional
11 free parameters for these normalizations.

We float the overall normalization of the generic $B\to X_u\ell\nu$ feed--down
sample, determining it from the fit.  To help in determining that
normalization, we take advantage of CLEO's recent
measurement \cite{bb:new_endpoint} of the branching
fraction for $b \to u \ell \nu$ decays with leptons in the $2.2 - 2.6$ GeV/$c$
momentum range: ${\cal B}(B \to X_u \ell \nu, 2.2 \le P_\ell \le 2.6 \text{ GeV}/c)
= (2.30 \pm 0.38)\times 10^{-4}$ (the ``end-point branching fraction'').
We constrained the $B\to X_u\ell\nu$ feed--down normalization by
 adding a $\chi^2$ term to the log likelihood of the fit:
\begin{equation}
-2\ln{\cal L} \to -2\ln{\cal L} + 
\frac{({\cal B}_{\text{em}}-{\cal B}_{\text{ep}})^2}{\sigma^2_{\text{em}}},
\end{equation}
where
${\cal B}_{\text{em}}$ is the measured end-point branching fraction,
$\sigma_{\text{em}}$ is the total experimental uncertainty on that measurement
and ${\cal B}_{\text{ep}}$ is the branching fraction
implied by the fit parameters.  The fit prediction in each iteration is given by
\begin{equation}
{\cal B}_{\text{ep}} =  {\cal B}_{u\ell\nu} f_{u\ell\nu} + 
\sum_{m} \sum_{i=1}^{N_{q^2}(m)} {\cal B}_{m,i} f_{m,i},
\end{equation}
where  $m\in (\pi^+,\pi^0,\rho^+,\rho^0,\omega,\eta)$,
${\cal B}_{m,i}$ is the branching fraction for the decay mode 
$m$ and $q^2$ interval $i$ in that iteration, $ f_{m,i}$ is the 
fraction of charged leptons for that mode and $q^2$ interval that are predicted
by the form-factor calculation to lie in the end-point region, ${\cal B}_{u\ell\nu}$
is the branching fraction for the $B \to X_u \ell \nu$ feed down background in
that iteration, and $f_{u\ell\nu}$ is the fraction of charged leptons in the end-point
momentum range obtained from our model.

The systematic error evaluation for the  $B\to X_u\ell\nu$ feed down, and
checks using alternative procedures, are described below.  The normalization
is floated independently for each systematic variation of the various  Monte Carlo, continuum, or
fake samples described below so that the effect on the background normalization
of mismodeling within the simulation is properly assessed.

In summary, we have nineteen free parameters in the fit:  the seven signal rates, the eleven
generic $b\to c$ background normalizations, and the one generic $B\to X_u\ell\nu$ feed--down
background normalization.  The continuum background and fake-lepton background samples
are absolutely normalized and their rates do not float in the fit.   In fits discussed below for
which we include the $|\Delta Q|=1$ information in the vector-meson modes, there are
an additional 3 $b\to c$ background normalization parameters, for a total of 22 free
parameters.

\subsection{Checks and results}

\begin{figure}[tb]
\centering
\leavevmode
\epsfysize=4.5in
\epsfbox{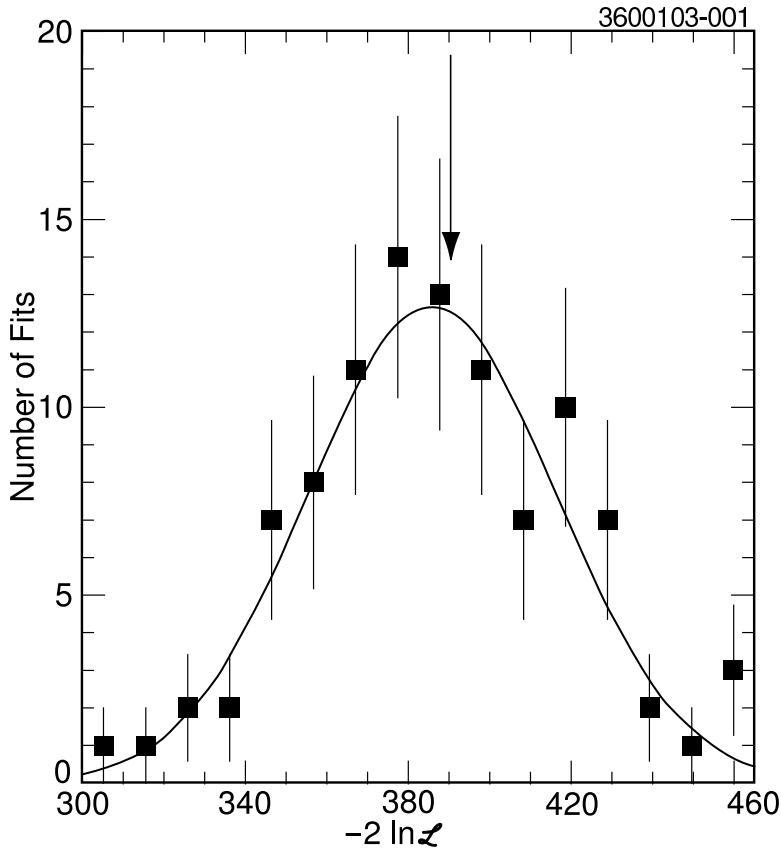}
\caption{Distribution of $-2\ln L$ from the bootstrap procedure
described in the text.  The arrow indicates the value obtained
from the corresponding fit to the data.}
\label{fig:bootstrap}
\end{figure}

We have examined the reliability of our fitting procedure via a bootstrap
technique.  We created 100 mock data samples by randomly choosing
a subset of events from each of our Monte Carlo samples.  From fits
to these samples we found that our procedure reproduces the branching
fractions without bias, and that the scatter of central values agrees with the
uncertainties reported by the fit to better than 15\%.
These studies were done with the $|\Delta Q|=1$ data included in the
vector modes as well as in the pseudoscalar modes.
The distribution of likelihoods that we obtained is shown in 
Figure~\ref{fig:bootstrap}.  At the time of the study, we
the $|\Delta Q|=1$ data included in the vector modes.
For comparison, the  likelihood obtained from a comparable fit
to the data is also shown.   As discussed above, this fit has $406-22=384$ degrees
of freedom.  The result from the fit to data is reasonable.

For the actual nominal fit to the data (no $|\Delta Q|=1$ data in the vector modes),
 we obtained a value $-2\ln L=240.3$ for  $259-19$ degrees of freedom.
Most bins in the data fit have sizable statistics, so interpretation
of $-2\ln L$ as a $\chi^2$ is reasonable.  The probability of $\chi^2$ for
the fit to the data is 0.48.

\begin{figure}[p]
\centering
\leavevmode
\epsfysize=7.5in
\epsfbox{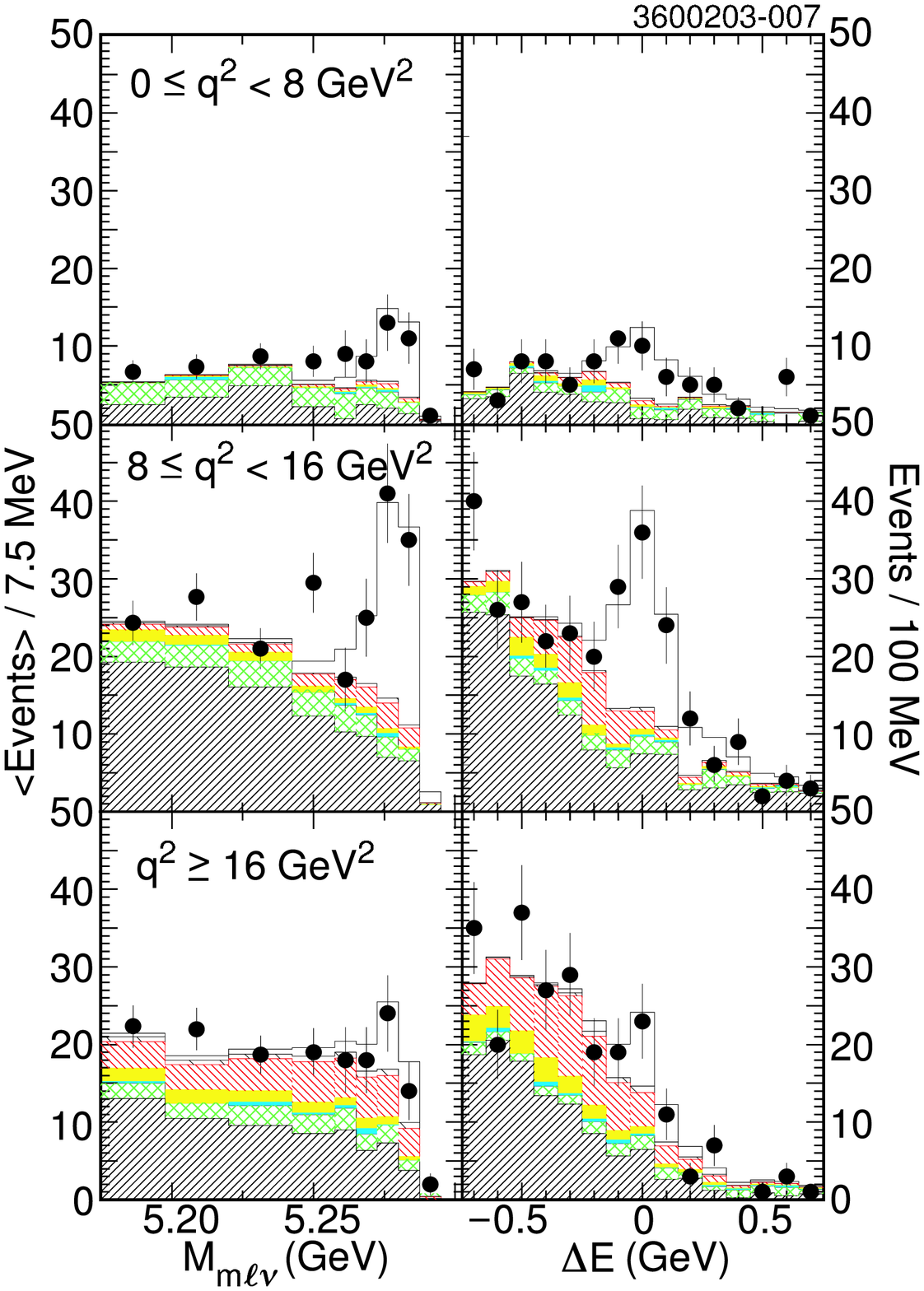}
\caption{$M_{m\ell\nu}$ (left) and $\Delta E$ (right) in
the $\Delta E$ and  $M_{m\ell\nu}$  ``signal'' band 
requiring $\Delta Q=0$ for the combined $\pi^\pm, \pi^0$ modes.  The points are
the on-resonance data.  The histogram components, from
bottom to top, are $b\to c$ (fine $45^\circ$ hatch), continuum (grey or green cross hatch), fake
leptons (cyan or dark grey), feed down from other $B\to X_u\ell\nu$ modes
(yellow or light grey), cross feed from the vector and $\eta$ modes into the reconstructed
modes (red or black fine $135^\circ$ hatch), cross feed among 
the $\pi$ modes (coarse $135^\circ$ hatch), and 
signal (open).  The normalizations are from the nominal fit.}
\label{fig:bmmasses_pi}
\end{figure}

\begin{figure}[p]
\centering
\leavevmode
\epsfysize=7.5in
\epsfbox{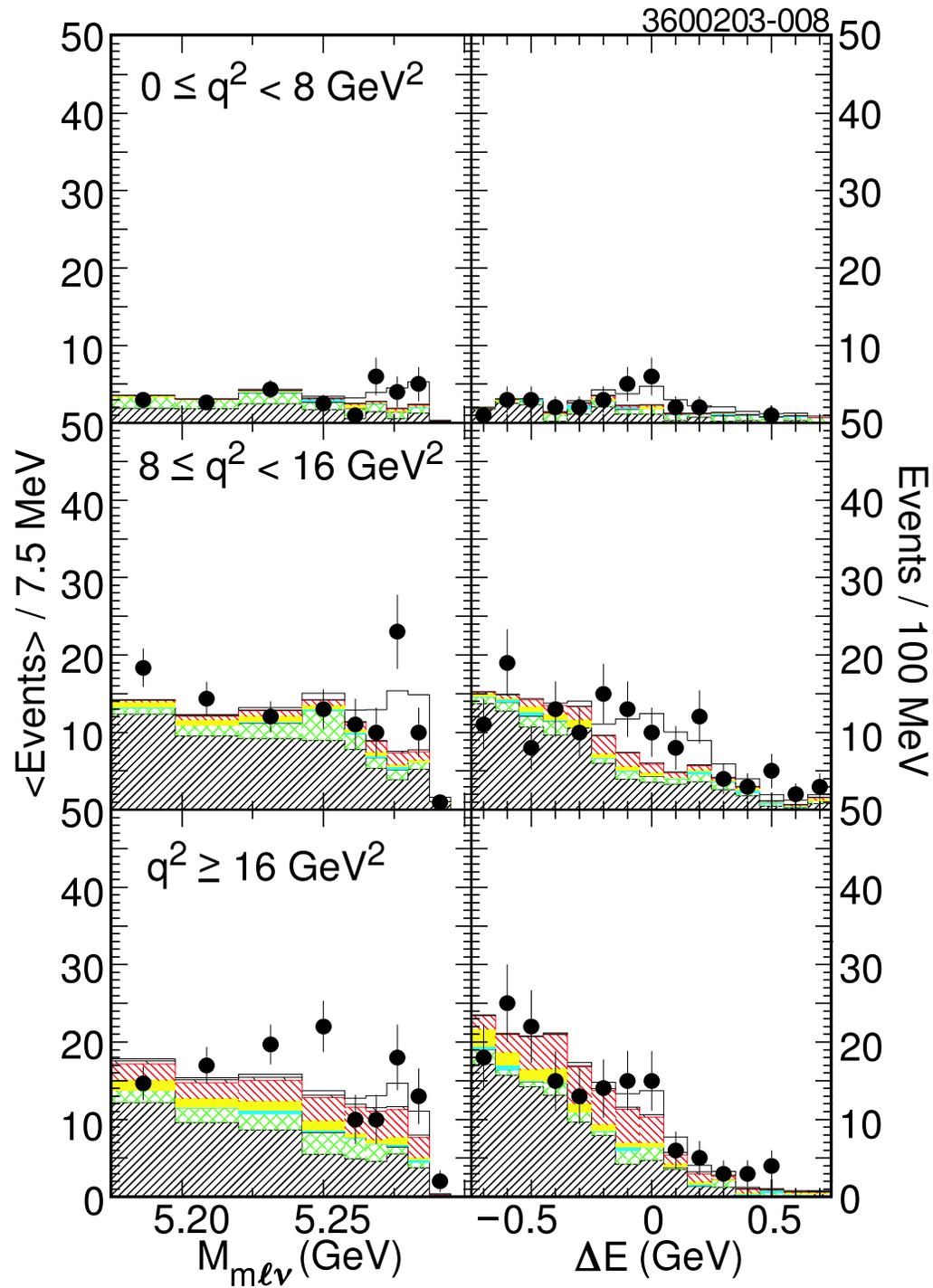}
\caption{$M_{m\ell\nu}$ (left) and $\Delta E$ (right) in
the $\Delta E$ and  $M_{m\ell\nu}$  ``signal'' band 
requiring $|\Delta Q|=1$ for the combined $\pi^\pm, \pi^0$ modes.  The points are
the on-resonance data.  See Fig.~\protect\ref{fig:bmmasses_pi} for component
and normalization descriptions.}
\label{fig:bmmasses_pi1}
\end{figure}

\begin{figure}[p]
\centering
\leavevmode
\epsfysize=7.5in
\epsfbox{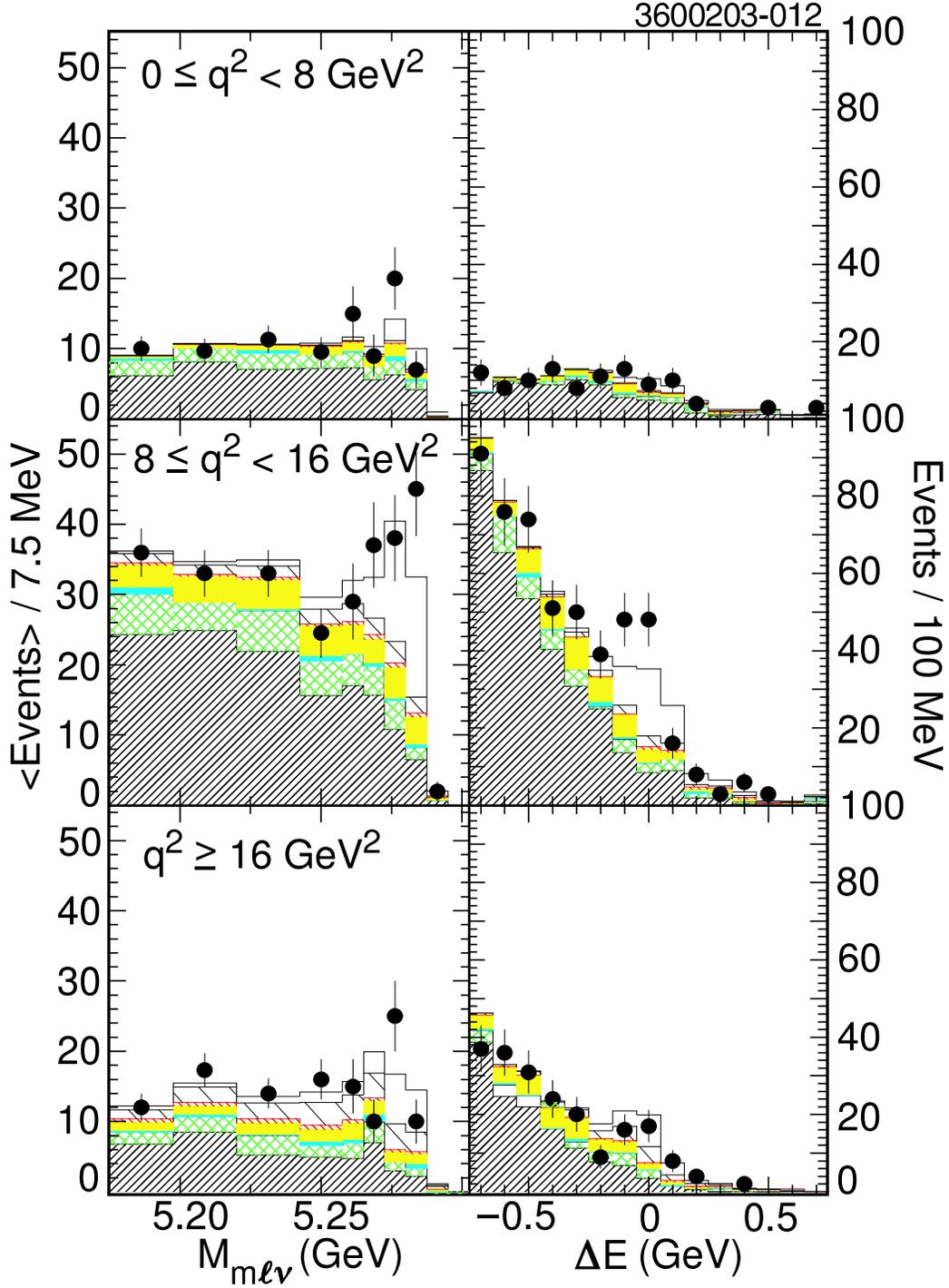}
\caption{$M_{m\ell\nu}$ (left) and $\Delta E$ (right) in
the $\Delta E$ and  $M_{m\ell\nu}$  ``signal'' band 
requiring $\Delta Q=0$ for the combined $\rho^\pm, \rho^0$ modes with
the requirement $p_\ell>1.5\ \mathrm{GeV}/c$ in the vector modes.  The points are
the on-resonance data.  The hatching and normalization are as in  
Fig.~\protect\ref{fig:bmmasses_pi} except that the red or black fine $135^\circ$ hatch 
cross feed  component is from $\pi$ and $\eta$ modes into the $\rho$ modes, and 
the coarse hatch
cross feed component is from among the vector modes.}
\label{fig:bmmasses_rho}
\end{figure}

\begin{figure}[p]
\centering
\leavevmode
\epsfysize=7.5in
\epsfbox{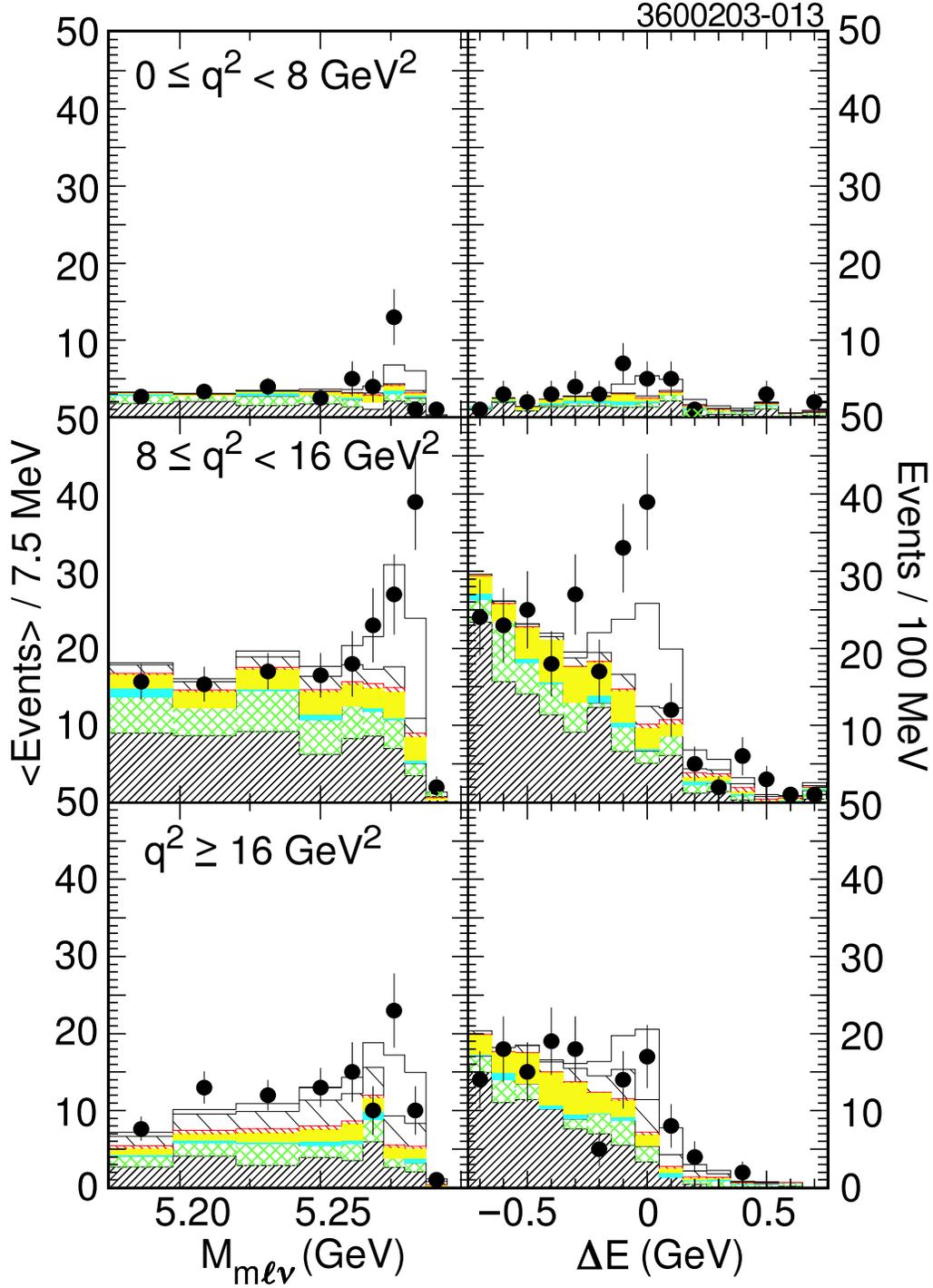}
\caption{$M_{m\ell\nu}$ (left) and $\Delta E$ (right) in
the $\Delta E$ and  $M_{m\ell\nu}$  ``signal'' band 
requiring $\Delta Q=0$ for the combined $\rho^\pm, \rho^0$ modes with
the requirement $p_\ell>2.0\ \mathrm{GeV}/c$ in the vector modes.  The points are
the on-resonance data.  The hatching is as in  
Fig.~\protect\ref{fig:bmmasses_rho}. The normalizations  come from the fit with
the corresponding lepton momentum requirement.}
\label{fig:bmmasses_rho_20}
\end{figure}

\begin{figure}[p]
\centering
\leavevmode
\epsfysize=7.5in
\epsfbox{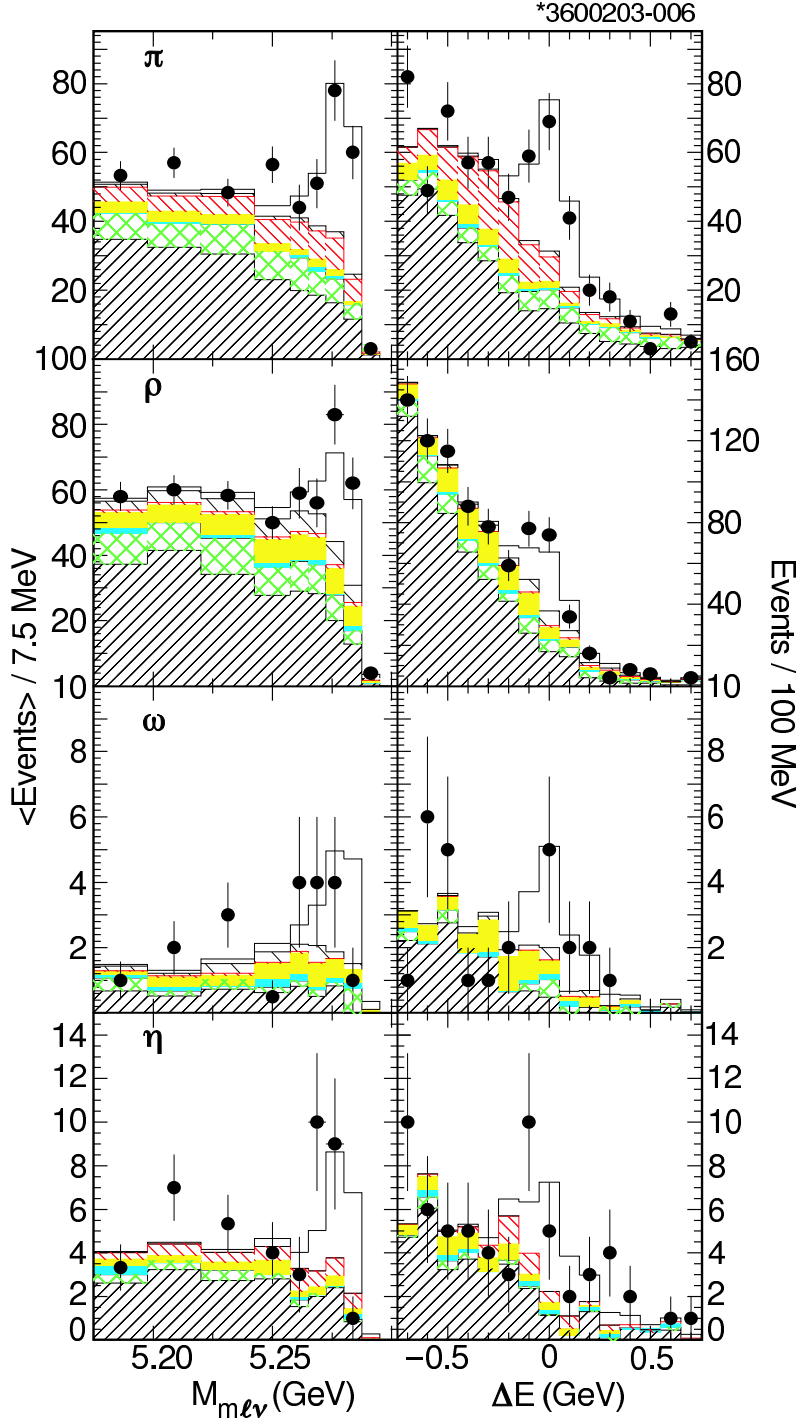}
\caption{$M_{m\ell\nu}$ (left), $\Delta E$ (right)  in
the $\Delta E$ and $M_{m\ell\nu}$ signal bands for $\Delta Q=0$
and summed over the entire $q^2$ range for the combined $\pi$ modes (top), 
$\rho$ modes (row 2), $\omega$ (row 3), and $\eta$ (bottom) modes.  
 See Figs.~\protect\ref{fig:bmmasses_pi} and~\protect\ref{fig:bmmasses_rho} 
for component and normalization descriptions.  For $\eta$ there is only
a single cross--feed component from the non--$\eta$ modes (red or black
fine $135^\circ$ hatch.}
\label{fig:bmmasses_summed}
\end{figure}

In Figures~\ref{fig:bmmasses_pi} through \ref{fig:bmmasses_rho_20} we show the $M_{m\ell\nu}$ ($\Delta E$)
distributions in the $\Delta E$ ($M_{m\ell\nu}$) signal band for the
individual $q^2$ regions examined for $\pi\ell\nu$ and for $\rho\ell\nu$.   For $\rho\ell\nu$, we
show both the distributions with the nominal 1.5 GeV$/c$ minimum lepton momentum requirement
and with the more restrictive 2.0 GeV$/c$ requirement of the original CLEO analysis.
The fits describe the data in these regions
well. The distributions summed over $q^2$ for the $\pi$ and $\rho$ modes and for
$\omega\ell\nu$ and $\eta\ell\nu$ are shown in Fig.~\ref{fig:bmmasses_summed}.
The
$\omega\ell\nu$ mode remains consistent both with the level expected given the
$\rho\ell\nu$ rate and with pure background.  Unless otherwise specified, the
normalizations in all figures derive from the fit with
the requirement $p_\ell>1.5 \text{ GeV}/c$ in the vector modes.

\begin{figure}[p]
\centering
\leavevmode
\epsfysize=7.5in
\epsfbox{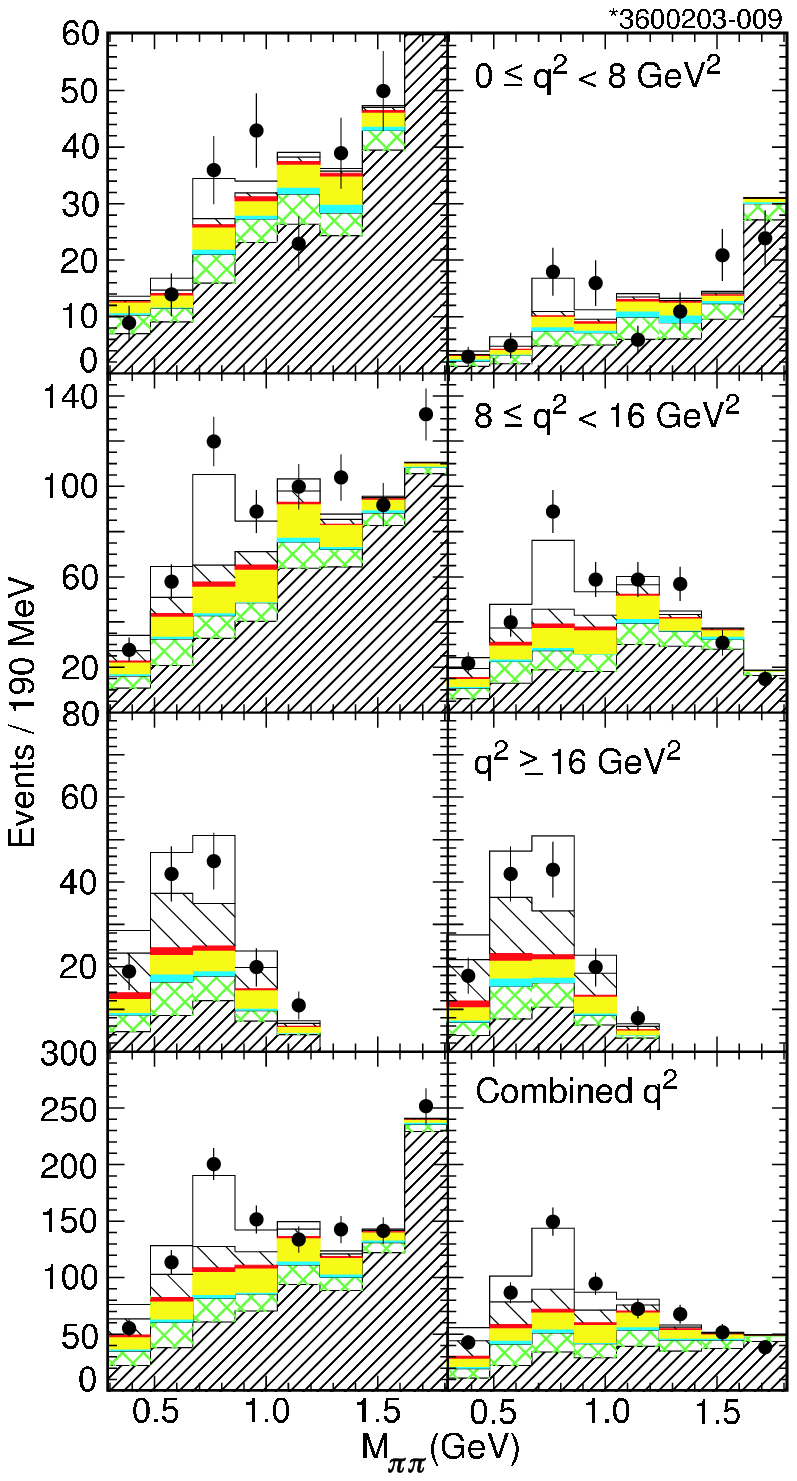}
\caption{Reconstructed mass distributions for $\rho\to\pi\pi$  in the $(M_{m\ell\nu},\Delta E)$ signal bin
for the two analyses with $p_\ell>1.5\ \mathrm{GeV}/c$ in the vector modes (left) and with 
$p_\ell>2.0\ \mathrm{GeV}/c$ (right).
See Fig.~\protect\ref{fig:bmmasses_rho} for component
and normalization descriptions.}
\label{fig:pimasses}
\end{figure}

\begin{figure}[p]
\centering
\leavevmode
\epsfysize=7.7in
\epsfbox{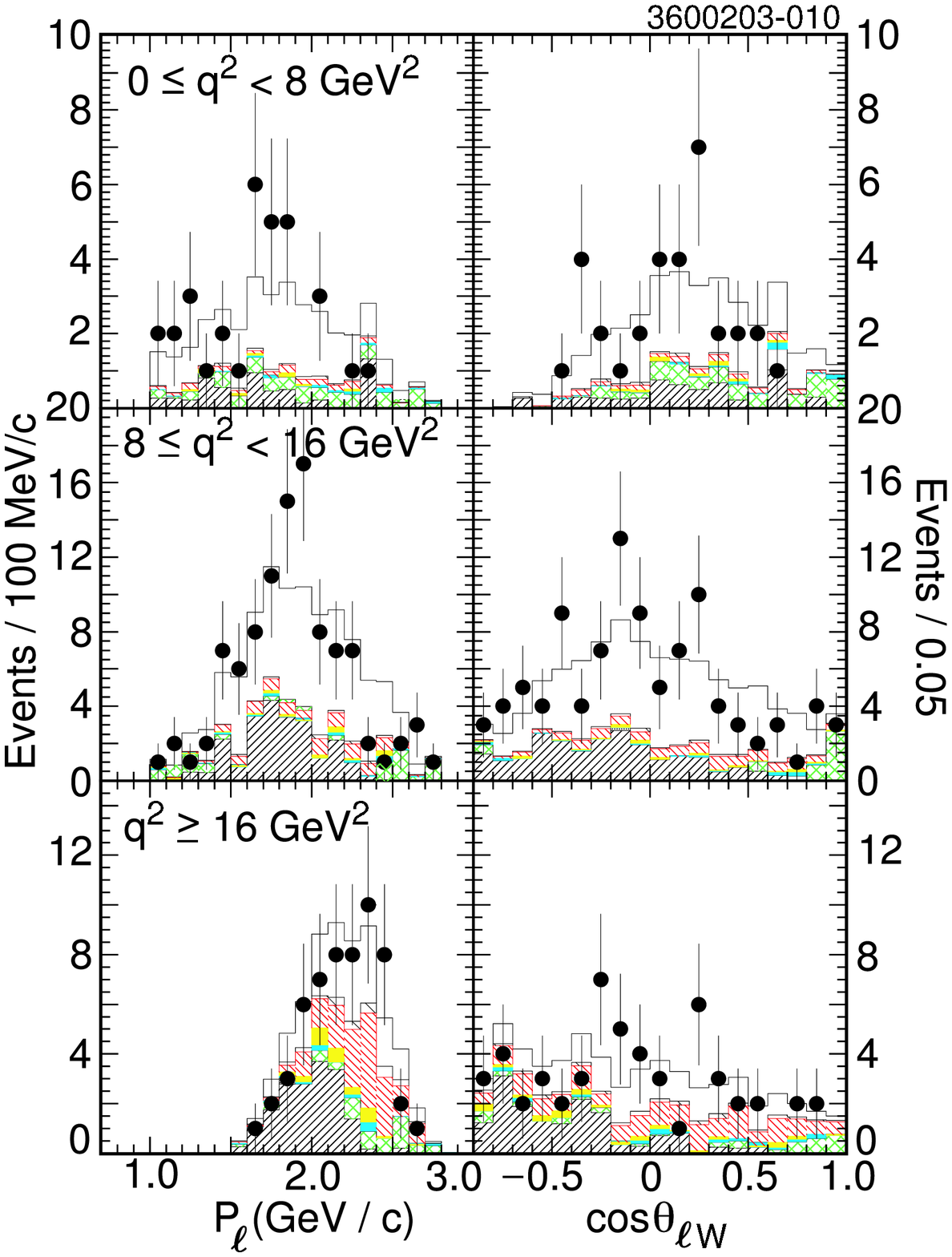}
\caption{Charged lepton momentum spectrum (left) and  $\cos\theta_{W\ell}$ (right)
 for the combined $\pi\ell\nu$ modes in the three $q^2$ intervals.
See Fig.~\protect\ref{fig:bmmasses_pi} for component
and normalization descriptions.}
\label{fig:other_variables_pi}
\end{figure}

\begin{figure}[tb]
\centering
\leavevmode
\epsfxsize=6.5in
\epsfbox{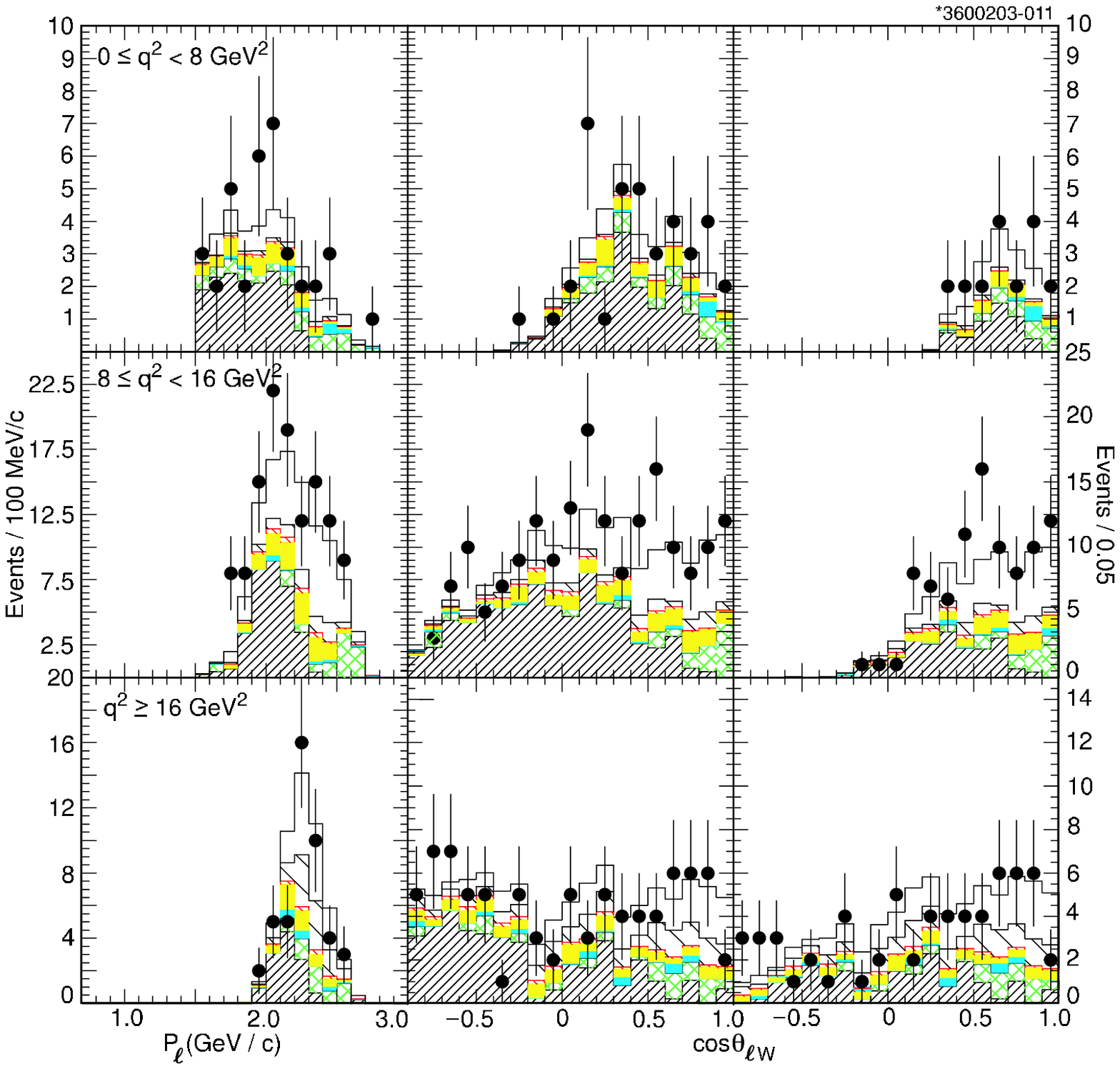}
\caption{Charged lepton momentum spectrum (left) and  $\cos\theta_{W\ell}$ distributions
 for the combined $\rho\ell\nu$ modes in the three $q^2$ intervals.  The $\cos\theta_{W\ell}$
distribution is shown for both the analysis with $p_\ell>1.5\ \mathrm{GeV}/c$ in the vector
modes (center) and for $p_\ell>2.0\ \mathrm{GeV}/c$ (right).
See Fig.~\protect\ref{fig:bmmasses_rho} for component
and normalization descriptions.}\label{fig:other_variables_rho}
\end{figure}

The lepton momentum spectra and $\cos\theta_{W\ell}$ distributions 
in the $(M_{m\ell\nu},\Delta E)$ signal bin are
shown in Figures~\ref{fig:other_variables_pi} and \ref{fig:other_variables_rho}.  This information
is not used in the fit, but shows good agreement with the signals preferred
in the fit.  
The $\pi\pi$ mass distribution for the
combined $\rho\ell\nu$ modes is shown in
Fig.~\ref{fig:pimasses}.  

The branching fractions from the nominal fit are
summarized in Table~\ref{tab:nomfit_br}.  The results are remarkably stable
as the lepton momentum requirement in the vector modes is varied.  The
greatest variation is observed in the lowest $q^2$ interval in the $\rho\ell\nu$
modes, which we expected because of the larger role that interference between the
form factors plays in that region.

Use of a $\chi^2$--based fitting procedure produced similar results, though we saw clearly 
that  low statistics bins had an undue influence on the results of that fitter.  Such sensitivity
was eliminated with the log likelihood minimization.

\begin{table}[tb]
\caption{Summary of branching fractions  from the nominal fit using the
Ball'01 and Ball'98 form factors for the $\pi$ and $\rho$ modes,
respectively.  The first uncertainties are statistical and the second
systematic (see Section~\ref{sec:systematics}).  The results
for the fits with more restrictive lepton momentum requirements
in the vector modes are also shown.  The $q^2$ intervals are specified in GeV$^2$.}
\ifthenelse{\boolean{cbxNote}}{
\begin{tabular}{ccccc}  \hline\hline
Mode & ${\cal B}_{q^2\,\mathrm{interval}}$ &\multicolumn{3}{c}{analysis requirement (vector modes)} \\ 
& $\times 10^4$ & $p_\ell>1.5\ \mathrm{GeV}/c$    & $p_\ell>1.75\ \mathrm{GeV}/c$  & $p_\ell>2.0 
\ \mathrm{GeV}/c$ \\
 \hline
$B^0\to\pi^-\ell^+\nu$  & ${\cal B}_{\text{total}}$ & 
                          $1.327\pm 0.177\pm 0.11$ &
                          $1.314\pm 0.177\pm 0.11$ &
                          $1.320\pm 0.177\pm 0.12$ \\
                        & ${\cal B}_{< 8}$ &
                          $0.431\pm 0.106\pm 0.05$ &
                          $0.428\pm 0.106\pm 0.05$ &
                          $0.418\pm 0.106\pm 0.05$ \\
                        & ${\cal B}_{8-16}$ &
                          $0.651\pm 0.105\pm 0.07$ &
                          $0.647\pm 0.105\pm 0.07$ &
                          $0.662\pm 0.105\pm 0.07$ \\
                        & ${\cal B}_{\ge16}$ &
                          $0.245\pm 0.094\pm 0.04$ &
                          $0.239\pm 0.094\pm 0.04$ &
                          $0.240\pm 0.094\pm 0.05$ \\ \hline
$B^0\to\rho^-\ell^+\nu$ & ${\cal B}_{\text{total}}$ &
                          $2.172\pm 0.338\;^{+0.47}_{-0.54}$ &
                          $2.340\pm 0.342\;^{+0.43}_{-0.51}$ &
                          $2.292\pm 0.349\;^{+0.40}_{-0.49}$\\
                        & ${\cal B}_{< 8}$ &
                          $0.429\pm 0.198\;^{+0.23}_{-0.23}$ &
                          $0.499\pm 0.203\;^{+0.21}_{-0.22}$ &
                          $0.622\pm 0.221\;^{+0.22}_{-0.23}$\\
                        & ${\cal B}_{8-16}$ &
                          $1.244\pm 0.256\;^{+0.27}_{-0.33}$ &
                          $1.321\pm 0.257\;^{+0.26}_{-0.29}$ &
                          $1.114\pm 0.253\;^{+0.23}_{-0.25}$\\
                        & ${\cal B}_{\ge16}$ &
                          $0.499\pm 0.097\;^{+0.08}_{-0.11}$ &
                          $0.520\pm 0.097\;^{+0.08}_{-0.10}$ &
                          $0.557\pm 0.097\;^{+0.07}_{-0.09}$\\ \hline
$B^+\to\eta\ell^+\nu$   & ${\cal B}_{\text{total}}$ &
                          $0.836\pm 0.307\pm 0.16$  &
                          $0.839\pm 0.308\pm 0.16$  &
                          $0.828\pm 0.307\pm 0.15$ \\ \hline\hline
\end{tabular}
}{
\begin{tabular}{ccccc}  \hline\hline
Mode & ${\cal B}_{q^2\,\mathrm{interval}}$ &\multicolumn{3}{c}{analysis requirement (vector modes)} \\ 
& $\times 10^4$ & $p_\ell>1.5\ \mathrm{GeV}/c$    & $p_\ell>1.75\ \mathrm{GeV}/c$  & $p_\ell>2.0 
\ \mathrm{GeV}/c$ \\
 \hline
$B^0\to\pi^-\ell^+\nu$  & ${\cal B}_{\text{total}}$ & 
                          $1.33\pm 0.18\pm 0.11$ &
                          $1.31\pm 0.18\pm 0.11$ &
                          $1.32\pm 0.18\pm 0.12$ \\
                        & ${\cal B}_{< 8}$ &
                          $0.43\pm 0.11\pm 0.05$ &
                          $0.43\pm 0.11\pm 0.05$ &
                          $0.42\pm 0.11\pm 0.05$ \\
                        & ${\cal B}_{8-16}$ &
                          $0.65\pm 0.11\pm 0.07$ &
                          $0.65\pm 0.11\pm 0.07$ &
                          $0.66\pm 0.11\pm 0.07$ \\
                        & ${\cal B}_{\ge16}$ &
                          $0.25\pm 0.09\pm 0.04$ &
                          $0.24\pm 0.09\pm 0.04$ &
                          $0.24\pm 0.09\pm 0.05$ \\ \hline
$B^0\to\rho^-\ell^+\nu$ & ${\cal B}_{\text{total}}$ &
                          $2.17\pm 0.34\;^{+0.47}_{-0.54}$ &
                          $2.34\pm 0.34\;^{+0.43}_{-0.51}$ &
                          $2.29\pm 0.35\;^{+0.40}_{-0.49}$\\
                        & ${\cal B}_{< 8}$ &
                          $0.43\pm 0.20\;^{+0.23}_{-0.23}$ &
                          $0.50\pm 0.20\;^{+0.21}_{-0.22}$ &
                          $0.62\pm 0.22\;^{+0.22}_{-0.23}$\\
                        & ${\cal B}_{8-16}$ &
                          $1.24\pm 0.26\;^{+0.27}_{-0.33}$ &
                          $1.32\pm 0.26\;^{+0.26}_{-0.29}$ &
                          $1.11\pm 0.25\;^{+0.23}_{-0.25}$\\
                        & ${\cal B}_{\ge16}$ &
                          $0.50\pm 0.10\;^{+0.08}_{-0.11}$ &
                          $0.52\pm 0.10\;^{+0.08}_{-0.10}$ &
                          $0.56\pm 0.10\;^{+0.07}_{-0.09}$\\ \hline
$B^+\to\eta\ell^+\nu$   & ${\cal B}_{\text{total}}$ &
                          $0.84\pm 0.31\pm 0.16$  &
                          $0.84\pm 0.31\pm 0.16$  &
                          $0.83\pm 0.31\pm 0.15$ \\ \hline\hline
\end{tabular}
}
\label{tab:nomfit_br}
\end{table}

 The increase in $-2\ln{\cal L}$ from best fit to ${\cal B}(B^+ \rightarrow
\eta \ell^+ \nu) = 0$ is 10.4, corresponding roughly to a 3.2$\sigma$ statistical
significance.

\section{Experimental systematics}
\label{sec:systematics}

\begin{table}[tb]
\caption{Contributions to the systematic error (\%) in each total and
partial branching fraction (${\cal B}$).  Simulation of the
detector and the second $B$ contribute to $\nu$ simulation.}
\begin{tabular}{r|cccc|cccc|c}  \hline\hline
 & \multicolumn{4}{c}{$\pi\ell\nu$} & \multicolumn{4}{c}{$\rho(\omega)\ell\nu$} & \\ 
  & &  \multicolumn{3}{c}{$q^2$ interval (GeV$^2$)} &  & \multicolumn{3}{c}{$q^2$ interval (GeV$^2$)} & \\
Systematic &${\cal B}_{\text{total}}$ & $<8$ & $8-16$ &$\ge16$ &${\cal B}_{\text{total}}$
& $<8$ & $8-16$ & $\ge16$ & $\eta$ \\ \hline
$\nu$ simulation                         &6.8&10.5&    9.2&   17.2&  18.7&   41.7&  19.4& 13.5&  17.3 \\ 
$B\to D/D^{*}/D^{**}/D^{\rm NR}X\ell\nu$ &1.7& 2.5&    1.9&    3.2&   2.0&   21.4&   4.7&  4.2&   5.5\\
$B\to X_u\ell\nu$ feed down                 &0.5& 3.0&    1.8&    1.9&   8.3&   23.8&   6.1&  5.6&   1.6\\
Continuum smoothing                      &1.0& 2.0&    0.2&    2.0&   3.0&   10.0&   1.0&  2.0&   2.0\\
Fakes                                    &3.0& 3.0&    3.0&    3.0&    3.0&   3.0&   3.0&  3.0&   3.0\\
Lepton ID                                &2.0& 2.0&    2.0&    2.0&    2.0&   2.0&   2.0&  2.0&   2.0\\
$f_{+-}/f_{00}$                          &2.4& 2.6&    2.3&    2.2&    0.0&   2.5&   1.0&  0.1&   4.1\\
$\tau_{B^+}/\tau_{B^0}$                  &0.2& 0.1&    0.3&    0.5&    2.1&   4.2&   1.4&  2.1&   1.4\\
Isospin                                  &0.0& 0.0&    0.0&    0.2&    2.4&   1.9&   2.7&  2.3&   0.1\\
Luminosity                               &2.0& 2.0&    2.0&    2.0&    2.0&   2.0&   2.0&  2.0&   2.0 \\ \hline
Upper                             &{\bf 8.6} &12.4   &10.7&   18.3&{\bf 21.4}&53.9& 21.5& 16.2& {\bf 19.3}\\ \hline
Non Resonant                             &--&   --&     --&     --&    -13&    -9&   -15&  -14&    \\ \hline
Lower                             &{\bf 8.6} &12.4&   10.7&   18.3&{\bf 25.1}&54.7& 26.2& 21.4& {\bf 19.3} \\ \hline\hline
\end{tabular}
\label{tab:systematics}
\end{table}

Table~\ref{tab:systematics} summarizes the contributions to the systematic
errors for the nominal analysis. The dominant contribution is from uncertainties
in ``$\nu$ simulation,'' which includes inaccuracies in detector simulation
and uncertainty in the decay model of the nonsignal $B$.  The breakdown of
``$\nu$ simulation''  into its component parts is given in
Table~\ref{tab:simsysa} (and with lepton momentum cuts for vector modes of
1.75 GeV/$c$ and 2.0 GeV/$c$, in Tables~\ref{tab:simsysb} and~\ref{tab:simsysc},
respectively).

We investigated the systematic uncertainties in ``$\nu$ simulation''  
by modifying, for each systematic contribution under consideration,
 the reconstruction output of all of
the Monte Carlo samples used in the fit. Using independent studies by CLEO
for this and other analyses, our modifications reflected the uncertainties
in charged-particle-finding and photon-finding efficiencies,
simulation of false charged particles
and photons, charged particle momentum resolution,
photon energy resolution, hadronic shower simulation, and charged particle identification.
In addition, we
reweighted the Monte Carlo samples to account for the uncertainties
in the rate and spectrum for $K^0_L$ production in $B$ decay
and in the process $b\to c\to s\ell\nu$, both of which affect the background
rate into the signal region.  The full MC samples were re-analyzed 
for each variation to allow for
leakage of events across the selection boundaries. The variations are described in more detail
in Appendix~\ref{app:systematics}.

For many of the variations in the simulation, we expect a cancellation between the 
change in the signal yield and the change in the efficiency.  (Note that we are
not changing the analysis -- the data yields remain unchanged.)
The cancellation arises as follows. If we degrade the reconstructed neutrino, the efficiency
for signal is reduced, but background tends to smear more readily into
the signal region.  Hence the signal yield also tends to be reduced, offsetting
the change in efficiency.  Because of the expected imperfections in our
simulation, we do not expect the observed cancellation to be perfectly
reliable.  For each variation, we therefore assign an additional uncertainty in
the branching fraction so that the total fractional uncertainty estimate is
\begin{equation}
\sigma = \sigma_{\mathrm{BR}} \oplus \frac{\sqrt{2}}{3}\min(\sigma_{\mathrm{yield}},\sigma_{\mathrm{eff}}).
\label{eqn:add_term}
\end{equation}
In this expression, $\sigma_{\mathrm{BR}}$ is the percentage change in the branching fraction from the
fit, $\sigma_{\mathrm{yield}}$ is the percentage change in the ``signal bin'' yield, and
$\sigma_{\mathrm{eff}}$ is the percentage change in the ``signal bin'' efficiency.  For complete
cancellation ($\sigma_{\mathrm{yield}}=\sigma_{\mathrm{eff}}$; $ \sigma_{\mathrm{BR}}=0$), 
the additional term amounts to the addition in quadrature
of one third of the change observed in the yield and in the efficiency.
When no cancellation is expected, the additional term is zero.  The values for
$\sigma_{\mathrm{yield}}$ and $\sigma_{\mathrm{eff}}$ are estimated by examining the
changes in the ``signal bin.''

Note that because of correlations between the three $q^2$ intervals in a given
mode, the sum of the modes tends to be less sensitive to the systematic
variations than the individual intervals themselves.

Consider now the items in Table~\ref{tab:systematics} other than ``$\nu$ simulation.''
We reweight the Monte Carlo to allow variation in the relative rates
for $D\ell\nu$, $D^*\ell\nu$, and $(Dn\pi)\ell\nu$, both for resonant $Dn\pi$ and
nonresonant $Dn\pi$.
We vary the rates by $\pm8\%$, $\pm6\%$, $\pm30\%$, and $\pm30\%$, respectively.  
Note that if we completely
eliminate any one of these charmed modes except $D^*\ell\nu$, the total branching
fractions for $\pi$ and $\rho$ remain stable within
4\% of themselves, which demonstrates that we are quite insensitive to the details of the
poorly measured nonresonant and resonant $(Dn\pi$) modes.  Zeroing
$D^*\ell\nu$ completely causes changes of only 15\%, further demonstrating
our insensitivity to the detailed modeling of the $b\to c\ell\nu$ process.

For the $B\to X_u\ell\nu$ background, we evaluate two contributions to the
systematic uncertainty.  First, we vary the nonperturbative parameters of the inclusive
spectrum used to drive the $X_u\ell\nu$ simulation within the uncertainties
obtained from the  $B\to X_s\gamma$ analysis that were used in the recent
end-point measurement \cite{bb:bsgamm_exp,bb:new_endpoint}.  That analysis
provides an error ellipse for the HQET parameters $\lambda_1$ versus $\overline{\Lambda}$,
and we choose the points on that ellipse that make the maximal
change.  The second contribution regards uncertainty in the hadronization of
the final state light quarks.  We change from our model that marries the ISGW II exclusive
and OPE inclusive calculations (see previous section) to a purely ``nonresonant''
hadronization procedure (similar to that of JETSET \cite{bb:jetset}).  The hadronization
is nonresonant in the sense that single hadron final states ({\it e.g.}, $a_1\ell\nu$) are
not produced. Resonances can appear in the multihadron final state ({\it e.g.}, $\rho\pi\ell\nu$).
To avoid overlap of the nonresonant sample with the signal modes, we eliminate
$B\to X_u\ell\nu$ events with a low mass $\pi\pi$ final state.  The uncertainties presented
correspond to a minimum $M_{\pi\pi}$ of 1 GeV. Variation of that threshold over the 0.9 -- 1.1
GeV range results in similar systematic estimates.  As a crosscheck, we have also used
the strictly resonant description of ISGW~II, which yields results consistent with
our uncertainty estimates.

We have used different normalization schemes for the $B\to X_u\ell\nu$ background
to check the sensitivity of the results under the normalization procedure.  If we drop
the end-point branching fraction constraint but still allow the normalization to float,
we see only minor shifts in the results and the end-point branching fraction predicted
by the fit is within one standard deviation of the measured value.  We have also
used an iterative procedure, where we fix the $B\to X_u\ell\nu$ normalization in the
fit, but update that normalization until the fit's predicted end-point branching fraction
converges to the central value (and then to $\pm1$ standard deviation) of the
CLEO measurement.  This procedure also gave consistent results.

As Table~\ref{tab:systematics} shows, uncertainty in the $B\to X_u\ell\nu$
feed down contributes little to the systematic error on
$\pi\ell\nu$ and $\eta\ell\nu$.
For the $\rho\ell\nu$ rate, however, the contribution is substantial.

Our nominal fit assumed equal production of charged
and neutral $B$ mesons: $f_{+-}/f_{00}= 1$. We varied this
fraction over the one standard deviation range indicated by
the recent CLEO result $f_{+-}/f_{00}= 1.04 \pm 0.08$ \cite{silvia}. 
The relationship enters both in the fit to implement the isospin constraint
and in the branching fraction calculation to calculate the number of $B^0$ 
mesons.
We used the measured ratio of $B$ meson lifetimes,
$\tau_{B^+}/\tau_{B^0}= 1.083\pm0.017$, which we varied by one standard deviation
to assess the associated uncertainty. The ratio
comes into the normalization of the neutral modes versus the charged
modes. We have also varied the isospin assumption.  In the nominal fit
we used a ratio of 2.  For the systematic estimate we lowered the 
$\rho^+:\rho^0$ ratio down to 1.43, as
suggested by Diaz-Cruz \cite{bb:DiazCruz}.  The deviation arises from 
$\rho^0--\omega$ mixing coupled with the large $\rho^0$ width.  Because
of the small $\eta$ and $\omega$ widths, we expect negligible deviation
from the the ideal factor of two for the other two ratios used.

The uncertainties related to lepton identification are estimated by varying
the measured hadronic fake rates within their uncertainties and by applying
the uncertainty in the measurement of the average lepton identification
efficiency.   Lepton--fake uncertainties are measured 
in the data as a function of momentum using cleanly tagged hadronic samples, including $K_S\to\pi^+\pi^-$
and $D^{*\pm}\to \pi^\pm D^0$, $D^0\to K^\pm \pi^\mp$.

Finally, we assessed our smoothing technique for the continuum data sample.
Recall that we use the off--resonance data distrubution with relaxed continuum--suppression
combined with the expected shape change 
over the fitted $\Delta E$ and $M_{m\ell\nu}$ region that is induced
by the relaxation.   We biased the  Monte Carlo prediction for the
shape change by the statistical uncertainty in
the parameterization for each of the fitted ($\Delta E$, $M_{m\ell\nu}$)
distributions.  The uncertainties come from fluctuating all distributions
coherently to induce the maximum change.

\begin{table}[tp]
  \centering 
  \caption{Percentage change in results for a fit with a modified simulation 
relative to a fit to the nominal MC simulation for each of the variations
contributing to the simulation systematic uncertainty.  The vector modes
were analyzed with the requirement $p_\ell>1.5\ \mathrm{GeV}/c$. The last
row shows the quadrature sum of the changes. }\label{tab:simsysa}
\ifthenelse{\boolean{cbxNote}}{
\begin{tabular}{cccccccccc}
\hline\hline
variation  & \multicolumn{4}{c}{$\pi^-\ell^+\nu$} & \multicolumn{4}{c}{$\rho^-\ell^+\nu$} & $\eta\ell\nu$ \\ \cline{2-5}\cline{6-9}
                               & total & $q^2<8$ & $8\le q^2<16 $ & $q^2\ge 16$
                                & total   & $q^2<8$ & $8\le q^2<16$ & $q^2\ge 16$ & total \\ \hline
$\gamma$ eff.      &  2.60 & 6.98 & 2.66 & 9.09 & 11.11 & 11.85 & 11.14 & 10.60 & 5.66 \\
$\gamma$ resol.    &  4.07 & 2.87 & 5.36 & 2.33 & 2.91 & 3.68 & 2.34 & 4.24 & 9.62 \\
$K_L$ shower       &  1.25 & 1.04 & 1.35 & 1.44 & 6.00 & 8.38 & 7.21 & 1.55 & 2.74 \\
particle ID        &  1.85 & 2.50 & 3.04 & 6.25 & 8.16 & 27.45 & 6.93 & 1.06 & 0.18 \\
split-off rejection &  1.52 & 2.92 & 3.05 & 5.03 & 1.23 & 9.44 & 1.80 & 2.54 & 5.50 \\
track eff.         &  3.69 & 4.45 & 4.19 & 2.56 & 8.64 & 13.27 & 9.53 & 3.39 & 9.46 \\
track resol.       &  1.01 & 1.77 & 2.44 & 11.19 & 6.22 & 12.74 & 6.01 & 2.66 & 0.92 \\
split-off sim.      &  0.40 & 1.35 & 0.53 & 2.31 & 1.03 & 10.37 & 0.97 & 4.67 & 6.02 \\
$K_L$ production   &  0.18 & 0.11 & 0.21 & 0.39 & 0.11 & 0.84 & 0.09 & 0.32 & 0.12 \\
$\nu$ production   &  0.52 & 3.46 & 2.18 & 1.95 & 0.59 & 15.06 & 4.12 & 0.88 & 2.93 \\ \hline
{\bf Total} & 6.76 & 10.46 & 9.20 & 17.19 & 18.73 & 41.72 & 19.43 & 13.53 & 17.25 \\
\hline\hline
\end{tabular}
}{
\begin{tabular}{cccccccccc}
\hline\hline
variation  & \multicolumn{4}{c}{$\pi^-\ell^+\nu$} & \multicolumn{4}{c}{$\rho^-\ell^+\nu$} & $\eta\ell\nu$ \\ \cline{2-5}\cline{6-9}
                               & total & $q^2<8$ & $8\le q^2<16 $ & $q^2\ge 16$
                                & total   & $q^2<8$ & $8\le q^2<16$ & $q^2\ge 16$ & total \\ \hline
$\gamma$ eff.      &  2.6 & 7.0  & 2.7  & 9.1  & 11.1 & 11.9 & 11.1 & 10.6 & 5.7  \\
$\gamma$ resol.    &  4.1 & 2.9  & 5.4  & 2.3  & 2.9  & 3.7  & 2.3  & 4.2  & 9.6  \\
$K_L$ shower       &  1.3 & 1.0  & 1.4  & 1.4  & 6.0  & 8.4  & 7.2  & 1.6  & 2.7  \\
particle ID        &  1.9 & 2.5  & 3.0  & 6.3  & 8.2  & 27.5 & 6.9  & 1.1  & 0.2  \\
split-off rejection &  1.5 & 2.9  & 3.0  & 5.0  & 1.2  & 9.4  & 1.8  & 2.5  & 5.5  \\
track eff.         &  3.7 & 4.5  & 4.2  & 2.6  & 8.6  & 13.3 & 9.5  & 3.4  & 9.5  \\
track resol.       &  1.0 & 1.8  & 2.4  & 11.2  & 6.2 & 12.7 & 6.0  & 2.7  & 0.9  \\
split-off sim.      &  0.4 & 1.4  & 0.5  & 2.3  & 1.0  & 10.4 & 1.0  & 4.7  & 6.0  \\
$K_L$ production   &  0.2 & 0.1  & 0.2  & 0.4  & 0.1  & 0.8  & 0.1  & 0.3  & 0.1  \\
$\nu$ production   &  0.5 & 3.5  & 2.2  & 2.0  & 0.6  & 15.1  & 4.1  & 0.9  & 2.9  \\ \hline
{\bf Total} & 6.8 & 10.4 & 9.2  & 17.2  & 18.7  & 41.7  & 19.4  & 13.5  & 17.3  \\
\hline\hline
\end{tabular}
}
\end{table}

\begin{table}[tp]
  \centering 
  \caption{Percentage change in results for a fit with a modified simulation 
relative to a fit to the nominal MC simulation for each of the variations
contributing to the simulation systematic uncertainty.  The vector modes
were analyzed with the requirement $p_\ell>1.75\ \mathrm{GeV}/c$. The last
row shows the quadrature sum of the changes. }\label{tab:simsysb}
\ifthenelse{\boolean{cbxNote}}{
\begin{tabular}{cccccccccc}
\hline\hline
variation  & \multicolumn{4}{c}{$\pi^-\ell^+\nu$} & \multicolumn{4}{c}{$\rho^-\ell^+\nu$} & $\eta\ell\nu$ \\ \cline{2-5}\cline{6-9}
                               & total & $q^2<8$ & $8\le q^2<16 $ & $q^2\ge 16$
                                & total   & $q^2<8$ & $8\le q^2<16$ & $q^2\ge 16$ & total \\ \hline
$\gamma$ eff. & 2.58 & 6.79 & 2.75 & 9.25 & 9.72 & 8.88 & 10.29 & 9.04 & 5.86 \\
$\gamma$ resol. & 4.01 & 2.66 & 5.38 & 2.37 & 3.16 & 4.63 & 2.68 & 4.07 & 9.71 \\
$K_L$ shower & 1.37 & 1.04 & 1.33 & 1.67 & 4.56 & 4.77 & 6.06 & 0.49 & 2.63  \\
particle ID  & 1.79 & 2.71 & 3.04 & 6.36 & 7.75 & 24.15 & 6.87 & 1.02 & 0.00  \\ 
split-off rejection & 1.51 & 2.48 & 3.06 & 4.74 & 0.50 & 1.70 & 0.87 & 2.43 & 5.03 \\ 
track eff. & 3.67 & 4.27 & 4.20 & 2.59 & 8.44 & 11.86 & 9.68 & 3.44 & 9.66 \\ 
track resol. & 0.99 & 1.76 & 2.59 & 11.41 & 4.61 & 8.10 & 4.93 & 1.79 & 0.82 \\
split-off sim. & 0.41 & 1.54 & 0.53 & 2.35 & 1.05 & 1.49 & 0.25 & 5.27 & 5.29 \\
$K_L$ production & 0.18 & 0.11 & 0.14 & 0.40 & 0.10 & 0.74 & 0.12 & 0.31 & 0.02 \\ 
$\nu$ production & 0.51 & 3.46 & 2.27 & 2.18 & 0.79 & 13.25 & 3.11 & 0.61 & 2.67  \\ \hline
{\bf Total}  &6.71 & 10.17 & 9.31 & 17.44 & 16.73 & 33.07 & 18.04 & 12.20 & 17.03 \\
\hline\hline
\end{tabular}
}{
\begin{tabular}{cccccccccc}
\hline\hline
variation  & \multicolumn{4}{c}{$\pi^-\ell^+\nu$} & \multicolumn{4}{c}{$\rho^-\ell^+\nu$} & $\eta\ell\nu$ \\ \cline{2-5}\cline{6-9}
                               & total & $q^2<8$ & $8\le q^2<16 $ & $q^2\ge 16$
                                & total   & $q^2<8$ & $8\le q^2<16$ & $q^2\ge 16$ & total \\ \hline
$\gamma$ eff.      & 2.6 & 6.8 & 2.8 & 9.3 & 9.7 & 8.9  & 10.3 & 9.0  & 5.9  \\
$\gamma$ resol.    & 4.0 & 2.7 & 5.4 & 2.4 & 3.2 & 4.6  & 2.7  & 4.1  & 9.7  \\
$K_L$ shower       & 1.4 & 1.0 & 1.3 & 1.7 & 4.6 & 4.8  & 6.1  & 0.5  & 2.6   \\
particle ID        & 1.8 & 2.7 & 3.0 & 6.4 & 7.8 & 24.2 & 6.9  & 1.0  & 0.0   \\ 
split-off rejection & 1.5 & 2.5 & 3.1 & 4.7 & 0.5 & 1.7  & 0.9  & 2.4  & 5.0  \\ 
track eff.         & 3.7 & 4.3 & 4.2 & 2.6 & 8.4 & 11.9 & 9.7  & 3.4  & 9.7  \\ 
track resol.       & 1.0 & 1.8 & 2.6 & 11.4& 4.6 & 8.1  & 4.9  & 1.8  & 0.8  \\
split-off sim.      & 0.4 & 1.5 & 0.5 & 2.4 & 1.1 & 1.5  & 0.3  & 5.3  & 5.3  \\
$K_L$ production   & 0.2 & 0.1 & 0.1 & 0.4 & 0.1 & 0.7  & 0.1  & 0.3  & 0.0  \\ 
$\nu$ production   & 0.5 & 3.5 & 2.3 & 2.2 & 0.8 & 13.3 & 3.1  & 0.6  & 2.7  \\ \hline
{\bf Total}        & 6.7 & 10.2 & 9.3 & 17.4 & 16.7 & 33.1  & 18.0  & 12.2  & 17.0 \\
\hline\hline
\end{tabular}
}
\end{table}

\begin{table}[tp]
  \centering 
  \caption{Percentage change in results for a fit with a modified simulation 
relative to a fit to the nominal MC simulation for each of the variations
contributing to the simulation systematic uncertainty.  The vector modes
were analyzed with the requirement $p_\ell>2.0\ \mathrm{GeV}/c$. The last
row shows the quadrature sum of the changes. }\label{tab:simsysc}
\ifthenelse{\boolean{cbxNote}}{
\begin{tabular}{cccccccccc}
\hline\hline
variation  & \multicolumn{4}{c}{$\pi^-\ell^+\nu$} & \multicolumn{4}{c}{$\rho^-\ell^+\nu$} & $\eta\ell\nu$ \\ \cline{2-5}\cline{6-9}
                               & total & $q^2<8$ & $8\le q^2<16 $ & $q^2\ge 16$
                                & total   & $q^2<8$ & $8\le q^2<16$ & $q^2\ge 16$ & total \\ \hline
$\gamma$ eff. & 2.60 & 6.84 & 2.70 & 8.75 & 12.34 & 12.29 & 14.55 & 8.32 & 5.91 \\
$\gamma$ resol. & 4.21 & 2.68 & 5.44 & 3.93 & 1.29 & 1.14 & 2.30 & 0.96 & 9.31 \\ 
$K_L$ shower & 1.34 & 0.87 & 1.72 & 1.67 & 2.39 & 1.84 & 3.16 & 1.20 & 2.65  \\ 
particle ID & 1.85 & 2.72 & 3.12 & 6.36 & 6.99 & 15.56 & 8.11 & 1.12 & 0.40 \\ 
split-off rejection & 1.73 & 2.71 & 3.03 & 5.91 & 1.82 & 1.52 & 2.92 & 0.74 & 5.65  \\ 
track eff. & 3.94 & 4.29 & 4.46 & 2.36 & 4.11 & 4.72 & 6.46 & 1.77 & 9.16  \\ 
track resol. & 0.98 & 1.55 & 2.98 & 11.80 & 3.57 & 8.20 & 2.39 & 2.69 & 1.04  \\ 
split-off sim. & 0.35 & 1.55 & 0.47 & 3.14 & 1.91 & 6.58 & 2.79 & 3.00 & 5.23  \\
$K_L$ production & 0.18 & 0.22 & 0.14 & 0.39 & 0.16 & 0.52 & 0.20 & 0.37 & 0.12  \\ 
$\nu$ production & 0.59 & 3.48 & 2.37 & 2.58 & 0.78 & 6.32 & 1.48 & 0.59 & 2.10  \\  \hline
{\bf Total} & 7.04 & 10.24 & 9.66 & 18.21 & 15.68 & 23.94 & 18.94 & 9.65 & 16.65  \\
\hline\hline
\end{tabular}
}{
\begin{tabular}{cccccccccc}
\hline\hline
variation  & \multicolumn{4}{c}{$\pi^-\ell^+\nu$} & \multicolumn{4}{c}{$\rho^-\ell^+\nu$} & $\eta\ell\nu$ \\ \cline{2-5}\cline{6-9}
                               & total & $q^2<8$ & $8\le q^2<16 $ & $q^2\ge 16$
                                & total   & $q^2<8$ & $8\le q^2<16$ & $q^2\ge 16$ & total \\ \hline
$\gamma$ eff.      & 2.6  & 6.8  & 2.7  & 8.8  & 12.3 & 12.3 & 14.6 & 8.3  & 5.9  \\
$\gamma$ resol.    & 4.2  & 2.7  & 5.4  & 3.9  & 1.3  & 1.1  & 2.3  & 1.0  & 9.3  \\ 
$K_L$ shower       & 1.3  & 0.9  & 1.7  & 1.7  & 2.4  & 1.8  & 3.2  & 1.2  & 2.6   \\ 
particle ID        & 1.9  & 2.7  & 3.1  & 6.4  & 7.0  & 15.6 & 8.1  & 1.1  & 0.4  \\ 
split-off rejection & 1.7  & 2.7  & 3.0  & 5.9  & 1.8  & 1.5  & 2.9  & 0.7  & 5.7   \\ 
track eff.         & 3.9  & 4.3  & 4.5  & 2.4  & 4.1  & 4.7  & 6.5  & 1.8  & 9.2   \\ 
track resol.       & 1.0  & 1.5  & 3.0  & 11.8 & 3.6  & 8.2  & 2.4  & 2.7  & 1.0   \\ 
split-off sim.      & 0.4  & 1.6  & 0.5  & 3.1  & 1.9  & 6.6  & 2.8  & 3.0  & 5.2   \\
$K_L$ production   & 0.2  & 0.2  & 0.1  & 0.4  & 0.2  & 0.5  & 0.2  & 0.4  & 0.1   \\ 
$\nu$ production   & 0.6  & 3.5  & 2.4  & 2.6  & 0.7  & 6.3  & 1.5  & 0.6  & 2.1   \\  \hline
{\bf Total}        & 7.0  & 10.2 & 9.7  & 18.2  & 15.7  & 23.9  & 18.9  & 9.7  & 16.7   \\
\hline\hline
\end{tabular}
}
\end{table}

In the $\rho$ modes, there is an additional uncertainty from the unknown contribution
of nonresonant $\pi\pi\ell\nu$ decays.  While little is known about these decays, we can
provide a framework for limiting those contributions through the study of reconstructed 
$\pi^0\pi^0\ell^\pm\nu$ decays and the consideration of Bose symmetry, isospin, and angular 
momentum.  The $B\to X_u\ell\nu$ decay results, before hadronization, in two final-state
light quarks.  These can have either isospin $I=0$ or $I=1$.  Because final-state interactions
preserve isospin, a final $\pi\pi$ state is also restricted to $I=0$ or $I=1$.  From Bose symmetry
considerations, the $\pi\pi$ state must have angular momentum $L$ even for $I=0$ and
$L$ odd for $I=1$.   Isospin considerations then imply
\[
\begin{array}{rr}
  I=1,L\ \textrm{odd}     & \pi^\pm\pi^0:\pi^+\pi^-:\pi^0\pi^0 = 2:1:0  \\
  I=0,L\ \textrm{even}   &  \pi^\pm\pi^0:\pi^+\pi^-:\pi^0\pi^0 = 0:2:1.
\end{array}
\]
Assuming that the $L=3,5,...$ configurations are suppressed relative
to the $L=1$ configuration, we can use $e^+e^-$ scattering data and
$\tau$ decay data to conclude that the $I=1,L\ \textrm{odd}$ component
is completely dominated by the $\rho$.  A significant nonresonant contribution
would therefore come via the $I=0,L\ \textrm{even}$ channel.  With the $I=0$
rate parameterized by $\alpha$, we expect partial widths in the ratios
\[
\pi^\pm\pi^0:\pi^+\pi^-:\pi^0\pi^0 = 2:1+2\alpha:\alpha.
\]

To estimate the systematic due to an unknown nonresonant
$\pi\pi\ell\nu$ contribution, we look for a component, after event selection, that could
mimic a $\rho\ell\nu$. To constrain such a contribution, we add the mode
$\pi^0\pi^0\ell\nu$ to the fit.
Procedurally, we generate $\pi^0\pi^0\ell\nu$ using the $\rho$ lineshape and the
$\rho\ell\nu$ form factors.  We then perform fits with the usual isospin constraint
on the partial widths ($\rho^\pm:\rho^0=2:1$) replaced with the $\pi\pi$ ratios given
above.  While the most relevant fit for the extraction of a systematic uncertainty number
has the parameter $\alpha$ floating, we also fix $\alpha=0$ to test the fit quality under
the assumption that observed $\pi^0\pi^0\ell\nu$ yields are consistent with cross feed
from other modes and the other standard backgrounds. 

%: pi0pi0 masses
\begin{figure}[t]
\begin{center}
\epsfxsize=6.0in
\epsfbox{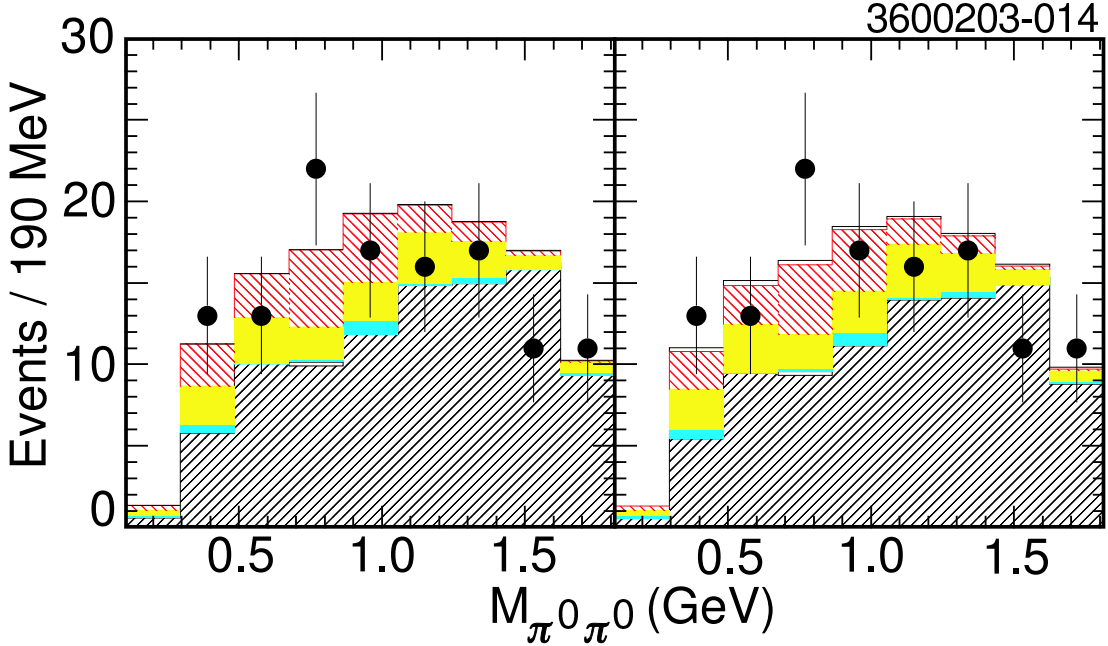}
\caption{The $\pi^0\pi^0$ mass distribution from the reconstructed
$\pi^0\pi^0\ell^\pm\nu$ ``signal bin'' from the nominal fit (left) and from
the fit including a $\pi^0\pi^0\ell^\pm\nu$ signal component (right) as described
in the text. The points are
the on-resonance data.  The histogram components, from
bottom to top, are $b\to c$ (fine $45^\circ$ hatch), continuum (grey or green cross hatch), fake
leptons (cyan or dark grey), feed down from other $B\to X_u\ell\nu$ modes
(yellow or light grey), cross feed from the signal modes into the reconstructed
modes (red or black fine $135^\circ$ hatch), and 
signal (open).  The normalizations are from the corresponding fits. }
\label{fig:pi0pi0}
\end{center}
\end{figure}

In the fits, the $\pi^0\pi^0\ell\nu$ mode is treated like the $\omega$ mode.  Only the sum of
the three $q^2$ intervals contributes to the likelihood, but the signal Monte Carlo is scaled
in each $q^2$ interval separately to maintain the above $\pi\pi$ ratios from one interval 
to the next.   Figure~\ref{fig:pi0pi0} shows the projection onto the
$m_{\pi^0\pi^0}$ distribution for fits with and
without a $\pi^0\pi^0\ell\nu$ signal component.  Note that the fit included data only from 
the three bins in the range $0.485 \le m_{\pi^0\pi^0} < 1.055$ GeV.
The fit quality is excellent when  the 
reconstructed $\pi^0\pi^0\ell\nu$  mode is included but the $\pi^0\pi^0\ell\nu$ signal
is forced to zero.  Table~\ref{tab:p0p0} summarizes the observed changes in the
$\rho^-\ell^+\nu$ branching fraction when we float the $\pi^0\pi^0\ell\nu$ signal
component.  The resulting $\pi^0\pi^0\ell\nu$ yield is consistent with zero.  The shifts
in the various $\rho\ell\nu$ branching fractions are larger effects than the increase in
their  errors due to correlations with the $\pi^0\pi^0\ell\nu$.  We thus take the shifts as
the estimate of the uncertainty. The pseudoscalar modes shift negligibly. 

\begin{table}[tb]
  \centering 
  \caption{Comparisons of the $\rho^-\ell^+\nu$ branching fractions when the
 $\pi^0\pi^0\ell\nu$ mode and component are added. The parameter $\alpha$
that normalizes the $\pi^0\pi^0\ell\nu$ component is described in the text.  The percentage change
relative to the standard fits in Table~\protect\ref{tab:nomfit_br} are indicated in parentheses below
the branching fractions.}\label{tab:p0p0}
\begin{tabular}{ccccccc}
\hline\hline
analysis & $\alpha$ & ${\cal B}(B\to\rho\ell\nu)$ & ${\cal B}_{q^2<8\ \mathrm{GeV}^2}$ & ${\cal B}_{8\le q^2<16\ \mathrm{GeV}^2}$ & ${\cal B}_{q^2\ge 16\ \mathrm{GeV}^2}$ & $\chi^2$/dof \\
               &                   &  ($10^{-4}$)                        &  ($10^{-4}$)                                          &  ($10^{-4}$)          & \\  \hline
$p_\ell>1.5\ \mathrm{GeV}/c$    & $0.25\pm0.21$ & $1.88\pm0.35$ & $0.393\pm0.209$ & $1.055\pm0.261$ & $0.428\pm0.099$ & 273.7 / (280-21) \\
& & (-13\%) & (-9\%) &  (-15\%) & (-14\%) & \\
$p_\ell>1.75\ \mathrm{GeV}/c$  & $0.22\pm0.18$ & $2.06\pm0.35$ & $0.455\pm0.216$ & $1.153\pm0.259$ & $0.455\pm0.099$ & 271.6 / (280-21) \\
 & & (-12\%) & (-8\%) &  (-9\%) & (-11\%) & \\
$p_\ell>2.0\ \mathrm{GeV}/c$    & $0.18\pm0.13$ & $2.17\pm0.36$ & $0.669\pm0.249$ & $1.009\pm0.241$ & $0.496\pm0.097$ & 281.1 / (280-21) \\
& & (-5\%) & (8\%) &  (-9\%) & (-11\%) & \\
\hline
\end{tabular}
\end{table}

In addition to the variations above, we have performed numerous systematic checks, including
variation of the selection criteria and investigation of electron and muon samples separately.
We have also investigated tighter momentum requirements in the pseudoscalar modes.
The observed variations were in general consistent within the uncertainties resulting from the statistical 
changes.  

\section{Dependence of branching fractions on form factors}
\label{sec:formfacts}

\begin{table}[tp]
\caption{Branching fractions ${\cal B}(B^0\to\pi^-\ell^+\nu)$ obtained under 
variation of the $\pi$  and $\rho/\omega\ell\nu$ form-factor
models.  Shown are the results for the total branching fraction,
the partial branching fraction in each $q^2$ bin, and the $-2\ln L$ for the
fit.  Branching fraction uncertainties are statistical only.  The estimated
model dependence is indicated after each set of variations.  All branching fractions
are in units of $10^{-4}$. The $\pi$ model variations are all presented
for the analysis with the $p_\ell>1.5\ \mathrm{GeV}/c$ requirement on the vector
modes.}
\label{tab:bf_pi_fits_model}
\begin{tabular}{llccccc}  \hline\hline
 & & & \multicolumn{3}{c}{$q^2$ interval (GeV$^2$)}  & \\
$\pi$ Model & $\rho$ Model & ${\cal B}_{\text{total}}$ &
${\cal B}_{< 8}$ & ${\cal B}_{8-16}$ & ${\cal B}_{\ge16}$ &
$-2\ln L$ \\ \hline
Ball'01 & Ball'98     & $1.327\pm0.177$ & $0.431\pm0.106$ & $0.651\pm0.105$ & $0.245\pm0.094$ & 240.3  \\
ISGW2   & Ball'98     & $1.327\pm0.176$ & $0.431\pm0.107$ & $0.660\pm0.106$ & $0.236\pm0.09 $ & 240.7  \\
SPD     & Ball'98     & $1.315\pm0.173$ & $0.436\pm0.106$ & $0.650\pm0.105$ & $0.229\pm0.088$ & 239.8  \\ 
\multicolumn{2}{l}{$1.7\times \mathrm{RMS}_{\pi\,\mathrm{FF}}$}  & 0.01 & 0.004 & 0.01 & 0.01  \\ \hline
Ball'01 & Ball'98     & $1.33\pm0.18$ & $0.43\pm0.11$ & $0.65\pm0.11$ & $0.25\pm0.09$ & 240.3 \\
Ball'01 & ISGW2       & $1.41\pm0.18$ & $0.45\pm0.11$ & $0.69\pm0.10$ & $0.27\pm0.09$ & 239.4 \\
Ball'01 & Melikhov'00 & $1.30\pm0.18$ & $0.43\pm0.11$ & $0.65\pm0.11$ & $0.22\pm0.09$ & 240.2 \\
Ball'01 & UKQCD'98    & $1.36\pm0.18$ & $0.44\pm0.11$ & $0.66\pm0.11$ & $0.26\pm0.09$ & 239.3 \\
\multicolumn{2}{l}{$1.7\times \mathrm{RMS}_{\rho\,\mathrm{FF}}$}  & 0.07 & 0.01 & 0.03 & 0.03 \\ 
 \hline \hline
\end{tabular}
\end{table}

\begin{table}[tp]
\caption{Branching fractions ${\cal B}(B^0\to\rho^-\ell^+\nu)$ obtained under 
variation of the $\pi\ell\nu$ and $\rho/\omega\ell\nu$ form-factor models.
Shown are the results for the total branching fraction,
the partial branching fraction in each $q^2$ bin, and the $-2\ln L$ for the
fit.  Branching fraction uncertainties are statistical only.  The estimated
model dependence is indicated after each set of variations.  
All branching fractions are in units of $10^{-4}$.  The $\pi$ model variations are all presented
for the analysis with the $p_\ell>1.5\ \mathrm{GeV}/c$ requirement on the vector
modes.  For the vector mode form-factor variation, we present the results
for all three momentum requirements.}
\label{tab:bf_rho_fits_model}
\begin{tabular}{llccccc}  \hline\hline
 & & & \multicolumn{3}{c}{$q^2$ interval (GeV$^2$)}  &   \\
$\pi$ Model & $\rho$ Model & ${\cal B}_{\text{total}}$ &
${\cal B}_{< 8}$ & ${\cal B}_{8-16}$ & ${\cal B}_{\ge16}$ & $-2\ln L$ \\ \hline
Ball'01 & Ball'98     & $2.172\pm0.338$ & $0.429\pm0.198$ & $1.244\pm0.256$ & $0.499\pm0.097$ & 240.3  \\ 
ISGW2 & Ball'98     & $2.176\pm0.338$ & $0.430\pm0.198$ & $1.248\pm0.256$ & $0.499\pm0.098$ & 240.7  \\ 
SPD     & Ball'98     & $2.169\pm0.338$ & $0.420\pm0.198$ & $1.250\pm0.256$ & $0.499\pm0.098$ & 239.8  \\ 
\multicolumn{2}{l}{$1.7\times \mathrm{RMS}_{\pi\,\mathrm{FF}}$}  & 0.01 & 0.01 & 0.004 & 0.004 \\ \hline
\multicolumn{7}{l}{$p_\ell> 1.5\ \mathrm{GeV}/c$} \\
Ball'01 & Ball'98     & $2.17\pm0.34$ & $0.43\pm0.20$ & $1.24\pm0.26$ & $0.50\pm0.10$  & 240.3 \\
Ball'01 & ISGW2       & $1.91\pm0.28$ & $0.30\pm0.13$ & $1.14\pm0.23$ & $0.47\pm0.10$  & 239.4 \\
Ball'01 & Melikhov'00 & $2.56\pm0.37$ & $0.33\pm0.15$ & $1.49\pm0.31$ & $0.75\pm0.14$  & 240.2 \\
Ball'01 & UKQCD'98    & $2.08\pm0.32$ & $0.39\pm0.17$ & $1.21\pm0.25$ & $0.49\pm0.10$  & 239.3 \\
\multicolumn{2}{l}{$1.7\times \mathrm{RMS}_{\rho\,\mathrm{FF}}$} & 0.41 & 0.09 & 0.22 & 0.19 \\ \hline
\multicolumn{7}{l}{$p_\ell> 1.75\ \mathrm{GeV}/c$} \\
Ball'01 & Ball'98     & $2.34\pm0.34$ &$ 0.50\pm0.20$ & $1.32\pm0.26$ & $0.52\pm0.10$ & 241.6 \\
Ball'01 & ISGW2       & $2.03\pm0.28$ &$ 0.34\pm0.13$ & $1.20\pm0.23$ & $0.49\pm0.10$ & 240.3 \\
Ball'01 & Melikhov'00 & $2.74\pm0.37$ &$ 0.38\pm0.16$ & $1.58\pm0.31$ & $0.78\pm0.14$ & 241.4 \\
Ball'01 & UKQCD'98    & $2.23\pm0.32$ &$ 0.45\pm0.18$ & $1.28\pm0.25$ & $0.51\pm0.10$ & 240.4 \\
\multicolumn{2}{l}{$1.7\times \mathrm{RMS}_{\rho\,\mathrm{FF}}$} & 0.44  & 0.11  & 0.24  & 0.20 \\ \hline
\multicolumn{7}{l}{$p_\ell> 2.0\ \mathrm{GeV}/c$} \\
Ball'01 & Ball'98     & $2.29\pm0.35$ & $0.62\pm0.22$ & $1.11\pm0.25$ & $0.56\pm0.10$ & 244.2 \\
Ball'01 & ISGW2       & $1.89\pm0.27$ & $0.38\pm0.13$ & $0.98\pm0.22$ & $0.54\pm0.09$ & 243.4 \\
Ball'01 & Melikhov'00 & $2.66\pm0.38$ & $0.48\pm0.17$ & $1.36\pm0.31$ & $0.83\pm0.14$ & 244.6 \\
Ball'01 & UKQCD'98    & $2.15\pm0.32$ & $0.54\pm0.19$ & $1.07\pm0.24$ & $0.55\pm0.09$ & 243.3 \\
\multicolumn{2}{l}{$1.7\times \mathrm{RMS}_{\rho\,\mathrm{FF}}$} & 0.47 &  0.15 &  0.24 &  0.21 \\ 
\hline\hline
\end{tabular}
\end{table}

In the original measurement of the exclusive charmless branching fractions \cite{bb:lkg_cleo_exclusive},
there were two roughly comparable contributions to the branching fraction errors from
the form-factor uncertainties.  The first contribution resulted because the efficiency varied
as a function of $q^2$ (inescapable with a lepton momentum cut), and the data were
lumped into a single $q^2$ bin.  Because we now extract the rates independently in
three separate $q^2$ ranges, this analysis should see a significant reduction in
this effect.  The second contribution resulted because there was significant $q^2$
dependence to the cross--feed rates between the pseudoscalar and the vector modes.
Again, since we extract the rates independently as a function of $q^2$, this
dependence should be reduced.

We have estimated the model
dependence based on changes of the branching fractions under variation
of the form-factor calculation.  The previous analysis~\cite{bb:lkg_cleo_exclusive} found that the
error on the branching fraction obtained from comparison of models was larger than
that obtained by variation of a particular form-factor parameterization within the
published uncertainties (when given).  Tables~\ref{tab:bf_pi_fits_model} and
\ref{tab:bf_rho_fits_model} show the variation in ${\cal B}(B^0\to \pi^-\ell^+\nu)$ and
${\cal B}(B^0\to \rho^-\ell^+\nu)$, respectively, as
the $\pi$ and vector form factors are varied.  We have included in the set of models
those which have the most extreme variations in shape of $d\Gamma/dq^2$.
For $\pi\ell\nu$, we find that our method results in almost no sensitivity to the
form factor used for the signal mode efficiencies.  We find a larger sensitivity
to the variation of the vector mode form factors because of cross feed from
those modes.   For $\rho \ell \nu$, there is almost no sensitivity to the
$\pi\ell\nu$ form factors, but significant sensitivity to the
$\rho \ell \nu$ form factors.

To assign uncertainties, we use an empirical observation from the original
analysis~\cite{bb:lkg_cleo_exclusive}.  For that analysis, for any given model, we varied the internal
parameters to determine an error on the rates extracted within that model.  We
then defined a range of potential branching fractions by taking the model with
the lowest result and subtracting one standard deviation from the variations
within that model, and taking the model with the highest result and adding one
standard deviation.  Our assigned uncertainty covered 70\% of this range.
(Note that this procedure gave us a more conservative range than taking
one half the spread among the central value of the models.)  Empirically, we found that this procedure
agreed with taking 1.7 times the RMS spread among models for all quantities
that we examined.  For these  results, we therefore apply this latter
procedure.  The results are also summarized in Tables~\ref{tab:bf_pi_fits_model} and
\ref{tab:bf_rho_fits_model}.
%:model errors

For purposes of direct comparison, had we adopted the procedure used in
recent $\rho\ell\nu$ analyses by the BABAR Collaboration \cite{bb:newBabar}
and by CLEO 2000 \cite{bb:lange_cleo_exclusive},
we would assign (absolute) uncertainties of $0.06\times10^{-4}$ 
(rather than $0.07\times10^{-4}$) and $0.33\times10^{-4}$ 
(rather than $0.41\times10^{-4}$) 
for the $\rho^-\ell^+\nu$ form--factor dependence on the
total branching fraction for $\pi^-\ell^+\nu$ and $\rho^-\ell^+\nu$, respectively.
The $\rho^-\ell^+\nu$ number, $0.33\times10^{-4}$, is 
about half of the size seen in the recent BABAR measurement, which,
like the CLEO 2000 measurement, is mainly sensitive to the end-point region 
$p_\ell>2.3\ \mathrm{GeV}/c$.

We stress that the form factors from any given model are {\em not} used to constrain
the relative rates extracted in each of the three $q^2$ regions.  Only the efficiencies
within each $q^2$ range are modified. Hence the quality of the fit used to extract the rates
does not discriminate among different form-factor descriptions.  This discrimination
is discussed in the following section.

Overall, our procedure has drastically reduced the sensitivity of the $\pi\ell\nu$
result to both the $\pi\ell\nu$ and the vector-mode form factors.  There is
essentially no dependence on the $\pi\ell\nu$ form factors themselves.  The combined
sensitivity to both the $\pi$ and $\rho$ form factors is about one third that of
the previous CLEO $\pi\ell\nu$ analysis.

The $\rho\ell\nu$ variation remains significant, though again this analysis shows
essentially no dependence on the $\pi\ell\nu$ form factor.  The overall uncertainty
of the form factors has reduced to about 80\% of the original CLEO $\rho\ell\nu$
measurement~\cite{bb:lkg_cleo_exclusive} (which had a smaller form-factor dependence than the 2000 CLEO $\rho\ell\nu$
analysis~\cite{bb:lange_cleo_exclusive}).  As one tightens the lepton momentum requirement,  the model dependence
increases slightly over the range we have studied.  As expected, the lowest $q^2$
interval shows the greatest sensitivity (fractionally) to the variation in the range.
For a given model, the variation of the total branching fraction as the lepton
momentum requirement is varied is small compared to the variation among models
for a given momentum requirement.  (The RMS variation of the former is
about 30\% of the RMS variation of the latter.)  We speculate that the
dominant model dependence likely arises from our $\cos\theta_{W\ell}>0$
requirement, which we applied to suppress $b\to c$ background.  Either finer $q^2$ 
binning or an alternate means of background suppression would provide a route for
further reduction of the form-factor dependence.

For the $\eta\ell\nu$ branching fraction, we find a dependence of $0.04\times 10^{-4}$
from variation of the $\pi\ell\nu$ form factors and $0.01\times 10^{-4}$ from variation
of the $\rho\ell\nu$ form factors.  The only $\eta\ell\nu$
form factor that we consider is ISGW II~\cite{Scora:1995ty}.  However,
the $\eta\ell\nu$ analysis presented
here is almost identical to the original $\pi\ell\nu$ analysis.  We therefore take the form-factor
dependence of 10\% found in that analysis as an estimate of the uncertainty from the $\eta$
form factors.  As the $\rho\ell\nu$ form factors contributed substantially to the 10\%
uncertainty in the previous analysis, yet contribute negligibly to $\eta\ell\nu$,
the 10\% should be a conservative estimate.

The results presented here agree well with the previous CLEO measurements and the
recent BABAR $\rho\ell\nu$ measurement.  The results of the original CLEO
measurement~\cite{bb:lkg_cleo_exclusive}
are superseded by this measurement.  The results of the CLEO 2000
measurement~\cite{bb:lange_cleo_exclusive}
are essentially statistically independent of those presented here.

%-----------------------------

\section{\boldmath Extraction of $|V_{ub}|$ and discrimination of models}

We extract $|V_{ub}|$ from the measured rates for $\pi\ell\nu$ only,
for  $\rho\ell\nu$ only, and then by using the combined information from
those two modes.   In all cases, the $|V_{ub}|$ extraction is based on
the results from the the analysis requiring $p_\ell>1.5\ \mathrm{GeV}/c$
in the vector modes.  We use a $B^0$ lifetime of
$(1.542 \pm 0.016)$ ps \cite{bb:pdg2002}.

\subsection{\boldmath $|V_{ub}|$ from $B \to \pi \ell \nu$}

\begin{figure}[t]
\centering
\leavevmode
\epsfxsize=4.5in
\epsfbox{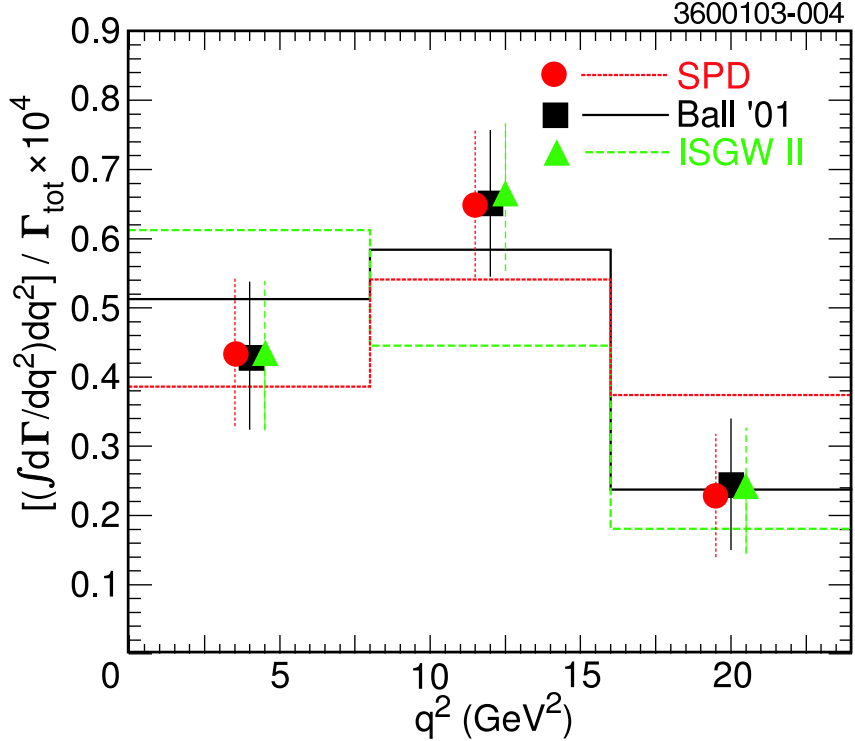}
\caption{Measured branching fractions in the restricted $q^2$ intervals for $B^0\to\pi^-\ell^+\nu$
(points) and the best fit to the predicted $d\Gamma/dq^2$ (histograms) for the
three models used to extract both rates and $|V_{ub}|$.  The data points have 
small horizontal offsets introduced
for clarity.  The last bin has been artificially truncated at 24 GeV$^2$ in the plot -- the
information out to $q^2_{\text{max}}$ has been included in the work.}
\label{fig:Vub_fits}
\end{figure}

For $\pi\ell\nu$, we first explore fitting $q^2$ distributions from various
form-factor predictions to the measured rates in the three $q^2$ bins.  To
be self-consistent, we extract $|V_{ub}|$ for a particular form factor using the rates 
from the fit with that model.  In practice, as we have seen, this makes little difference
in the $\pi$ modes in this analysis.  Since each model predicts the total rate modulo $|V_{ub}|$,
$|V_{ub}|$  becomes the one free parameter for the fit that normalizes the
prediction to the observed rates.  The quality of the fit measures how well the
form-factor shape describes the data, so provides one means of discrimination
among form factors.  The results of this procedure are summarized in 
Table~\ref{tab:Vub_fits_one}.  For the three calculations that have been used for
both efficiency and $|V_{ub}|$ extraction, the data rates with the best fits
for each predicted form factor are shown in Figure~\ref{fig:Vub_fits}. 
The probability of $\chi^2$ in our various fits for the
 ISGW~II model varies between one to three percent, indicating that this model is likely
to be less reliable for determination of $|V_{ub}|$ from $\pi\ell\nu$.  Note
further that the spread among the central values from the various calculations
is fairly small relative to the uncertainties quoted in the calculations themselves.

Because the extracted rates in the $q^2$ intervals are now essentially independent
of the $\pi\ell\nu$ form factor, one can extract $|V_{ub}|$ from our results for form
factors not considered here.  We provide in Appendix~\ref{app:method} a detailed 
methodology for doing so.

To determine the effect of the systematic uncertainties, we repeat the above fit
using the three $q^2$ rates obtained from the branching ratio fit after each
systematic variation.  This procedure automatically accounts for correlations
among the three intervals.  We then increase the uncertainty for each variation
by one half of the fractional error introduced by the second term in
Equation~\ref{eqn:add_term}.  The factor of one half arises from the square
root involved in extraction of $|V_{ub}|$ from the rate.

% Entered 2003.01.19, A. Warburton
\begin{table}[tb]
\caption{$|V_{ub}|$ extracted from fits to the rates measured in the
three $q^2$ intervals for a variety of form factors for $\pi\ell\nu$.
The table indicates form-factor calculation, $|V_{ub}|$ with
statistical error only, predicted $\Gamma^{\text{th}}_\pi/|V_{ub}|^2$
with the estimated theoretical uncertainty, the $\chi^2$ for the fit,
and the probability of $\chi^2$ given the two degrees of freedom.}
\label{tab:Vub_fits_one}
\begin{minipage}{6.0 in}
\renewcommand{\footnotesep}{7 pt}
\renewcommand{\baselinestretch}{0.7}
\renewcommand{\footnoterule}{\vspace{ -20 pt} }
\begin{tabular}{llcccc}  \hline\hline
$\pi$ Model & $\rho$ Model & $|V_{ub}| \times 10^3$ & $\Gamma^{\text{th}}_\pi /
|V_{ub}|^2$ (ps$^{-1}$)& Fit $\chi^2$ & $P(\chi^2)$ \\ \hline
Ball'01 & Ball'98          & $3.21 \pm 0.21$ & $8.4^{+3.5}_{-2.4}$ & 1.0  & 0.61 \\
KRWWY\footnote{Uses rates determined with the Ball'01 form factor.}
             & Ball'98          & $3.40 \pm 0.23$ & $7.3\pm 2.5$ & 5.3  & 0.07 \\
ISGW2 & Ball'98          & $2.90 \pm 0.20$ & $9.6\pm 4.8$ & 7.3  & 0.03 \\
SPD     & Ball'98          & $2.96 \pm 0.19$ & $9.6\pm 2.9$ & 4.0  & 0.14 \\
\hline
Ball'01 & Ball'98          & $3.21 \pm 0.21$ & $8.4^{+3.5}_{-2.4}$ & 1.0  & 0.61 \\
Ball'01 & ISGW2          & $3.31 \pm 0.20$ & $8.4^{+3.5}_{-2.4}$ & 1.2  & 0.55 \\
% The following line has been updated, but commented out (A.W., 2003.01.19)
%ISGW2 & ISGW2          & $2.99 \pm 0.19$ & $9.6\pm 4.8$ & 8.5 & 0.01 \\
%\hline
Ball'01 & Melikhov'00 & $3.18 \pm 0.21$ & $8.4^{+3.5}_{-2.4}$ & 0.9  & 0.63 \\
Ball'01 & UKQCD'98   & $3.24 \pm 0.20$ & $8.4^{+3.5}_{-2.4}$ & 1.1  & 0.59 \\ \hline\hline
\end{tabular}
\end{minipage}
\end{table}

As we discuss below, each of the form-factor calculations used to extract $|V_{ub}|$
from the full $q^2$ range has some measure of model dependence.  
We determine a systematic error in $|V_{ub}|$ from the quoted theoretical uncertainty
in form-factor normalizations, with the following procedure.  For each form
factor used, we
recalculate $|V_{ub}|$ when we increase or decrease the form-factor normalization
by one standard deviation.   Due to the poor agreement
of the ISGW~II form factor with the $\pi\ell\nu$ data in conjunction with the somewhat
{\em ad hoc} assumptions about the form-factor $q^2$--dependence in that mode,
we drop ISGW~II from consideration.  From the others, we find the minimum value
$V_{\text{min}}$ and the maximum value $V_{\text{max}}$.  We then
assign an asymmetric error of 70\% of the deviation relative to the nominal
central value -- that is, we take $0.7(V_{\text{max}}-V_{\text{nom}})$ and 
$0.7(V_{\text{nom}}-V_{\text{min}})$.  Because the 
result obtained using Ball'01 is close to the mean, we take that result as the nominal value.
Note that when a symmetric theory error is quoted on the rate, we re-interpret that error
as symmetric on the {\em amplitude}.  To be precise, we map $\gamma_{\text{th}}\pm\sigma_{\text{th}}$
to $\gamma_{\text{th}}\pm \sigma_{\text{th}}(1\pm\sigma_{\text{th}}/(4\gamma_{\text{th}}))$.
This procedure yields
\begin{equation}
|V_{ub}|=(3.21\pm 0.21\pm0.14 \;^{+0.62}_{-0.45}\pm 0.10)\times 10^{-3},
\end{equation}
 where the errors are
statistical, experimental systematic, the estimated uncertainties
from the $\pi\ell\nu$ form-factor shape and normalization,
 and the $\rho\ell\nu$ form-factor shape, respectively.
The $\rho\ell\nu$ form-factor contribution has been estimated
using the $1.7\sigma_{\mathrm{RMS}}$ prescription.

Again for direct comparison with other experiments, taking one half, 
rather than 70\%, as the scale factor for estimating the uncertainties yields
$|V_{ub}|=(3.21\pm 0.21\pm0.14\;^{+0.44}_{-0.32}\pm 0.07)\times 10^{-3}$.

Note that the error on $|V_{ub}|$ from the
uncertainty in the rates under variation of form factors is completely dwarfed by
the error arising from uncertainty in the theoretical normalization of the form factor.

Our second, preferred, method for determining $|V_{ub}|$ attempts to
reduce the number of modeling assumptions and hence to provide a more
robust uncertainty estimate. We therefore limit our consideration to
 form factors determined from LCSR and
from LQCD calculations are QCD-based calculations.  These calculations, 
however, are only valid over a restricted $q^2$ region.
The LCSR assumptions are expected to break down for $q^2\ge 16 \ \text{GeV}^2$,
while the current LQCD calculations are valid
only for $q^2\gtrsim \ 16\ \text{GeV}^2$.
Extrapolation outside of these ranges therefore introduces a dependence
on the form used for the extrapolation.  This introduces another uncertainty
that is difficult to asses.  To minimize this uncertainty, we
extract $|V_{ub}|$ from these more restricted regions.  For LQCD, we determine
$|V_{ub}|$ from the measured rate and the calculated rate  in the 
range $q^2\ge \ 16\ \text{GeV}^2$.  For LCSR, we determine $|V_{ub}|$
by fitting the calculated LCSR rates to the measured rates in the two $q^2$
intervals below $16\ \text{GeV}^2$.  The results are
shown in Table~\ref{tab:Vub_fits_two}.   

To produce a final LQCD result for the $q^2\ge \ 16\ \text{GeV}^2$ region, we
take a statistically--weighted average of the different LQCD results. To
the precision quoted, we obtained identical results if we based the statistical
weights on the upper, the lower, or the average of the asymmetric statistical
errors quoted in Table~\ref{tab:Vub_fits_two}.   We assume the systematic errors
are completely correlated among the different calculations: if $\alpha_i$ is the
statistical weight used in the average for  calculation $i$ and $\hat{\sigma}_i$ is the fractional
systematic error for that calculation, then the total fractional systematic error $\hat{\sigma}$ assigned to
the average is $\hat{\sigma} = \sum\alpha_i\hat{\sigma}_i$.  The theoretical errors quoted in
Table~\ref{tab:Vub_fits_two} do not include any uncertainty from the quenched approximation, which is estimated
to be in the 10\% to 20\% range.  We  add an additional 15\% in quadrature to the
systematic uncertainty just described to obtain the average theoretical systematic uncertainty
 quoted in the table.

From our average of the LQCD--based results, we estimate
\begin{equation}
|V_{ub}|_{q^2\ge 16\text{ GeV}^2}=(2.88\pm 0.55\pm 0.30\;^{+0.45}_{-0.35}\pm 0.18)\times 10^{-3},
\end{equation}
where the errors are statistical, experimental systematic, LQCD uncertainties, 
and $\rho\ell\nu$ form-factor dependence, respectively.  The LQCD uncertainties have
been combined in quadrature.

% Entered 2003.01.19, A. Warburton, modified 2/24/03
\begin{table}[tb]
\caption{Values for $|V_{ub}|$ obtained using form factors (FF) from
light-cone sum rules in the $q^2$ interval $0-16$ GeV$^{2}$ (top two
rows) and from LQCD for $q^2\ge16$ GeV$^2$ (bottom five rows).
Only the statistical errors on $|V_{ub}|$ are indicated. The data rates obtained
using Ball'01 for $\pi\ell\nu$ and Ball'98 for $\rho\ell\nu$ were used
as the input for all values obtained.}
\label{tab:Vub_fits_two}
\begin{minipage}{6.0 in}
\renewcommand{\footnotesep}{7 pt}
\renewcommand{\baselinestretch}{0.7}
\renewcommand{\footnoterule}{\vspace{ -20 pt} }
\begin{tabular}{lcccc}  \hline\hline
$\pi$ FF & $|V_{ub}| \times 10^3$ & $\Gamma^{\text{th}}_\pi /
|V_{ub}|^2$ (ps$^{-1}$)& Fit $\chi^2$ & $P(\chi^2)$ \\ \hline
Ball'01    & $3.20 \pm 0.22$ & $6.9^{+2.4}_{-1.8}$ & 1.0  & 0.32 \\
KRWWY & $3.46 \pm 0.24$ & $5.7\pm 1.9$ & 5.0  & 0.025 \\
\hline
FNAL\footnote{The authors of \protect\cite{El-Khadra:2001rv} have provided the rate integrated over this range and the corresponding uncertainty.} 
              & $2.88\pm 0.55$ & $1.91^{+0.46}_{-0.13}\pm0.31$ & --  & -- \\
JLQCD\footnote{The authors of \protect\cite{Aoki:2001rd} have provided the rate integrated over this range and the corresponding uncertainty.} 
              & $3.05\pm 0.58$ & $1.71^{+0.66}_{-0.56}\pm0.46$ & --  & -- \\
APE\footnote{We have integrated  over the restricted $q^2$
interval to obtain rates using the FF parameterization from the two APE methods, scaled the uncertainties accordingly, and performed a simple average of the two rates.} 
              & $2.97 \pm 0.57$ & $1.80^{+0.89}_{-0.71}\pm0.47$ & --& -- \\
UKQCD\footnote{We have integrated  the FF parameterization over the restricted $q^2$
interval to obtain the central value and have scaled the uncertainties accordingly.} 

              & $2.63 \pm 0.50$ & $2.3^{+0.77}_{-0.51}\pm0.51$ & --  & -- \\ 
average\footnote{See text.}
              &  $2.88\pm 0.55$ & $1.92^{+0.32}_{-0.12}\pm0.47$ & --  & -- \\ \hline\hline
\end{tabular}
\end{minipage}
\end{table}
%:vubTable2

Taking the
simple average of the two LCSR values and again using the 70\% range to estimate the 
theoretical uncertainty, we characterize the LCSR results as 
\begin{equation}
|V_{ub}|_{q^2< 16\text{ GeV}^2}=(3.33\pm 0.24 \pm0.15 \;^{+0.57}_{-0.40}\pm 0.06)\times 10^{-3}.
\end{equation}
Using the fractional errors from the LCSR calculations alone gives similar theoretical uncertainties.

We average the LQCD and LCSR results, with 
correlated experimental systematics taken into account, according to the procedure
laid out in Appendix~\ref{app:AvgProc}.   The LQCD value enters the average with
a weight of $\alpha_\pi=0.20$.  As noted in the appendix, we choose the weight to minimize
the total overall uncertainty.  To be conservative, we have treated the theoretical
uncertainties as if they were completely correlated.

\begin{equation}
|V_{ub}|=(3.24\pm 0.22 \pm 0.13 \;^{+0.55}_{-0.39}\pm 0.09)\times 10^{-3}.
\end{equation}
We take this as the more reliable determination of $|V_{ub}|$ from our complete data
in this mode.

\begin{figure}[t]
\centering
\leavevmode
\epsfysize=3.5in
\epsfbox{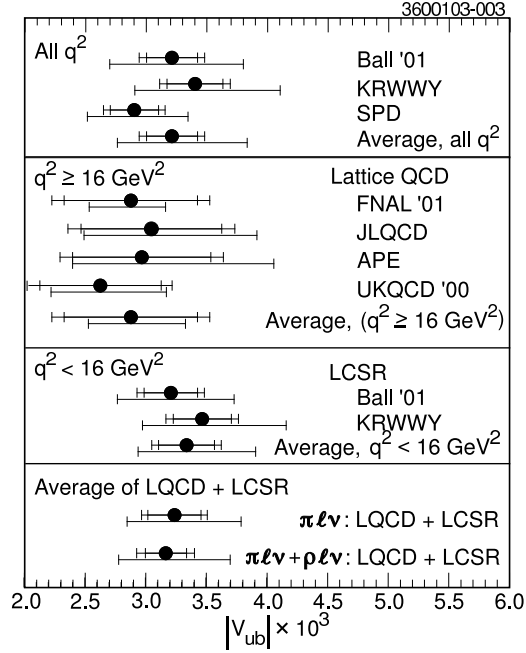}
\caption{Values for $|V_{ub}|$ obtained from $\pi\ell\nu$ using the entire $q^2$ range for the
various form-factor calculations (top block), using LQCD for $q^2\ge16$ GeV$^2$
(second block), using LCSR for $q^2<16$ GeV$^2$ (third block), and
our average of the last two (bottom block) for $\pi\ell\nu$ only and for
$\pi\ell\nu$ and $\rho\ell\nu$ combined.  In all cases, the top bar indicates the
statistical and all the experimental systematics (combined in quadrature), the
lower bar indicates the approximate ``one standard deviation'' range of 
motion due to the theoretical uncertainties.}\label{fig:vub_pi}
\end{figure}

The variations in $|V_{ub}|$ and our averages are illustrated in Figure~\ref{fig:vub_pi}.

\subsection{\boldmath $|V_{ub}|$ from $B \to \rho \ell \nu$}

\begin{figure}[t]
\centering
\leavevmode
\epsfxsize=4.5in
\epsfbox{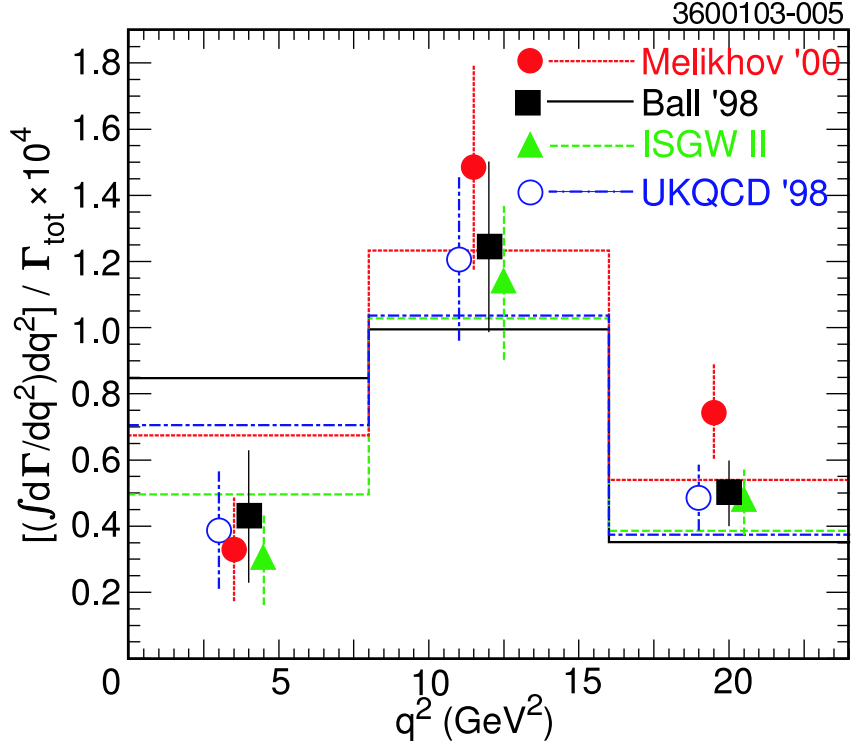}
\caption{Measured branching fractions in the restricted $q^2$ intervals for $B^0\to\rho^-\ell^+\nu$
(points) and the best fit to the predicted $d\Gamma/dq^2$ (histograms) for the
three models used to extract both rates and $|V_{ub}|$.  The data points have 
small horizontal offsets introduced
for clarity.}
\label{fig:Vub_rho_fits}
\end{figure}

We proceed with $B \to \rho \ell \nu$ in much the same fashion as with $B \to \pi \ell \nu$.
The fits of the different form factors  to the rates extracted from the three $q^2$
intervals in the data are illustrated in Figure~\ref{fig:Vub_rho_fits} and are
summarized in Table~\ref{tab:Vub_fits_one_rho}.
Because of the relatively large variation in the
rates extracted from the data using the different form-factor calculations, 
we again perform the extraction of $|V_{ub}|$ entirely
within the context of a given form-factor calculation.  
In general, the theoretical predictions do not match the data as well as we saw for
the $\pi\ell\nu$ mode.  In spite of some of the poor fits, we consider all four sets of form
factors as we estimate $|V_{ub}|$ with this mode.  As we expected from
the branching fraction results, the $|V_{ub}|$ extracted
from the  $\rho\ell\nu$ information does not depend
on the  $\pi\ell\nu$ form factor used in the analysis.

% Entered 2003.01.19, A. Warburton
\begin{table}[tb]
\caption{$|V_{ub}|$ extracted from fits to the rates measured in the
three $q^2$ intervals for a variety of form factors for $\rho\ell\nu$.
The table indicates form-factor calculation, $|V_{ub}|$ with
statistical error only, predicted $\Gamma^{\text{th}}_\rho/|V_{ub}|^2$ with the
estimated theoretical uncertainty, the $\chi^2$ for the fit, and the
probability of $\chi^2$ given the two degrees of freedom.}
\label{tab:Vub_fits_one_rho}
\begin{minipage}{6.0 in}
\renewcommand{\footnotesep}{7 pt}
\renewcommand{\baselinestretch}{0.7}
\renewcommand{\footnoterule}{\vspace{ -20 pt} }
\begin{tabular}{llcccc}  \hline\hline
$\pi$ Model & $\rho$ Model & $|V_{ub}| \times 10^3$ & $\Gamma^{\text{th}}_\rho /
|V_{ub}|^2$ (ps$^{-1}$)& Fit $\chi^2$ & $P(\chi^2)$ \\ \hline
Ball'01 & Ball'98 &     $2.90 \pm 0.21$ & $16.9 \pm 5.1$ &  7.6 & 0.02 \\
Ball'01 & ISGW2 &       $2.96 \pm 0.21$ & $14.2 \pm 7.1$ &  3.3 & 0.19 \\
Ball'01 & Melikhov'00 & $2.46 \pm 0.17$ & $26.2 \pm 5.2$ &  8.1 & 0.02 \\
Ball'01 & UKQCD'98   &  $2.88 \pm 0.20$ & $16.5^{+3.5}_{-2.3}$ &  5.2 & 0.08 \\
\hline
Ball'01 & Ball'98 &     $2.90 \pm 0.21$ & $16.9 \pm 5.1$ &  7.6 & 0.02 \\
ISGW2 & Ball'98 &       $2.90 \pm 0.21$ & $16.9 \pm 5.1$ &  7.6 & 0.02 \\
SPD & Ball'98 &         $2.90 \pm 0.21$ & $16.9 \pm 5.1$ &  7.8 & 0.02 \\
\hline\hline
\end{tabular}
\end{minipage}
\end{table}

For an estimate of $|V_{ub}|$ based on the models and fits in Table~\ref{tab:Vub_fits_one_rho},
we take the Ball'98 results as the central value.  Estimating the uncertainties
as described in the previous section, we obtain
\begin{equation}
|V_{ub}|=(2.90\pm 0.21\;^{+0.31}_{-0.36} \;^{+0.73}_{-0.46})\times 10^{-3},
\end{equation}
 where the errors are
statistical, experimental systematic, and the estimated uncertainties
from 70\% of the total spread in the results as we vary the $\rho\ell\nu$ form-factor
calculations over $\pm1$ standard deviation, respectively.  This estimate is similar to, though
somewhat larger than, that obtained from the quoted Ball'98 uncertainty.

% Entered 2003.01.19, A. Warburton
\begin{table}[tb]
\caption{Values for $|V_{ub}|$ obtained using form factors (FF) from
light-cone sum rules in the $q^2$ interval $0-16$ GeV$^{2}$ (first
row) and from LQCD for $q^2\ge16$ GeV$^2$ (second row).  Only
the statistical errors are indicated. The data rates obtained using
Ball'01 for $\pi\ell\nu$ and Ball'98 for $\rho\ell\nu$ were used as
the input for all values obtained.}
\label{tab:Vub_fits_two_rho}
\begin{minipage}{6.0 in}
\renewcommand{\footnotesep}{7 pt}
\renewcommand{\baselinestretch}{0.7}
\renewcommand{\footnoterule}{\vspace{ -20 pt} }
\begin{tabular}{lcccc}  \hline\hline
$\rho$ FF & $|V_{ub}| \times 10^3$ & $\Gamma^{\text{th}}_\rho /
|V_{ub}|^2$ (ps$^{-1}$)& Fit $\chi^2$ & $P(\chi^2)$ \\ \hline
Ball'98    & $2.67 \pm 0.27$ & $14.2 \pm 4.3$ & 4.5  & 0.03 \\
\hline
UKQCD'98   & $3.34 \pm 0.32$ & $2.9 ^{+0.62}_{-0.40}$ & --  & -- \\ \hline\hline
\end{tabular}
\end{minipage}
\end{table}

Restricting ourselves to the theoretically more reliable use of LQCD for
$q^2\ge16$ GeV$^2$ and LCSR for $q^2<16$ GeV$^2$, we have only the
two results listed in Table~\ref{tab:Vub_fits_two_rho}.   In addition to the theoretical
uncertainty quoted for UKQCD'98, we add an additional 20\% in quadrature as an
estimate of the quenching uncertainty.  This is  larger than for the $\pi\ell\nu$ case
both because the $\rho$ is a broad resonance and because of the potential for
larger biases from quenching given the interference between the various form
factors.  We also apply our reinterpretation of symmetrical theoretical
errors on the rate as symmetric errors on the amplitude.  The results in the two
$q^2$ intervals are thus
\begin{equation}
|V_{ub}|_{q^2\ge 16\text{ GeV}^2}=(3.34\pm 0.32\;^{+0.27}_{-0.36}\;^{+0.50}_{-0.40})\times 10^{-3},
\end{equation}
and
\begin{equation}
|V_{ub}|_{q^2< 16\text{ GeV}^2}=(2.67\pm 0.27\;^{+0.38}_{-0.42} \;^{+0.47}_{-0.35})\times 10^{-3}.
\end{equation}

We average the LQCD and LCSR results, with 
correlated experimental systematics taken into account.  We again employ the procedure
described in Appendix~\ref{app:AvgProc}.  The optimal weight for combining
the two intervals, treating the systematic uncertainties as completely
correlated, is $\alpha_\rho=0.5$.
\begin{equation}
|V_{ub}|=(3.00\pm 0.21 \;^{+0.29}_{-0.35} \;^{+0.49}_{-0.38}\pm0.28)\times 10^{-3}.
\end{equation}
The errors are statistical, experimental systematic, theoretical systematic based on the
LQCD and LCSR uncertainties, and $\rho\ell\nu$ form-factor shape uncertainty.  To be conservative,
we have assigned the latter error based on the variation seen in the total branching fraction
in this mode.  The contribution from the $\pi\ell\nu$ form-factor shape is negligible.
Again, we take this as our preferred method of extracting $|V_{ub}|$ from our $\rho\ell\nu$ data.

\begin{figure}[t]
\centering
\leavevmode
\epsfysize=3.5in
\epsfbox{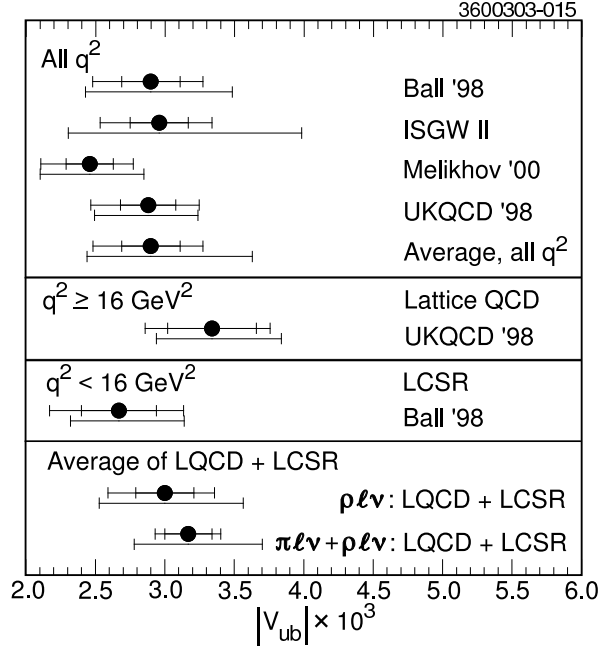}
\caption{Values for $|V_{ub}|$ obtained from $\rho\ell\nu$ using the entire $q^2$ range for the
various form-factor calculations (top block), using LQCD for $q^2\ge16$ GeV$^2$
(second block), using LCSR for $q^2<16$ GeV$^2$ (third block), and
our average of the last two (bottom block) for $\rho\ell\nu$ only and for
$\pi\ell\nu$ and $\rho\ell\nu$ combined.  In all cases, the top bar indicates the
statistical and all the experimental systematics (combined in quadrature), the
lower bar indicated the approximate ``one standard deviation'' range of 
motion due to the theoretical uncertainties.}\label{fig:vub_rho}
\end{figure}

The $|V_{ub}|$ results obtained from $\rho\ell\nu$ are shown in Figure~\ref{fig:vub_rho}.

\subsection{\boldmath $|V_{ub}|$ from a Combination of $B \to \pi \ell \nu$ and $B \to \rho \ell \nu$}

We have averaged the $|V_{ub}|$ determinations obtained separately from the $B \to \pi \ell \nu$ and 
$B \to \rho \ell \nu$ modes.  For this average, we 
considered only the results obtained using the LCSR and LQCD calculations
applied to the $q^2<16\ \textrm{GeV}^2$ and $q^2\ge 16\ \textrm{GeV}^2$ results, respectively.
The averaging procedure amounts to the determination of 
the optimal weight  $\beta$ to be applied to the LCSR 
and LQCD average obtained from $B \to \pi \ell \nu$  relative to that obtained 
from $B \to \rho \ell \nu$ (see Appendix~\ref{app:AvgProc}).  
We held the values $\alpha_\pi$ and $\alpha_\rho$, each of which determines 
the weight of the LQCD result relative to the LCSR result in the individual mode, fixed at
the optimal values found in the preceding subsections.
The weight $\beta=0.7$ provided the optimal combination.  With this weighting, we find
\begin{equation}
|V_{ub}|=(3.17\pm 0.17 \;^{+0.16}_{-0.17} \;^{+0.53}_{-0.39}\pm0.03)\times 10^{-3}.
\end{equation}
The errors are statistical, experimental systematic, theoretical systematic based on the
LQCD and LCSR uncertainties, and $\rho\ell\nu$ form-factor shape uncertainty, respectively. 
Note that because of cross feed among the modes considered, the $\pi\ell\nu$
and $\rho\ell\nu$ modes are anticorrelated, resulting, in particular, in the minimal
dependence of the average result on the $\rho\ell\nu$ form-factor shape.

\section{Summary}

With a sample of $9.7\times 10^6$ $B\bar{B}$ pairs,
we have studied $B$ decays to $\pi\ell\nu$, $\rho\ell\nu$,
$\omega\ell\nu$, and $\eta\ell\nu$, where $\ell=e\text{ or } \mu$.  
From the combination of a broad momentum range for the 
charged lepton momentum and independent extraction
of rates in three separate $q^2$ intervals, we were able to
reduce the uncertainties from modeling within the form-factor calculations.
For the decay $B^0\to\pi^-\ell^+\nu$, we have determined the branching fractions
\begin{eqnarray}
{\cal B}(0\le q^2< 8 \,\text{GeV}^2)& = &
(0.43\pm 0.11 \pm0.05 \pm 0.004\pm 0.01)\times 10^{-4} \\ \nonumber
{\cal B}(8\le q^2< 16 \,\text{GeV}^2) & = & 
(0.65\pm 0.11 \pm0.07\pm 0.01\pm 0.03 )\times 10^{-4} \\ \nonumber
{\cal B}(q^2\ge 16 \,\text{GeV}^2) & = & 
(0.25\pm 0.09 \pm0.04 \pm 0.01\pm 0.03)\times 10^{-4}. 
\end{eqnarray}
Combining these rates and taking into account correlated systematic uncertainties, we obtain
\begin{equation}
{\cal B}(B^0\to\pi^-\ell^+\nu)=(1.33\pm 0.18 \pm0.11 \pm 0.01\pm 0.07)\times 10^{-4},
\end{equation}
where the errors are
statistical, experimental systematic, the estimated uncertainties
from the $\pi\ell\nu$ form factor, and those from the $\rho\ell\nu$ form factors, respectively.   

For the decay $B^0\to\rho^-\ell^+\nu$, we have determined the branching fractions
\begin{eqnarray}
{\cal B}(0\le q^2< 8 \,\text{GeV}^2)& = &
(0.43\pm 0.20\pm0.23\pm0.09\pm0.01)\times 10^{-4} \\ \nonumber
{\cal B}(8\le q^2< 16 \,\text{GeV}^2) & = & 
(1.24\pm 0.26\;^{+0.27}_{-0.33}\pm 0.22\pm 0.004 )\times 10^{-4} \\ \nonumber
{\cal B}(q^2\ge 16 \,\text{GeV}^2) & = & 
(0.50\pm 0.10\;^{+0.08}_{-0.11}\pm 0.19\pm 0.004)\times 10^{-4}. 
\end{eqnarray}
Combining these rates, again taking into account correlated systematic uncertainties, we obtain
\begin{equation}
{\cal B}(B^0\to\rho^-\ell^+\nu)=(2.17\pm 0.34\;^{+0.47}_{-0.54} \pm 0.41\pm 0.01)\times 10^{-4},
\end{equation}
where the errors are
statistical, experimental systematic, the estimated uncertainties
from the $\rho\ell\nu$ form factors, and those from the $\pi\ell\nu$ form factor, respectively.

When the theoretical uncertainties that result from form-factor $q^2$--dependence are
evaluated in a common fashion, the branching fractions obtained in this analysis
have uncertainties from the form-factor $q^2$--dependence that are reduced by 
about a factor of two compared to previous $\rho\ell\nu$ analyses
\cite{bb:lkg_cleo_exclusive,bb:lange_cleo_exclusive,bb:newBabar}.  These uncertainties
are almost eliminated for the $\pi\ell\nu$ branching fraction.

We see evidence for the decay $B^+\to\eta\ell^+\nu$ with a statistical significance
corresponding roughly to $3.2\sigma$.  The rate we obtain,
\begin{equation}
{\cal B}(B^+\to\eta\ell^+\nu)=(0.84 \pm 0.31 \pm 0.16 \pm0.09)\times 10^{-4},
\end{equation}
is consistent, within sizable errors, with that expected
from the measured pion rate and isospin relations.  Only an ISGW~II form factor
has been examined, and a 10\% model dependence uncertainty has been assigned
based on the previous CLEO $\pi\ell\nu$ analysis. The final error quoted combines
this estimate with the dependence on the $\pi\ell\nu$ and $\rho\ell\nu$ form factors.

From the $\pi\ell\nu$ $q^2$ behavior that we have observed, we find the ISGW~II
form factor for $\pi\ell\nu$ consistent with data at only the 3\% level.  

By fitting LQCD and LCSR calculations to the observed $q^2$ behavior in $\pi\ell\nu$, 
restricting each calculation to its valid $q^2$ range, and then combining the results,
we extract
\begin{equation}
|V_{ub}|=(3.24\pm 0.22 \pm 0.13 \;^{+0.55}_{-0.39}\pm 0.09)\times 10^{-3},
\end{equation}
 where the errors are statistical, experimental systematic, the estimated uncertainties
from the $\pi\ell\nu$ form-factor shape and normalization, and those from the $\rho\ell\nu$
form factors' shapes, respectively.
From a similar analysis of the $\rho\ell\nu$ mode, we obtain
\begin{equation}
|V_{ub}|=(3.00\pm 0.21 \;^{+0.29}_{-0.35} \;^{+0.49}_{-0.38}\pm0.28)\times 10^{-3}.
\end{equation}
The errors are statistical, experimental systematic, theoretical systematic based on the
LQCD and LCSR uncertainties, and $\rho\ell\nu$ form-factor shape uncertainty, respectively.
In general, the $\rho\ell\nu$ form-factor calculations did not agree as well with the
observed $\rho\ell\nu$ data as did the $\pi\ell\nu$ form-factor calculations 
with the $\pi\ell\nu$ data.

Combining these two modes for an overall result from this analysis, we obtain 
\begin{equation}
|V_{ub}|=(3.17\pm 0.17 \;^{+0.16}_{-0.17} \;^{+0.53}_{-0.39}\pm0.03)\times 10^{-3}.
\end{equation}
Given the manner with which  the theoretical uncertainties have been estimated,
the quoted values should be interpreted as being closer in spirit to ``one standard deviation''
than to ``the allowed range''.

These results trade off the potential statistical gain over the previous CLEO
analyses in favor of relaxation of theoretical constraints.  Had we fixed the
relative rate in the three $q^2$ intervals in the $\pi\ell\nu$ and $\rho\ell\nu$
modes, a more pronounced improvement in statistical precision would have
resulted.   By relaxing the constraint, on the other hand, we have minimized
our reliance on modeling in extraction of rates and of $|V_{ub}|$.  

These results supersede the $\pi\ell\nu$ and $\rho\ell\nu$ results obtained in reference 
\cite{bb:lkg_cleo_exclusive}. They agree, within measurement uncertainties, with the CLEO 2000 
$\rho\ell\nu$ result \cite{bb:lange_cleo_exclusive} and with the recent BABAR 
$\rho\ell\nu$ analysis \cite{bb:newBabar}.

The results for $|V_{ub}|$ obtained here are compatible with the results obtained
from the recent CLEO end-point measurement \cite{bb:new_endpoint}. 
The estimated theoretical uncertainties remain sizable for both $\pi\ell\nu$
and $\rho\ell\nu$, and there remain uncertainties in the estimates themselves.
We therefore do not average
these results, but view the compatibility as an indication that the uncertainties have
not been appreciably underestimated.  Significant progress in extraction of $|V_{ub}|$
from exclusive decays will require a major improvement in theory.

We thank A. Kronfeld, J. Simone,
T. Onogi,
T. Feldmann, P. Kroll,
C. Maynard, and
D. Melikhov
for assistance with form factors.
We gratefully acknowledge the effort of the CESR staff 
in providing us with
excellent luminosity and running conditions.
M. Selen thanks the Research Corporation, 
and A.H. Mahmood thanks the Texas Advanced Research Program.
This work was supported by the 
National Science Foundation 
and the
U.S. Department of Energy.

\appendix

\section{Description of experimental systematic uncertainty determination}
\label{app:systematics}
The techniques employed in this analysis rest fundamentally
on complete, accurate reconstruction of all particles from both $B$ decays
in an event.  As a result,
systematic uncertainty estimates that reflect uncertainties in the detector
simulation must account for the reliability with which an entire event can be
reconstructed, not just the signal particles.  For example, if there is a residual
uncertainty in the track reconstruction efficiency,  the signal efficiency will 
not only be affected by incorrectly assessing the loss of the signal mode
particles, it will also be affected by ``misreconstruction'' of the neutrino four--momentum.
Furthermore, the rate at which background samples can smear into the
signal region is also affected by the overall misreconstruction.   

We therefore estimate the systematic uncertainties due to detector modeling
by modifying each reconstructed Monte Carlo event in
each signal and background sample.  For each study, the size of the
variation has generally been determined by independent comparisons of data
and Monte Carlo. The following list describes the
variations that enter the systematic determination:
\begin{description}
  \item[tracking efficiency]  We have limited our uncertainty in track--finding efficiency for high (above 250 MeV/$c$) and
        low momentum tracks to be under 0.5\% and 2.6\%, respectively.   These limits were obtained with hadronic samples, and
        therefore include any discrepancies in the interaction cross sections.    To determine the systematic error from
        the uncertainty in tracking efficiency, we apply an additional inefficiency of 0.75\% and
       2.6\% to each high momentum track and to each low momentum track, respectively, in the simulation.
  \item[tracking resolution] We increase the mismeasurement of each momentum component for each reconstructed charged particle by 10\% of itself, which is
        outside the range for which core distributions agree, but compensates for discrepancies in the tails.
  \item[$\gamma$ efficiency]  We have limited our uncertainty in photon reconstruction efficiency to 2\%.  In our studies, we have
       actually applied an additional 3\% efficiency loss per photon, then scaled the observed shifts back by $2/3$.
  \item[$\gamma$ resolution]  We also degrade the photon energy resolution by 10\% of itself.
 \item[split-off simulation] Studies of $\gamma\gamma\to K_S K_S$ have indicated that the combination of  mismodeling the physics
      processes and hadronic showers leads to an excess of isolated reconstructed showers (split offs) at the rate of $0.03$/hadron in data
      relative to the Monte Carlo.  To estimate the potential effect on our analysis, we interpret the entire excess as mismodeling of
      the hadronic showers, and add showers at this rate to each of our Monte Carlo samples.
 \item[split-off rejection]  We bias  our neural net parameter, which is derived from
    the distribution of energy within the crystals in the shower relative to the primary impact point of a ``parent'' charged hadron, to
    move photon--like results in the Monte Carlo towards hadronic--shower--like results.  We limit the variations based on
    data and Monte Carlo comparisons of the parameter as a function of shower energy.
 \item[$K_L$ showers] In our simulation of $K_L$ showers, we increase the energy deposited in our CsI calorimeter.  The
    variation is based on data and Monte Carlo comparisons of the energy deposited by $K^\pm$ showers after correction for
    the minimum--ionizing component.
 \item[$K_L$ production] By comparing the data and Monte Carlo $K_S$ energy spectrum and yield, we found that our $K_L$ rate needed
    to be decreased by $(7.2\pm 1.0)$\%, and that no correction was needed for the spectrum.  The nominal analysis reweights events
    with $K_L$ accordingly, and we vary the weight according to its uncertainty to estimate the systematic contribution.
 \item[extra $\nu$ production]  An important source of background is events that contain both a $b\to c\ell\nu$ decay and a $c\to s\ell\nu$
   decay, where the latter can originate with either $B$ meson in the event.   We reweight the Monte Carlo so that the lepton momentum
   spectrum from secondary charm decay agrees with a spectrum obtained by convoluting a recent measurement of the charm meson
   momentum spectrum from $B$ decay \cite{bb:moneti}
   % PRD 56, 3783 (1997)
   with the MARK~III measurement of the inclusive lepton momentum spectrum from charm decay \cite{bb:Delco}. 
   %PRL 54, 1976 (1985))
   The nominal result is corrected based on this procedure.  To estimate the systematic uncertainty, we define spectrum ``envelopes'' 
   and reweight our Monte Carlo samples to match this spectrum.  The envelopes were defined by throwing
    500 toy Monte Carlo spectra in which all experimental inputs were varied according to their uncertainties and
    finding the variation within each momentum bin that contained 68\% of the toy spectra.
 \item[particle ID]  We simultaneously shift all dE/dx and time-of-flight distributions in the simulation by 1/4 and 1/2 of the intrinsic
   resolution, respectively.  We take the full variation we observe as our uncertainty, even though this procedure leads to a very 
   conservative systematic estimate.
\end{description}

For each of these variations, we modify or reweight each event in each Monte Carlo sample
in a full reanalysis of these samples.  The set of modified samples for each variation replaces
the nominal samples input to the branching fraction fit.  For each variation, the shifts in the
fit results provide the first input into the systematic estimates on the branching fractions
 for that variation.   We can view the shifts in results as arising from two components: a
change in the signal efficiency and a change in the predicted background level.  These
changes tend to cancel in the total shift: a variation that reduces the signal reconstruction
efficiency also simultaneously increases the background level (and reduces the signal
yield from the fit). As the main text describes, we increase our systematic estimate to allow for 
imperfections in the predicted cancellation.

\section{\boldmath Extraction of $|V_{ub}|$ from the measured $d\Gamma(B^0\to\pi^-\ell^+\nu)/dq^2$ data with future form-factor
calculations}
\label{app:method}

The branching fractions in the three $q^2$ ranges for $B\to\pi\ell\nu$ exhibit
very little dependence on the precise form factors used to extract the branching
fractions.  The results can therefore be reliably used to obtain values for
$|V_{ub}|$ using future $B\to\pi\ell\nu$ form-factor calculations that are improved
over those used in this paper.   This appendix provides the detail needed to
ascertain the proper experimental uncertainties for such an extraction using
the same fitting technique presented above.  The main difficulty stems from
proper evaluation of the experimental uncertainties because of correlations
(both positive and negative) among the results for the three ranges.  The 
correlations arise both statistically  from the fitting procedure used to extract
the three rates and systematically as we vary the details of the simulation.

\begin{table}[tb]
  \centering 
  \caption{Central values and statistical uncertainties for $B^0\to\pi^-\ell^+\nu$ branching fractions for the nominal fit and for each systematic variation of the Monte Carlo samples 
 input to the fit.  The detector--related systematic uncertainties in $|V_{ub}|$ are obtained by
 fitting the results from the relevant set of $q^2$ intervals for each systematic study.  The total branching
 fraction is shown as well for completeness.
 All results were obtained using the Ball'01 form factor for the $\pi\ell\nu$
 modes and the Ball'98 form factors for the $\rho\ell\nu$ modes.}\label{tab:BallKnobs}
\begin{tabular}{ccccc}
\hline\hline
systematic     & \multicolumn{4}{c}{$10^{4}\times \mathcal{B}(B^0\to\pi^-\ell^+\nu)$}   \\
change          & total &   $0\le q^2< 8 \,\text{GeV}^2$ & $8\le q^2< 16 \,\text{GeV}^2$ & $16 \,\text{GeV}^2 \le q^2 < q^2_{\mathrm{max}} $
                                                                                                                              \\ \hline
nominal            & $1.327\pm 0.177$ & $0.431\pm 0.106$ & $0.651\pm 0.105$ & $0.245\pm 0.094$ \\ \hline
$\gamma$ eff.      & $1.348\pm 0.194$ & $0.476\pm 0.117$ & $0.674\pm 0.117$ & $0.198\pm 0.103$ \\
$\gamma$ resol.    & $1.379\pm 0.183$ & $0.445\pm 0.111$ & $0.686\pm 0.109$ & $0.249\pm 0.096$ \\
$K_L$ shower       & $1.311\pm 0.173$ & $0.426\pm 0.104$ & $0.642\pm 0.104$ & $0.242\pm 0.091$ \\
particle ID        & $1.342\pm 0.180$ & $0.414\pm 0.108$ & $0.668\pm 0.107$ & $0.260\pm 0.096$ \\
split-off rejection & $1.338\pm 0.179$ & $0.415\pm 0.108$ & $0.667\pm 0.107$ & $0.255\pm 0.095$ \\
track eff.         & $1.357\pm 0.185$ & $0.446\pm 0.112$ & $0.669\pm 0.110$ & $0.242\pm 0.097$ \\
track resol.       & $1.317\pm 0.179$ & $0.438\pm 0.108$ & $0.664\pm 0.108$ & $0.215\pm 0.094$ \\
split-off sim.      & $1.326\pm 0.178$ & $0.432\pm 0.108$ & $0.655\pm 0.106$ & $0.240\pm 0.093$ \\
$K_L$ production  $\uparrow$ & $1.325\pm 0.176$ & $0.431\pm 0.106$ & $0.651\pm 0.105$ & $0.244\pm 0.094$ \\
$K_L$ production   $\downarrow$& $1.330\pm 0.177$ & $0.432\pm 0.107$ & $0.653\pm 0.105$ & $0.246\pm 0.094$ \\
$\nu$ production  $\uparrow$ & $1.344\pm 0.178$ & $0.425\pm 0.106$ & $0.669\pm 0.106$ & $0.251\pm 0.095$ \\
$\nu$ production   $\downarrow$ & $1.322\pm 0.175$ & $0.439\pm 0.106$ & $0.641\pm 0.104$ & $0.242\pm 0.093$ \\
\hline\hline
\end{tabular}
\end{table}

To extract a central value of $|V_{ub}|$, we perform a $\chi^2$ fit to the nominal branching fractions from the
three $q^2$ intervals listed in Table~\ref{tab:BallKnobs}.  This $|V_{ub}|$ fit includes the correlation coefficients among
the rates from the branching fraction fit to the data: $\rho_{12}=-0.035$, $\rho_{13}=0.003$, and
$\rho_{23}= -0.037$.

To evaluate the error arising from simulation uncertainties (``$\nu$ simulation'' in Table~\ref{tab:systematics})
on the results, we redo our $\chi^2$ fit for $|V_{ub}|$ using the new rates listed in Table~\ref{tab:BallKnobs}
for each variation.  For the results presented here, we have used the correlation coefficients from the branching
fraction fit to the data for each variation.  In practice, the coefficients remain stable enough that using
the nominal coefficients in all fits is sufficient.  The change relative to the nominal $|V_{ub}|$ result provides the first input to
the uncertainty estimate.  For the uncertainty estimate in $K_L$ production and secondary $\nu$ production,
we take the average of the ``up'' and ``down'' shifts as our overall estimate.
To allow for misestimation of correlated changes between background levels and signal efficiencies
in the results (see main text), we increase the fractional uncertainty on $|V_{ub}|$ from each variation
by adding in quadrature the quantities listed in Table~\ref{tab:correls}.   Finally, the $\gamma$ efficiency
uncertainty should be scaled back to $2/3$ of the value found above.  We combine all of the
uncertainties in quadrature to arrive at the total ``$\nu$ simulation'' systematic for $|V_{ub}|$.

\begin{table}[tb]
  \centering 
  \caption{Fractional uncertainties to be added in quadrature to systematic shifts in $|V_{ub}|$
           to account for uncertainty in cancellations arising from correlated efficiency and
           background changes.  The correction is shown for the various different $q^2$ ranges
           used in this analysis.}\label{tab:correls}
\begin{tabular}{cccccc}
\hline\hline
systematic     & \multicolumn{5}{c}{additional systematic (\%)}  \\
change          & full range &   $0\le q^2< 16 \,\text{GeV}^2$ &   $0\le q^2< 8 \,\text{GeV}^2$ & $8\le q^2< 16 \,\text{GeV}^2$ & $16 \,\text{GeV}^2 \le q^2 < q^2_{\mathrm{max}} $ \\ \hline
$\gamma$ eff.      & 1.67 & 0.51 & 0.72 & 1.22 & 1.49 \\
$\gamma$ resol.    & 0.19 & 0.28 & 0.14 & 0.43 & 0.30 \\
$K_L$ shower       & 0.25 & 0.30 & 0.14 & 0.16 & 0.46 \\
particle ID        & 0.25 & 1.09 & 0.29 & 0.27 & 0.58 \\
split-off rejection & 0.00 & 0.56 & 0.24 & 0.21 & 0.35 \\
track eff.         & 0.99 & 1.62 & 0.72 & 0.90 & 1.17 \\
track resol.       & 0.49 & 0.25 & 0.14 & 0.11 & 0.44 \\
split-off sim.      & 0.23 & 0.39 & 0.24 & 0.11 & 0.17 \\
$K_L$ production   & 0.01 & 0.02 & 0.01 & 0.00 & 0.01 \\
$\nu$ production   & 0.12 & 0.43 & 0.28 & 0.19 & 0.13 \\
\hline\hline
\end{tabular}
\end{table}

We evaluate the uncertainty from our modeling of the $B\to X_u\ell\nu$ backgrounds
in much the same fashion.  The fit variations that we have used for this purpose
are listed in Table~\ref{tab:buotherfits}.  An earlier version of our $B\to X_u\ell\nu$ 
generator was used in the study, and the table also shows the ``nominal'' result
obtained with that version.  We did not expect large differences from our change,
and indeed the results obtained are very similar to the nominal results in
Table~\ref{tab:BallKnobs}.  To obtain the uncertainty estimate resulting from
the hadronization model, we compare the results using purely nonresonant hadronization
to that using our nominal mixture of resonant and nonresonant modes.  To obtain
the uncertainty resulting from our choice of parameters for the OPE-based
inclusive differential rate calculation, we take the average of the shift from
the last two lines in the table relative to the above nonresonant result.  Note
that these variations do not affect our signal Monte Carlo samples.

\begin{table}[tb]
  \centering 
  \caption{Central values and statistical uncertainties for $B^0\to\pi^-\ell^+\nu$ branching fractions for the reference fit 
  and for each systematic variation of the $B\to X_u\ell\nu$ background simulation
 input to the fit.  The associated systematic uncertainties in $|V_{ub}|$ are obtained by
 fitting the results from the relevant set of $q^2$ intervals for each systematic study.  The total branching
 fraction is shown as well for completeness.
 All results were obtained using the Ball'01 form factor for the $\pi\ell\nu$
 modes and the Ball'98 form factors for the $\rho\ell\nu$ modes.}\label{tab:buotherfits}
\begin{tabular}{cccccc}
\hline\hline
OPE                   & hadron- & \multicolumn{4}{c}{$10^{4}\times \mathcal{B}(B^0\to\pi^-\ell^+\nu)$}  \\
parameters       &ization               & total &   $0\le q^2< 8 \,\text{GeV}^2$ & $8\le q^2< 16 \,\text{GeV}^2$ & $16 \,\text{GeV}^2 \le q^2 < q^2_{\mathrm{max}} $
                                                                                                                              \\ \hline
nominal             & nominal     & $1.324\pm 0.177$ & $0.423\pm 0.107$ & $0.655\pm 0.105$ & $0.246\pm 0.094$ \\
nominal             & nonres. & $1.322\pm 0.177$ & $0.431\pm 0.106$ & $0.639\pm 0.105$ & $0.251\pm 0.094$ \\
``High'' & nonres. & $1.311\pm 0.176$ & $0.428\pm 0.106$ & $0.637\pm 0.105$ & $0.246\pm 0.094$ \\
``Low''  & nonres. & $1.329\pm 0.177$ & $0.434\pm 0.106$ & $0.646\pm 0.105$ & $0.248\pm 0.095$ \\
\hline\hline
\end{tabular}
\end{table}

For the remainder of the systematic uncertainties, we take one half of the fractional
uncertainties  listed in Table~\ref{tab:systematics}.  The factor of one half arises
because of the square root involved in extraction of $|V_{ub}|$ from the rates.

\section{\boldmath Averaging $|V_{ub}|$ results}
\label{app:AvgProc}

In each of the $\pi\ell\nu$ and $\rho\ell\nu$ modes, we have extracted two
results for $|V_{ub}|$ that are largely free from modeling assumptions:
a value based on the application of LCSR--derived form factors for
$q^2 < 16\ \textrm{GeV}^2$, and a value based on the application of
LQCD--derived form factors for $q^2\ge 16\ \textrm{GeV}^2$.   We therefore
have three averages to be calculated: the combination of the two results
within the $\pi\ell\nu$ mode and within the $\rho\ell\nu$ mode, and
the combination of the two modes.   The averaging procedure should
take into account, in particular, the correlations present in the systematic
uncertainties in the result.   This appendix describes our averaging
procedure.

The statistical correlations have been taken
into account in the LCSR--derived results.  An evaluation of
remaining statistical correlations found that they had little impact on 
the final statistical error, and we have not included them in the final
procedure.  Proper treatment would have led to a decrease in
the overall uncertainty that would be hidden at the quoted
precision.

Regarding theoretical uncertainties, while the two techniques have different
systematic effects, both approaches currently have systematic issues that
are difficult to evaluate.  For example, there is a quark--hadron duality assumption in the
LCSR approach, and the current LQCD calculations have been evaluated
in the ``quenched'' approximation.   Treating the uncertainties as uncorrelated
would therefore be likely to underestimate the ``true'' theoretical uncertainty.
To be conservative, we treat the theoretical uncertainties as if they were fully correlated.

Let us first consider the two results obtained within a given mode. We
wish to combine the results with a weight that minimizes the overall
uncertainty and preserves the systematic correlation information.   
Defining the weight of the LQCD--derived result (denoted $|V_{ub}|^{\ge 16}$)
by $\alpha$, the LCSR--derived result (denoted $|V_{ub}|^{<16}$) enters
with a weight $1-\alpha$:
\begin{equation}
|V_{ub}|_\alpha = \alpha |V_{ub}|^{\ge 16} + (1-\alpha)|V_{ub}|^{<16}.
\label{eq:vub_alpha}
\end{equation}
The statistical uncertainties are uncorrelated, and are combined as
\begin{equation}
\sigma_{\mathrm{stat}}^2 = (\alpha\sigma_{\mathrm{stat}}^{\ge 16})^2 + ((1-\alpha)\sigma_{\mathrm{stat}}^{<16})^2.
\end{equation}
Correlated uncertainties, such as the theoretical uncertainties, are combined as
\begin{equation}
\sigma_{\mathrm{corr}} = (\alpha\sigma_{\mathrm{corr}}^{\ge 16}) + ((1-\alpha)\sigma_{\mathrm{corr}}^{<16}).
\end{equation}

For  each simulation variation (labelled $i$),  we perform the full analysis to obtain
$|V_{ub}|^{\ge 16}_i$) and $|V_{ub}|^{<16}_i$.  The systematic uncertainty defined
for the variation is
\begin{equation}
\sigma_i = |V_{ub}|_\alpha^{\mathrm{nom}} - (\alpha |V_{ub}|^{\ge 16}_i + (1-\alpha)|V_{ub}|^{<16}_i),
\end{equation}
where $|V_{ub}|_\alpha^{\mathrm{nom}}$ is the average resulting from Equation~\ref{eq:vub_alpha}.
This procedure preserves the systematic correlation.  We combine this estimate in quadrature
with the additional uncertainty contribution to allow for imperfect modeling of the correlated changes 
between signal efficiency and raw yield (see Section~\ref{sec:systematics}).

Finally, for each value of $\alpha$ the experimental and theoretical uncertainties are combined in
quadrature (taking the average theoretical uncertainty in the case of asymmetric uncertainties).  
We scan over $\alpha$ and choose the value that minimizes the total uncertainty.  

We perform a similar procedure to combine the results from the two modes.  The weights
obtained individually for the different $q^2$ regions in each mode are fixed.  The uncorrelated,
correlated, and anticorrelated uncertainties are combined in exact analogy to the above descriptions.
Taking $\beta$ as the weight of the $\pi\ell\nu$ mode in the average, we have
\begin{equation}
|V_{ub}|_\beta = \beta |V_{ub}|^{\pi} + (1-\beta)|V_{ub}|^{\rho}.
\end{equation}
For each simulation variation, the systematic estimate becomes
\begin{equation}
\sigma_i = |V_{ub}|_\beta^{\mathrm{nom}} - \left[ \beta (\alpha_\pi |V_{ub}|^{\ge 16,\pi}_i + (1-\alpha_\pi)|V_{ub}|^{<16,\pi}_i)
+ (1-\beta)(\alpha_\rho |V_{ub}|^{\ge 16,\rho}_i + (1-\alpha_\rho)|V_{ub}|^{<16,\rho}_i)\right].
\end{equation}
These uncertainties are, as before, combined in quadrature, along with the contribution for imperfect
modeling of the correlated efficiency and yield changes.

We scan over the weight $\beta$ to find the value that minimizes the overall combined uncertainty.
Once again we treat the theoretical uncertainties in the $\pi\ell\nu$ and $\rho\ell\nu$ form factors
as correlated in this procedure.

\nopagebreak


\begin{thebibliography}{99}
\bibitem{bb:CKM} {N. Cabibbo, Phys. Rev. Lett. {\bf 10}, 531 (1963);
    M. Kobayashi and T. Maskawa, Prog. Theor. Phys. {\bf 49}, 652 
(1973).}
\bibitem{bb:PDG_vub_minireview}
   M. Battaglia and L. Gibbons, {\em Determination of $|V_{ub}|$},  in
   {\em Review of Particle Properties}, K.~Hagiwara {\it et al.}, Phys. 
Rev. D {\bf 66}, 010001 (2002).
\bibitem{bb:lkg_cleo_exclusive}
   J.~P. Alexander  {\it et al.} [CLEO Collaboration],
   Phys. Rev. Lett.  {\bf 77}, 5000 (1996).
\bibitem{bb:lange_cleo_exclusive}
   B. H. Behrens  {\it et al.} [CLEO Collaboration],
   Phys. Rev. D {\bf 61}, 052001 (2000)
   [arXiv:hep-ex/9905056].
\bibitem{bb:vb_thesis} V.~Boisvert, Ph.~D. Thesis, Cornell University, 
2002 (unpublished).

%
% Lattice QCD
%

%\cite{Abada:1993dh}
\bibitem{Abada:1993dh}
A.~Abada {\it et al.},
%``Semileptonic decays of heavy flavors on a fine grained lattice,''
Nucl.\ Phys.\ B {\bf 416}, 675 (1994)
[arXiv:hep-lat/9308007].
%%CITATION = HEP-LAT 9308007;%%


%\cite{Allton:1994ui}
\bibitem{Allton:1994ui}
C.~R.~Allton {\it et al.}  [APE Collaboration],
%``Lattice calculation of D and B meson semileptonic decays using the 
%Clover action at beta = 6.0 on APE,''
Phys.\ Lett.\ B {\bf 345}, 513 (1995)
[arXiv:hep-lat/9411011].
%%CITATION = HEP-LAT 9411011;%%


%\cite{DelDebbio:1997kr}
\bibitem{DelDebbio:1997kr}
L.~Del Debbio, J.~M.~Flynn, L.~Lellouch, and J.~Nieves  [UKQCD 
Collaboration],
%``Lattice-constrained parametrizations of form factors for 
%semileptonic  and rare radiative B decays,''
Phys.\ Lett.\ B {\bf 416}, 392 (1998)
[arXiv:hep-lat/9708008].
%%CITATION = HEP-LAT 9708008;%%


%\cite{Hashimoto:1997sr}
\bibitem{Hashimoto:1997sr}
S.~Hashimoto, K.~I.~Ishikawa, H.~Matsufuru, T.~Onogi, and N.~Yamada,
%``Lattice study of B $\to$ pi semileptonic decay using nonrelativistic 
%  lattice QCD,''
Phys.\ Rev.\ D {\bf 58}, 014502 (1998)
[arXiv:hep-lat/9711031].
%%CITATION = HEP-LAT 9711031;%%


%\cite{Ryan:1998tj}
\bibitem{Ryan:1998tj}
S.~Ryan, A.~El-Khadra, S.~Hashimoto, A.~Kronfeld, P.~B.~Mackenzie, and
J.~Simone,
%``Anti-B $\to$ pi l anti-nu at three lattice spacings,''
Nucl.\ Phys.\ Proc.\ Suppl.\  {\bf 73}, 390 (1999)
[arXiv:hep-lat/9810041].
%%CITATION = HEP-LAT 9810041;%%


%\cite{Ryan:1999kx}
\bibitem{Ryan:1999kx}
S.~M.~Ryan, A.~X.~El-Khadra, A.~S.~Kronfeld, P.~B.~Mackenzie, and
J.~N.~Simone,
%``B and D semileptonic decays to light mesons,''
Nucl.\ Phys.\ Proc.\ Suppl.\  {\bf 83}, 328 (2000)
[arXiv:hep-lat/9910010].
%%CITATION = HEP-LAT 9910010;%%


%\cite{Lellouch:1999dz}
\bibitem{Lellouch:1999dz}
L.~Lellouch,
%``Exclusive semileptonic B decays: Lattice results and dispersive 
%bounds,''
[arXiv:hep-ph/9912353].
%%CITATION = HEP-PH 9912353;%%


%\cite{Bowler:1999xn}
\bibitem{Bowler:1999xn}
K.~C.~Bowler {\it et al.}  [UKQCD Collaboration],
%``Improved B $\to$ pi l nu/l form factors from the lattice,''
Phys.\ Lett.\ B {\bf 486}, 111 (2000)
[arXiv:hep-lat/9911011].
%%CITATION = HEP-LAT 9911011;%%


%\cite{Becirevic:1999kt}
% BK Parameterization (not a lattice calculation per se)
\bibitem{Becirevic:1999kt}
D.~Becirevic and A.~B.~Kaidalov,
%``Comment on the heavy $\to$ light form factors,''
Phys.\ Lett.\ B {\bf 478}, 417 (2000)
[arXiv:hep-ph/9904490].
%%CITATION = HEP-PH 9904490;%%


%\cite{Aoki:2000by}
\bibitem{Aoki:2000by}
S.~Aoki {\it et al.}  [JLQCD Collaboration],
%``Differential decay rate for B $\to$ pi l nu semileptonic decays,''
Nucl.\ Phys.\ Proc.\ Suppl.\  {\bf 94}, 329 (2001)
[arXiv:hep-lat/0011008].
%%CITATION = HEP-LAT 0011008;%%


%\cite{El-Khadra:2001rv}
\bibitem{El-Khadra:2001rv}
A.~X.~El-Khadra, A.~S.~Kronfeld, P.~B.~Mackenzie, S.~M.~Ryan, and
J.~N.~Simone,
%``The semileptonic decays B $\to$ pi l nu and D $\to$ pi l nu from 
%lattice  QCD,''
Phys.\ Rev.\ D {\bf 64}, 014502 (2001)
[arXiv:hep-ph/0101023].
%%CITATION = HEP-PH 0101023;%%


%\cite{Aoki:2001rd}
\bibitem{Aoki:2001rd}
S.~Aoki {\it et al.}  [JLQCD Collaboration],
%``Differential decay rate of B $\to$ pi l nu semileptonic decay with  
%lattice NRQCD,''
Phys.\ Rev.\ D {\bf 64}, 114505 (2001)
[arXiv:hep-lat/0106024].
%%CITATION = HEP-LAT 0106024;%%


%\cite{Abada:2000ty}
\bibitem{Abada:2000ty}
A.~Abada, D.~Becirevic, P.~Boucaud, J.~P.~Leroy, V.~Lubicz, and
F.~Mescia,
%``Heavy $\to$ light semileptonic decays of pseudoscalar mesons from 
%lattice  QCD,''
Nucl.\ Phys.\ B {\bf 619}, 565 (2001)
[arXiv:hep-lat/0011065].
%%CITATION = HEP-LAT 0011065;%%


%
% QCD Sum Rules
%

%\cite{Ball:1997rj}
\bibitem{Ball:1997rj}
P.~Ball and V.~M.~Braun,
%``Use and misuse of QCD sum rules in heavy-to-light transitions: The 
%decay B $\to$ rho e nu reexamined,''
Phys.\ Rev.\ D {\bf 55}, 5561 (1997)
[arXiv:hep-ph/9701238].
%%CITATION = HEP-PH 9701238;%%


%\cite{Ball:1998kk}
\bibitem{Ball:1998kk}
P.~Ball and V.~M.~Braun,
%``Exclusive semileptonic and rare B meson decays in {QCD},''
Phys.\ Rev.\ D {\bf 58}, 094016 (1998)
[arXiv:hep-ph/9805422].
%%CITATION = HEP-PH 9805422;%%


%\cite{Khodjamirian:1997ub}
\bibitem{Khodjamirian:1997ub}
A.~Khodjamirian, R.~R\"{u}ckl, S.~Weinzierl, and O.~I.~Yakovlev,
%``Perturbative QCD correction to the B $\to$ pi transition form 
%factor,''
Phys.\ Lett.\ B {\bf 410}, 275 (1997)
[arXiv:hep-ph/9706303].
%%CITATION = HEP-PH 9706303;%%


%\cite{Khodjamirian:2000ds}
\bibitem{Khodjamirian:2000ds}
A.~Khodjamirian, R.~R\"{u}ckl, S.~Weinzierl, C.~W.~Winhart, and
O.~I.~Yakovlev,
%``Predictions on B $\to$ pi anti-l nu/l, D $\to$ pi anti-l nu/l and 
%D$\to$ K  anti-l nu/l from QCD light-cone sum rules,''
Phys.\ Rev.\ D {\bf 62}, 114002 (2000)
[arXiv:hep-ph/0001297].
%%CITATION = HEP-PH 0001297;%%


%\cite{Bakulev:2000fb}
\bibitem{Bakulev:2000fb}
A.~P.~Bakulev, S.~V.~Mikhailov, and R.~Ruskov,
%``New shapes of the rho meson light cone distribution amplitudes: How 
%can  they influence the B $\to$ rho e nu decay form factors,''
[arXiv:hep-ph/0006216].
%%CITATION = HEP-PH 0006216;%%


%\cite{Huang:2000hs}
\bibitem{Huang:2000hs}
T.~Huang, Z.~Li, and X.~Wu,
%``Improved approach to the heavy-to-light form factors in the 
%light-cone  QCD sum rules,''
[arXiv:hep-ph/0011161].
%%CITATION = HEP-PH 0011161;%%


%\cite{Wang:2001mi}
\bibitem{Wang:2001mi}
W.~Y.~Wang and Y.~L.~Wu,
%``B $\to$ pi l nu decay and $|$V(ub)$|$,''
Phys.\ Lett.\ B {\bf 515}, 57 (2001)
[arXiv:hep-ph/0105154].
%%CITATION = HEP-PH 0105154;%%


%\cite{Wang:2001bh}
\bibitem{Wang:2001bh}
W.~Y.~Wang and Y.~L.~Wu,
%``B $\to$ rho l nu decay and $|$V(ub)$|$,''
Phys.\ Lett.\ B {\bf 519}, 219 (2001)
[arXiv:hep-ph/0106208].
%%CITATION = HEP-PH 0106208;%%


%\cite{Ball:2001fp}
\bibitem{Ball:2001fp}
P.~Ball and R.~Zwicky,
%``Improved analysis of B $\to$ pi e nu from QCD sum rules on the  
%light-cone,''
JHEP {\bf 0110}, 019 (2001)
[arXiv:hep-ph/0110115].
%%CITATION = HEP-PH 0110115;%%


%
% Quark-Model Papers
%

%\cite{Wirbel:1985ji}
\bibitem{Wirbel:1985ji}
M.~Wirbel, B.~Stech, and M.~Bauer,
%``Exclusive Semileptonic Decays Of Heavy Mesons,''
Z.\ Phys.\ C {\bf 29}, 637 (1985).
%%CITATION = ZEPYA,C29,637;%%


%\cite{Korner:1987kd}
\bibitem{Korner:1987kd}
J.~G.~K\"orner and G.~A.~Schuler,
%``Exclusive Semileptonic Decays Of Bottom Mesons In The Spectator 
%Quark Model,''
Z.\ Phys.\ C {\bf 38}, 511 (1988)
[Erratum-ibid.\ C {\bf 41}, 690 (1988)].
%%CITATION = ZEPYA,C38,511;%%


%\cite{Isgur:gb}
\bibitem{Isgur:gb}
N.~Isgur, D.~Scora, B.~Grinstein, and M.~B.~Wise,
%``Semileptonic B And D Decays In The Quark Model,''
Phys.\ Rev.\ D {\bf 39}, 799 (1989).
%%CITATION = PHRVA,D39,799;%%


%\cite{Scora:1995ty}
\bibitem{Scora:1995ty}
D.~Scora and N.~Isgur,
%``Semileptonic meson decays in the quark model: An update,''
Phys.\ Rev.\ D {\bf 52}, 2783 (1995)
[arXiv:hep-ph/9503486].
%%CITATION = HEP-PH 9503486;%%


%\cite{Melikhov:1995xz}
\bibitem{Melikhov:1995xz}
D.~Melikhov,
%``Form Factors of Meson Decays in the Relativistic Constituent Quark 
%Model,''
Phys.\ Rev.\ D {\bf 53}, 2460 (1996)
[arXiv:hep-ph/9509268].
%%CITATION = HEP-PH 9509268;%%


%\cite{Beyer:1998ka}
\bibitem{Beyer:1998ka}
M.~Beyer and D.~Melikhov,
%``Form factors of exclusive b $\to$ u transitions,''
Phys.\ Lett.\ B {\bf 436}, 344 (1998)
[arXiv:hep-ph/9807223].
%%CITATION = HEP-PH 9807223;%%


%\cite{Faustov:1995bf}
\bibitem{Faustov:1995bf}
R.~N.~Faustov, V.~O.~Galkin, and A.~Y.~Mishurov,
%``Relativistic description of exclusive heavy to light semileptonic 
%decays B $\to$ pi (rho) e neutrino,''
Phys.\ Rev.\ D {\bf 53}, 6302 (1996)
[arXiv:hep-ph/9508262].
%%CITATION = HEP-PH 9508262;%%


%\cite{Demchuk:1997uz}
\bibitem{Demchuk:1997uz}
N.~B.~Demchuk, P.~Y.~Kulikov, I.~M.~Narodetsky, and P.~J.~O'Donnell,
%``Light-front model for exclusive semileptonic B and D decays,''
Phys.\ Atom.\ Nucl.\  {\bf 60}, 1292 (1997)
[Yad.\ Fiz.\  {\bf 60N8}, 1429 (1997)]
[arXiv:hep-ph/9701388].
%%CITATION = HEP-PH 9701388;%%


%\cite{Grach:1996nz}
\bibitem{Grach:1996nz}
I.~L.~Grach, I.~M.~Narodetsky, and S.~Simula,
%``Weak decay form factors of heavy pseudoscalar mesons within a  
%light-front constituent quark model,''
Phys.\ Lett.\ B {\bf 385}, 317 (1996)
[arXiv:hep-ph/9605349].
%%CITATION = HEP-PH 9605349;%%


%\cite{:2000ae}
% NB: This was submitted to PRD in 2000, but does not yet appear as 
published.
\bibitem{:2000ae}
Riazuddin, T.~A.~Al-Aithan, and A.~H.~Gilani,
%``Form factors for B $\to$ pi l nu decay in a model constrained by 
%chiral  symmetry and quark model,''
Int.\ J.\ Mod.\ Phys.\ {\bf A17}, 4927 (2002)
[arXiv:hep-ph/0007164].
%%CITATION = HEP-PH 0007164;%%


%\cite{Melikhov:2000yu}
\bibitem{Melikhov:2000yu}
D.~Melikhov and B.~Stech,
%``Weak form factors for heavy meson decays: An update,''
Phys.\ Rev.\ D {\bf 62}, 014006 (2000)
[arXiv:hep-ph/0001113].
%%CITATION = HEP-PH 0001113;%%


%\cite{Feldmann:1999sm}
\bibitem{Feldmann:1999sm}
T.~Feldmann and P.~Kroll,
%``Skewed parton distributions for B $\to$ pi transitions,''
Eur.\ Phys.\ J.\ C {\bf 12}, 99 (2000)
[arXiv:hep-ph/9905343].
%%CITATION = HEP-PH 9905343;%%


%\cite{Flynn:2000gd}
\bibitem{Flynn:2000gd}
J.~M.~Flynn and J.~Nieves,
%``Form factors for semileptonic B $\to$ pi and D $\to$ pi decays from 
%the  Omnes representation,''
Phys.\ Lett.\ B {\bf 505}, 82 (2001)
[arXiv:hep-ph/0007263].
%%CITATION = HEP-PH 0007263;%%


%\cite{Beneke:2000wa}
\bibitem{Beneke:2000wa}
M.~Beneke and T.~Feldmann,
%``Symmetry-breaking corrections to heavy-to-light B meson form factors 
%at  large recoil,''
Nucl.\ Phys.\ B {\bf 592}, 3 (2001)
[arXiv:hep-ph/0008255].
%%CITATION = HEP-PH 0008255;%%


%\cite{Choi:1999nu}
\bibitem{Choi:1999nu}
H.~M.~Choi and C.~R.~Ji,
%``Light-front quark model analysis of exclusive 0- $\to$ 0- 
%semileptonic  heavy meson decays,''
Phys.\ Lett.\ B {\bf 460}, 461 (1999)
[arXiv:hep-ph/9903496].
%%CITATION = HEP-PH 9903496;%%



%
% Perturbative QCD
%

%\cite{Kurimoto:2001zj}
\bibitem{Kurimoto:2001zj}
T.~Kurimoto, H.-n.~Li, and A.~I.~Sanda,
%``Leading power contributions to B $\to$  pi, rho transition form 
%factors,''
Phys.\ Rev.\ D {\bf 65}, 014007 (2002)
[arXiv:hep-ph/0105003].
%%CITATION = HEP-PH 0105003;%%


%
% ``Experimentally Motivated'' Form Factor Determinations
%

%\cite{Ligeti:1995yz}
\bibitem{Ligeti:1995yz}
Z.~Ligeti and M.~B.~Wise,
%``$$|$V_{ub}$|$$ from exclusive $B$ and $D$ decays,''
Phys.\ Rev.\ D {\bf 53}, 4937 (1996)
[arXiv:hep-ph/9512225].
%%CITATION = HEP-PH 9512225;%%


%\cite{Aitala:1997cm}
% NB: Experimental input to Ligeti:1995yz
\bibitem{Aitala:1997cm}
E.~M.~Aitala {\it et al.}  [E791 Collaboration],
%``Measurement of the form-factor ratios for D+ $\to$ anti-K*0 e+ 
%nu/e,''
Phys.\ Rev.\ Lett.\  {\bf 80}, 1393 (1998)
[arXiv:hep-ph/9710216].
%%CITATION = HEP-PH 9710216;%%


%
% Dispersion Relations
%

%\cite{Burdman:1996kr}
\bibitem{Burdman:1996kr}
G.~Burdman and J.~Kambor,
%``Dispersive Approach to Semileptonic Form-Factors in Heavy-to-Light 
%Meson Decays,''
Phys.\ Rev.\ D {\bf 55}, 2817 (1997)
[arXiv:hep-ph/9602353].
%%CITATION = HEP-PH 9602353;%%


%\cite{Lellouch:1995yv}
\bibitem{Lellouch:1995yv}
L.~Lellouch,
%``Lattice-Constrained Unitarity Bounds for $\bar 
%B~0\to\pi~+\ell~-\bar\nu_\ell$ Decays,''
Nucl.\ Phys.\ B {\bf 479}, 353 (1996)
[arXiv:hep-ph/9509358].
%%CITATION = HEP-PH 9509358;%%


%\cite{Mannel:1998kp}
\bibitem{Mannel:1998kp}
T.~Mannel and B.~Postler,
%``Improved unitarity bounds for anti-B0 $\to$ pi+ l- anti-nu/l 
%decays,''
Nucl.\ Phys.\ B {\bf 535}, 372 (1998)
[arXiv:hep-ph/9805425].
%%CITATION = HEP-PH 9805425;%%

\bibitem{bb:CLEO-nim} Y. Kubota {\it et al.}, Nucl. Instrum. Methods 
Phys.
    Res., Sect. A {\bf 320}, 66 (1992).
\bibitem{bb:silicon-nim} T. S. Hill, Nucl. Instrum. Methods Phys.
    Res., Sect. A {\bf 418}, 32 (1998).
\bibitem{bb:emissres} Spurious CsI showers
from hadronic interactions add linearly to $E_{\text{miss}}$,
but tend to average out in the vector sum for $\vec{P}_{\text{miss}}$.
Incorrect mass assignment also smears $E_{\text{miss}}$.
\bibitem{bb:fox_wolfram} G.~C. Fox and S. Wolfram,
   Phys. Rev. Lett. {\bf 41}, 1581 (1978).

\bibitem{bb:GEANT} R. Brun {\it et al.}, GEANT 3.15, CERN DD/EE/84-1.

\bibitem{bb:inclusive_theory}
   F.~De Fazio and M.~Neubert,
   JHEP {\bf 9906}, 017 (1999)
   [arXiv:hep-ph/9905351]

\bibitem{bb:bsgamm_theor}
   A.~L.~Kagan and M.~Neubert,
   \epjC7, 5 (1999)
   [arXiv:hep-ph/9805303].

\bibitem{bb:bsgamm_exp}
   S. Chen {\em et al.} [CLEO Collaboration],
   Phys. Rev. Lett. {\bf 87}, 251807 (2001)
   [arXiv:hep-ex/0108032].

\bibitem{bb:new_endpoint}
   A. Bornheim {\em et al.} [CLEO Collaboration],
   Phys. Rev. Lett. {\bf 88}, 231803 (2002)
   [arXiv:hep-ex/0202019].

\bibitem{bb:pdg2002}
K. Hagiwara  {\em et al.} [Particle Data Group], Phys.\ Rev.\ D {\bf66},
010001 (2002).

\bibitem{bb:BarlowBeeston}
R.~J.~Barlow and C.~Beeston, Comput. Phys. Commun. {\bf 77}, 219 (1993).

\bibitem{bb:jetset}
T.~Sj\"ostrand,
%``PYTHIA 5.7 and JETSET 7.4: Physics and manual,''
[arXiv:hep-ph/9508391].

\bibitem{silvia}
J. P. Alexander  {\em et al.}  [CLEO collaboration], Phys.\ Rev.\ 
Lett.\  {\bf 86}, 2737 (2001)
[arXiv:hep-ex/0006002].

\bibitem{bb:DiazCruz}
J.~L.~Diaz-Cruz, G.~Lopez Castro, and J.~H.~Munoz,
%``Isospin corrections to charmless semileptonic $B \to V$ 
%transitions,''
Phys.\ Rev.\ D {\bf 54}, 2388 (1996)
[arXiv:hep-ph/9605344].

\bibitem{bb:newBabar}
B.~Aubert {\em et al.}  [BABAR Collaboration],
%``Measurement of the CKM matrix element $|$V(ub)$|$ with B $\to$ rho e 
%nu  decays,''
[arXiv:hep-ex/0301001].  Submitted to PRL.

\bibitem{bb:moneti}
  L.~Gibbons {\em et al.}  [CLEO Collaboration],
%``The inclusive decays B $\to$ D X and B $\to$ D* X,''
Phys.\ Rev.\ D {\bf 56}, 3783 (1997)
[arXiv:hep-ex/9703006].


  \bibitem{bb:Delco}
  R.~M.~Baltrusaitis {\it et al.}  [MARK-III Collaboration],
%``A Direct Measurement Of Charmed D+ And D0 Semileptonic Branching 
%Ratios,''
Phys.\ Rev.\ Lett.\  {\bf 54}, 1976 (1985)
[Erratum-ibid.\  {\bf 55}, 638 (1985)].

\end{thebibliography}
\end{document}